\documentclass[aps,reprint,nofootinbib,nobibnotes,notitlepage,superscriptaddress,onecolumn,prd,
 amsmath,amssymb
]{revtex4-2}

\usepackage[caption=false]{subfig}
\usepackage{braket}
\usepackage{float}
\usepackage{lipsum}
\usepackage{graphicx}
\usepackage{dcolumn}
\usepackage{bm}
\usepackage{natbib}
\usepackage{hyperref}
\hypersetup{
	colorlinks = true,
    linkcolor = Red,
    urlcolor  = Red,
    citecolor = Red
}

\usepackage{microtype}

\usepackage{verbatim}
\usepackage[amssymb]{SIunits}
\usepackage{tabularx}
\usepackage[dvipsnames]{xcolor}
\usepackage{wasysym}
\usepackage[export]{adjustbox}
\usepackage{multirow}
\usepackage{tikz}

\def\be{\begin{equation}}
\def\ee{\end{equation}}
\usepackage{enumitem}

\usepackage{color}
\definecolor{darkgreen}{RGB}{0,120,0}

\definecolor{darkgreen}{RGB}{0,120,0}
\newcommand{\resub}[1]{{#1}}

\newcommand{\av}[1]{\left\langle{#1}\right\rangle} 

\newcommand{\vk}{\vec k}
\newcommand{\hk}{\hat{\vec k}}

\newcommand{\vx}{\vec x}

\newcommand{\Si}{\mathsf{S}^{-1}}
\newcommand{\F}{\mathcal{F}}

\newcommand{\tj}[6]{\begin{pmatrix} {#1} & {#2} & {#3}\\ {#4} & {#5} & {#6}\end{pmatrix}}

\def\beq{\begin{eqnarray}}
\def\eeq{\end{eqnarray}}
\let\vec\mathbf

\usepackage{empheq}

\def\mnras{\rm{MNRAS}}

\def\aap{\rm{A\&A}}

\begin{document}

\title{{\Large Non-Gaussianity Beyond the Scalar Sector:}\\
{\large A Search for Tensor and Mixed Tensor-Scalar Bispectra with \textit{Planck} Data}}

\author{Oliver~H.\,E.~Philcox}
\email{ohep2@cantab.ac.uk}
\affiliation{Department of Physics,
Columbia University, New York, NY 10027, USA}
\affiliation{Department of Physics, Stanford University, Stanford, CA 94305, USA}
\affiliation{Simons Society of Fellows, Simons Foundation, New York, NY 10010, USA}

\author{Maresuke Shiraishi}
\email{shiraishi\_maresuke@rs.sus.ac.jp}
\affiliation{School of General and Management Studies, Suwa University of Science, Chino, Nagano
391-0292, Japan}

\begin{abstract} 
    \noindent Primordial gravitational waves could be non-Gaussian, just like primordial scalar perturbations. Although the tensor two-point function has thus-far remained elusive, the three-point function could, in principle, be large enough to be detected in Cosmic Microwave Background temperature and polarization anisotropies. We perform a detailed analysis of tensor and mixed tensor-scalar non-Gaussianity through the \textit{Planck} PR4 bispectrum, placing constraints on eleven primordial templates, spanning various phenomenological and physical regimes including modifications to gravity, additional fields in inflation, and primordial magnetic fields. All analysis is performed using modern quasi-optimal binned estimators, and yields no evidence for tensor non-Gaussianity, with a maximum detection significance of \resub{$2\sigma$}. Our constraints are derived primarily from large-scales (except for tensor-scalar-scalar models), and benefit greatly from the inclusion of $B$-modes. Although we find some loss of information from binning, mask effects and residual foreground contamination, our $f_{\rm NL}$ bounds improve over those of previous analyses by \resub{$(20-700)\%$}, with six of the eleven models being analyzed for the first time. Unlike for scalar non-Gaussianity, future low-noise experiments such as LiteBIRD, the Simons Observatory and CMB-S4, will yield considerable improvement in tensor non-Gaussianity constraints.
\end{abstract}

\maketitle

\section{Introduction}

\noindent What happened during inflation? The Universe's earliest moments were mostly likely controlled by physical processes operating on energy scales vastly above those probed on Earth. To understand primordial physics, our only hope is to proceed indirectly, searching for low-energy inflationary remnants that persist in the Universe today. Two options present themselves: scalar perturbations, which seeded structure in the late Universe; and tensor perturbations in the form of primordial gravitational waves sourced by vacuum fluctuations or additional fields in inflation. To date, only scalar fluctuations have been detected \citep[e.g.,][]{2020A&A...641A...6P,Planck:2018jri,BICEP:2021xfz}.

A detection of primordial tensor modes is often heralded as a holy grail of modern inflationary cosmology \citep[e.g.,][]{Meerburg:2016nhs,Beutler:2021eqq}. A measurement of the primordial tensor power spectrum (usually parametrized by the tensor-to-scalar ratio, $r$) would reveal the energy scale of inflation itself and rule out many alternative primordial hypotheses \citep[e.g.,][]{Guzzetti:2016mkm}. Much more information could be extracted from the statistical properties of such a signal, in particular its non-Gaussianity. Analogously to the scalar sector, tensor non-Gaussianity is sourced by interactions between multiple inflationary fields and changes to the standard inflationary assumptions, such as slow-roll. At leading order, this information is encoded in the bispectrum, describing the cubic interactions between tensor modes or their cross-correlation with the known scalar degree of freedom. 

In the simplest models of inflation, such signals are too small to be observed \citep{Maldacena:2002vr}. Assuming a single-field scenario endowed with Einsteinian gravity, non-linearities source only a single type of three-point function whose amplitude is highly suppressed ($\sim r^2$) \citep{Maldacena:2011nz}. Furthermore, tensor consistency relations \citep{Maldacena:2002vr} restrict the squeezed limits of bispectra in generic single-field models, even for large $r$. Whilst this may feel a little depressing, it provides great motivation for further study: any detection of the tensor bispectrum would necessarily imply non-standard physics during inflation. A large variety of these type of models exist (as we describe below), which could lead to detectable levels of non-Gaussianity. Moreover, many of these feature bispectra with much larger amplitudes than the power spectra, \textit{i.e.}\ they could be detected \textit{without} a prior detection of the tensor two-point function.

Models containing additional fields (possibly with spin) provide an attractive source of tensor non-Gaussianity \citep{Dimastrogiovanni:2018gkl,Dimastrogiovanni:2022afr,Dimastrogiovanni:2018uqy,Raveendran:2016wjz,Lee:2016vti}. For example, many models of inflation involve couplings of gauge fields to the scalar sector, often through axion-gauge field interactions, which can source large tensor bispectra with peculiar features such as parity-violation and scale-dependent non-Gaussianity \citep{Watanabe:2010fh,Barnaby:2012xt,Lue:1998mq,Sorbo:2011rz,Barnaby:2011vw,Barnaby:2012xt,Cook:2013xea,Namba:2015gja,Dimastrogiovanni:2016fuu,Agrawal:2017awz,Maleknejad:2012fw,Komatsu:2022nvu,Thorne:2017jft,Agrawal:2017awz,Shiraishi:2013kxa,Shiraishi:2016yun,Niu:2022quw,Hiramatsu:2020jes,Chowdhury:2016yrh,Fujita:2018vmv,Ozsoy:2021onx}. An alternative proposition is to modify the gravitational sector by introducing, for example, higher-derivative Weyl terms in the inflationary action \citep{DeLuca:2019jzc,Maldacena:2011nz,Shiraishi:2011st}, Chern-Simons interactions \citep{Lue:1998mq,Alexander:2004wk,Soda:2011am,Shiraishi:2011st,Bartolo:2017szm,Bartolo:2018elp,Mylova:2019jrj,Bartolo:2020gsh,Christodoulidis:2024ric} (see also \citep{Huang:2013epa}), or non-vanishing graviton mass terms \citep{Domenech:2017kno,Fujita:2018ehq,Fujita:2019tov}. One can also invoke the most general inflationary Lagrangian involving only the metric tensor and a scalar field (with second-order field equations), equivalent to the generalized Galileon form \citep{Gao:2011vs,Gao:2012ib}. Furthermore, large bispectra could be formed via non-attractor phases of inflation \citep{Ozsoy:2019slf,Namjoo:2012aa,Martin:2012pe,Christodoulidis:2024ric}, primordial magnetic fields \citep{Shiraishi:2011dh,Shiraishi:2012sn}, non-standard \resub{and excited} vacuum states \citep{Gong:2023kpe,Kanno:2022mkx,Berezhiani:2014kga,Akama:2020jko,Naskar:2020vkd,Peng:2024eok,Christodoulidis:2024ric} (many of which violate the consistency relations \citep{Maldacena:2002vr,Duivenvoorden:2019ses,Meerburg:2016ecv}, but do not peak in flattened configurations, unlike for scalars \citep{Akama:2020jko}), phase transitions \citep{Adshead:2009bz}, (p)reheating \citep{Naskar:2019shl} and beyond. Beyond specific ultra-violet inflationary models, generic predictions for tensor non-Gaussianity have been obtained using both the Effective Field Theory of Inflation and bootstrap methods \citep{Cabass:2021fnw,Cabass:2022jda,Bordin:2020eui,Cabass:2021iii,Pajer:2020wxk,Baumann:2020dch,Naskar:2019shl,Naskar:2018rmu}. In all of these cases, previous papers have obtained specific predictions for the (three-point) inflationary correlators; the goal of this work is to compare such models to observational data.

From an experimental perspective, measuring tensor non-Gaussianity is a daunting task -- as such, it has received significantly less attention than that of scalar non-Gaussianity despite (arguably) equally interesting theoretical motivations. Due to their inherent decay with redshift, gravitational waves can be best detected in Cosmic Microwave Background (CMB) datasets \citep[e.g.,][]{Kamionkowski:2015yta} (though see \citep{Jeong:2012nu,Schmidt:2012nw,Dimastrogiovanni:2014ina,Philcox:2023uor} for examples of late-time probes). Moreover, to avoid contamination from the scalar sector, it is beneficial to search for such signatures in polarization data, particularly $B$-modes, which are not sourced by scalar fluctuations at leading order. Though CMB polarization is more difficult to measure than CMB temperature and comes with its own difficulties such as lensing and galactic dust contamination, it is a fast-evolving area of experimental research, and one which will gain tremendously from future experiments such as the Simons Observatory \citep{SimonsObservatory:2018koc} and LiteBIRD \citep{LiteBIRD:2022cnt}. Constraints on tensor and mixed tensor-scalar non-Gaussianity can be best wrought by considering the full set of $T$-, $E$- and $B$-mode CMB correlators; previous works (using WMAP and \textit{Planck} data \citep{Shiraishi:2014ila,Planck:2015zrl,Planck:2015zfm,Planck:2019kim}, see \citep{Shiraishi:2019yux} for a review) have focused almost exclusively on the first two observables. Inclusion of $B$-modes is crucial if we wish to obtain tight bounds on parity-violating models and will become progressively more important in future experiments, since these are not subject to cosmic variance at leading order (particularly if delensing is applied) \citep[e.g.,][]{Meerburg:2016ecv,Duivenvoorden:2019ses,Shiraishi:2019yux,Shiraishi:2012rm,Shiraishi:2013vha,DeLuca:2019jzc,Shiraishi:2011st,Namba:2015gja,Shiraishi:2016yun,Bartolo:2018elp,Shiraishi:2010kd,Domenech:2017kno,SimonsObservatory:2018koc,Shiraishi:2013kxa,Shiraishi:2012sn,LiteBIRD:2024twk,Tahara:2017wud,Coulton:2019odk}.

In this work, we perform a comprehensive analysis of tensor-tensor-tensor, tensor-tensor-scalar and tensor-scalar-scalar bispectra using the full \textit{Planck} PR4 dataset \citep{Planck:2020olo}, analyzed with modern binned bispectrum estimators, developed in \citep{Philcox:2023uwe,Philcox:2023psd,PolyBin,Philcox4pt2}. This builds on a number of previous works, including the $T$-mode analyses of WMAP \citep{Shiraishi:2013wua,DeLuca:2019jzc,Shiraishi:2014ila,Shiraishi:2017yrq} and $T+E$-mode analyses of \textit{Planck} \citep{Planck:2015zfm,Planck:2019kim,Planck:2015zrl}. In particular, we extend the treatment of \citep{Philcox:2023xxk}, which searched for axion signatures in the \textit{Planck} dataset and, for the first time, included $B$-modes (as well as performing a blind model-independent analysis). In contrast to previous studies, we aim to constrain a large number of models in a consistent manner: here, we employ a set of eleven physical templates that model different inflationary scenarios, with a range of different phenomenological properties, including equilateral and squeezed templates, parity-conservation and parity-violation, and different combinations of scalar and tensor fields. For clarity, these are summarized in Tab.\,\ref{tab: summary}. To our knowledge, only five of the models have been constrained previously and only one with polarization data; our work thus vastly expands the constraints on the tensor sector. Such results, as well as their future extensions, can yield significant insights into primordial physics, and, one hopes, could set the stage for future detections of tensor non-Gaussianity.

\vskip 4pt
The remainder of this text is structured as follows. \S\ref{sec: models} presents the eleven primordial bispectrum templates used in this analysis along with their theoretical motivations and relation to the observed CMB three-point functions. In \S\ref{sec: fisher}, we forecast the constraining power of CMB data on each model using Fisher analyses and derive \resub{(quasi-)}optimal binning and weighting schemes for \textit{Planck} data. \S\ref{sec: analysis} contains details of our analysis pipeline, including the dataset, estimator, and likelihoods, before we present the main results in \S\ref{sec: results}. We conclude in \S\ref{sec: conclusions} with a summary and comparison to previous work. Additional data analysis results are presented in \resub{Appendices \ref{app: extra-plots}\,\&\,\ref{app: weights}}. Our main results are given in Tab.\,\ref{tab: all-results} \resub{\& \ref{tab: all-results-smica}}, which may be compared to the literature constraints listed in Tab.\,\ref{tab:fNL_prev}.

\begin{table}[]
    \centering
    \begin{tabular}{ll||l|l|lr}
    \textbf{Name}& & \textbf{Type} & \textbf{Parity} & \textbf{Motivation} &\\\hline 
    \textbf{\,\,\,\textit{Tensor-Tensor-Tensor}}&&&&&\\
    Squeezed & \eqref{eq:h3_sq} & Squeezed & Even & Phenomenological, e.g., primordial magnetic fields & \citep{Shiraishi:2011dh,Shiraishi:2012rm,Shiraishi:2012sn}\\
    Equilateral & \eqref{eq:h3_eq} & Equilateral & Even & Phenomenological, e.g., general single-field inflation & \citep{Gao:2011vs,Gao:2012ib}\\
    ${W^3(n_{\rm NL}=+1)}$ & \eqref{eq:h3_W3} & Equilateral & Even & Weyl gravity (blue-tilted $k^{-5}$ scaling) & \citep{Maldacena:2011nz,Shiraishi:2011st,DeLuca:2019jzc}\\
    ${W^3(n_{\rm NL}=0)}$ & \eqref{eq:h3_W3} & Equilateral & Even & Weyl gravity (scale-invariant $k^{-6}$ scaling)& \citep{Maldacena:2011nz,Shiraishi:2011st,DeLuca:2019jzc}\\
    ${W^3(n_{\rm NL}=-1)}$ & \eqref{eq:h3_W3} & Equilateral & Even & Weyl gravity (red-tilted $k^{-7}$ scaling)& \citep{Maldacena:2011nz,Shiraishi:2011st,DeLuca:2019jzc}\\
    ${\widetilde{W}W^2(n_{\rm NL}=+1)}$ & \eqref{eq:h3_*WW2} & Equilateral & Odd & Weyl gravity  (blue-tilted $k^{-5}$ scaling)& \citep{Shiraishi:2011st,Soda:2011am}\\
    ${\widetilde{W}W^2(n_{\rm NL}=0)}$ & \eqref{eq:h3_*WW2} & Equilateral & Odd & Weyl gravity (scale-invariant $k^{-6}$ scaling)& \citep{Shiraishi:2011st,Soda:2011am}\\
    ${\widetilde{W}W^2(n_{\rm NL}=-1)}$ & \eqref{eq:h3_*WW2} & Equilateral & Odd & Weyl gravity  (red-tilted $k^{-7}$ scaling) & \citep{Shiraishi:2011st,Soda:2011am}\\
    ${\widetilde{F}F}$ & \eqref{eq:h3_*FF} & Equilateral & Both & Axion -- gauge-field couplings &
   \citep{Cook:2013xea,Shiraishi:2013kxa,Namba:2015gja,Agrawal:2017awz} 
    \\\hline
    \textbf{\,\,\,\textit{Tensor-Tensor-Scalar}}&&&&\\
    ${\widetilde{W}W}$ & \eqref{eq:h2zeta} & Squeezed & Odd & Chern-Simons inflation & \citep{Bartolo:2017szm,Bartolo:2018elp}\\\hline
    \textbf{\,\,\,\textit{Tensor-Scalar-Scalar}}&&&&\\
    Squeezed & \eqref{eq:hzeta2} & Squeezed & Even & Phenomenological, e.g., massive gravity & \citep{Maldacena:2002vr,Domenech:2017kno}
    \end{tabular}
    \caption{Summary of the tensor and mixed tensor-scalar bispectrum templates considered in this work. For each correlator of interest we give its dominant form (equilateral or squeezed), parity properties, theoretical motivation and key references. Previous and new constraints on the characteristic $f_{\rm NL}$ amplitudes are given in Tab.\,\ref{tab:fNL_prev}\,\&\,\ref{tab: all-results}.}\label{tab: summary}
\end{table}

\section{Theoretical Models}\label{sec: models}

\noindent In the simplest models of inflation, namely single-field slow-roll with Einsteinian gravity, non-linear interactions during inflation can produce non-zero tensor-tensor-tensor and tensor-tensor-scalar bispectra. However, slow-roll suppression forces these to have undetectably small amplitudes \cite{Maldacena:2002vr,Adshead:2009bz}. Relaxing these inflationary assumptions, for example through the addition of multiple fields or a non-standard gravitational sector, gives rise to a rich array of primordial bispectra whose amplitudes could, quite plausibly, be measurable. 

In this section, we consider a variety of previously-derived primordial models (both physical and phenomenological, building on \citep{Shiraishi:2019yux}), and present their corresponding bispectrum templates. Later we will constrain their amplitudes using the latest {\it Planck} $T$, $E$ and $B$ datasets. Several models have been analyzed also in previous works; a compilation of existing constraints is shown in Tab.\,\ref{tab:fNL_prev}. All previous constraints will be updated in this work, and we will place the first bounds on a variety of other models. For reference, a summary of the models, their key properties, and notable references is given in Tab.\,\ref{tab: summary}.

\subsection{General Properties of Primordial and CMB Bispectra}

\noindent The scalar primordial curvature perturbation ($\zeta$) and tensor gravitational wave (GW; $h^{(\pm2)}$) can be expressed in Fourier space as
\beq
   \zeta(\vx) \equiv  \int_{\vk}
 \zeta_\vk e^{i \vk \cdot \vx}, \qquad h_{ij}(\vx) \equiv \frac{\delta g_{ij}^{\rm TT}(\vx)}{a^2} = \int_{\vk}
 \sum_{\lambda = \pm 2} h_\vk^{(\lambda)} e_{ij}^{(\lambda)}(\hk)
 e^{i \vk \cdot \vx} ,
\eeq
where $a$ is the scale factor, $\delta g$ is the perturbed metric tensor, $\int_{\vk}\equiv (2\pi)^{-3}\int d^3 \vk$, and the helicity-$\pm 2$ polarization tensor $e_{ij}^{(\pm 2)}$ obeys $e_{ii}^{(\lambda)}(\hk) = \hat{k}_i e_{ij}^{(\lambda)}(\hk) = 0$, $e_{ij}^{(\lambda) *}(\hk) = e_{ij}^{(-\lambda)}(\hk) = e_{ij}^{(\lambda)}(- \hk)$ and $e_{ij}^{(\lambda)}(\hk) e_{ij}^{(\lambda')}(\hk) = 2 \delta_{\lambda, -\lambda'}^{\rm K}$ \cite{Shiraishi:2010kd}. In the following, we focus on inflationary models preserving statistical homogeneity; thus, the bispectrum of curvature perturbations ($\xi_{\vk}^{(0)} \equiv \zeta_\vk$) and GWs ($\xi_{\vk}^{(\pm 2)} \equiv h_\vk^{(\pm 2)}$) can be written
\begin{align}
  \Braket{\prod_{i=1}^3 \xi_{\vk_i}^{(\lambda_i)}} = (2\pi)^3 \delta^{(3)}\left(\sum_{i=1}^3 \vk_i \right) B_{\vk_1 \vk_2 \vk_3}^{\lambda_1 \lambda_2 \lambda_3} . \label{eq:xi3_homo}
\end{align}
where $B_{\vk_1 \vk_2 \vk_3}^{\lambda_1\lambda_2\lambda_3}$ is the bispectrum of three helicity states $\lambda_1,\lambda_2,\lambda_3\in\{0,\pm2\}$.

Under a parity transformation $\vx \to \mathbb{P}[\vx] \equiv - \vx$, the scalar and tensor perturbations transform as $\xi_\vk^{(\lambda)} \to \mathbb{P}[\xi_\vk^{(\lambda)}] = \xi_{- \bf k}^{(-\lambda)}$. Combining this with the reality conditions for $\zeta(\vx)$ and $h_{ij}(\vx)$ (which imply $\xi_\vk^{(\lambda) *} = \xi_{- \bf k}^{(\lambda)}$), one can derive a parity transformation rule for the bispectrum:
\begin{align}
  B_{\vk_1 \vk_2 \vk_3}^{\lambda_1 \lambda_2 \lambda_3}
  \to \mathbb{P}[B_{\vk_1 \vk_2 \vk_3}^{\lambda_1 \lambda_2 \lambda_3}]
  = B_{- \vk_1 - \vk_2 - \vk_3}^{- \lambda_1 - \lambda_2 - \lambda_3}
  = [ B_{\vk_1 \vk_2 \vk_3}^{- \lambda_1 - \lambda_2 - \lambda_3}]^* .
\end{align}
The bispectrum can be decomposed into parity-even and parity-odd components as $B_{\vk_1 \vk_2 \vk_3}^{\lambda_1 \lambda_2 \lambda_3} = B_{\vk_1 \vk_2 \vk_3}^{\lambda_1 \lambda_2 \lambda_3, +} + B_{\vk_1 \vk_2 \vk_3}^{\lambda_1 \lambda_2 \lambda_3, -}$. By definition, these obey $\mathbb{P}[B_{\vk_1 \vk_2 \vk_3}^{\lambda_1 \lambda_2 \lambda_3, \pm }] = \pm B_{\vk_1 \vk_2 \vk_3}^{\lambda_1 \lambda_2 \lambda_3, \pm }$ implying that 
\begin{align}
   B_{\vk_1 \vk_2 \vk_3}^{\lambda_1 \lambda_2 \lambda_3, \pm }
  = \pm B_{- \vk_1 - \vk_2 - \vk_3}^{- \lambda_1 - \lambda_2 - \lambda_3, \pm}
  = \pm [ B_{\vk_1 \vk_2 \vk_3}^{- \lambda_1 - \lambda_2 - \lambda_3, \pm}]^*,
\end{align}
\textit{i.e.}\ parity-even (parity-odd) bispectra are related to those of opposite helicities by a conjugate and a factor of $+1$ ($-1$). An important conclusion is that any asymmetry of the bispectrum under a helicity flip (\textit{i.e.}\ $B_{\vk_1 \vk_2 \vk_3}^{\lambda_1\lambda_2\lambda_3}\neq \left[B_{\vk_1 \vk_2 \vk_3}^{-\lambda_1–\lambda_2-\lambda_3}\right]^*$) gives rise to parity violation in the bispectrum, sourcing non-zero $B_{\vk_1 \vk_2 \vk_3}^{\lambda_1 \lambda_2 \lambda_3, -}$.\footnote{In correlators involving only scalars (whence $\lambda_i=0$), parity violation would correspond to an imaginary bispectrum component; such a signal, however, cannot be present in two and three-point correlators if statistical isotropy is preserved \cite{Shiraishi:2016mok}.}

\begin{table}[t]
\begin{center}
  \begin{tabular}{ll||c||c|c||c|c|c} 
  \multirow{2}{*}{} & \multirow{2}{*}{} & WMAP & \multicolumn{2}{c||}{\textit{Planck} 2015, 2018} &  \multicolumn{3}{c}{\textit{Planck} PR4}  \\ \cline{3-8}
  & & T & T & T+E & T & T+E & T+E+B \\ \hline
  \textbf{\,\,\,\textit{Tensor-Tensor-Tensor}}&&&&&&\\
  Squeezed & ($\times 10^{-1}$) & $ 22 \pm 17$ \cite{Shiraishi:2013wua} & $29 \pm 18$ \cite{Planck:2015zrl} & - & - & - & - \\
   $W^3( n_{\rm NL} = 0)$ & ($\times 10^{-2}$) &$ 3 \pm 9$ \cite{DeLuca:2019jzc} & - & - & - & - & - \\
  $\widetilde{W}W^2(n_{\rm NL} = -1)$ & ($\times 10^{0}$) & $140 \pm 140$ \cite{Shiraishi:2014ila} & - & - & - & - & - \\
  $\widetilde{F}F$ & ($\times 10^{-2}$) & $6 \pm 15$ \cite{Shiraishi:2014ila} & $ 6 \pm 16$ \cite{Planck:2019kim} & $ 8 \pm 11$ \cite{Planck:2019kim} & $5 \pm 20$ \cite{Philcox:2023xxk} & $10 \pm 10$ \cite{Philcox:2023xxk} & $9 \pm 7$ \cite{Philcox:2023xxk} \\
  \hline
  \textbf{\,\,\,\textit{Tensor-Scalar-Scalar}}&&&&&&\\
  Squeezed & ($\times 10^{0}$) & $ 84 \pm 49$ \cite{Shiraishi:2017yrq} & - & - & - & - & - 
  \end{tabular} 
\end{center}
\caption{Previous constraints on tensor and mixed bispectrum amplitudes from WMAP and {\it Planck} data (with some combination of $T$-, $E$- and $B$-modes). We chose scaling exponents to allow for direct comparison with Tab.\,\ref{tab: all-results}, which contains the new results derived in this work. A summary of each model can be found in Tab.\,\ref{tab: summary} and \S\ref{sec: models}. Almost all models are analyzed using modal bispectrum decompositions \citep[e.g.,][]{Shiraishi:2019exr,Shiraishi:2014roa,2009PhRvD..80d3510F}, except for the PR4 analyses, which use binned estimators \citep{Philcox:2023psd} (with information loss leading to slightly wide $T$-only constraints).}\label{tab:fNL_prev}
\end{table}

The curvature perturbation and GW source CMB temperature and polarization anisotropies. At linear order, their harmonic coefficients, $a_{\ell m}^{X}$ for $X \in \{T,E,B\}$, are given as \cite{Shiraishi:2010sm,Shiraishi:2010kd}
\beq
  a_{\ell m}^{T/E} &=& 4\pi i^{\ell} \int_{\vk}
\left[ {}_s {\cal T}_{\ell}^{T/E}(k)\zeta_\vk  Y_{\ell m}^*(\hk) 
+ {}_t {\cal T}_{\ell}^{T/E}(k) \left(h_\vk^{(+2)}{}_{-2} Y_{\ell m}^*(\hk)+h_\vk^{(-2)}{}_{+2} Y_{\ell m}^*(\hk)\right)\right]\\\nonumber 
    a_{\ell m}^{B} &=& 
4\pi i^{\ell} \int_{\vk}
 {}_t {\cal T}_{\ell}^{B}(k) \left(h_\vk^{(+2)} {}_{-2} Y_{\ell m}^*(\hk)-h_\vk^{(-2)} {}_{+2} Y_{\ell m}^*(\hk)\right) , 
\label{eq:alm}
\eeq
where ${}_{s/t} {\cal T}_{\ell}^{X}$ is the scalar/tensor-mode linear transfer function, and ${}_s Y_{\ell m}$ is a spin-weighted spherical harmonic. Notably, $T$- and $E$-modes trace both scalars and the sum of tensor helicity states, whilst $B$-modes trace only the difference of tensor helicity states. Under a parity transformation, the harmonic coefficients transform as $a_{\ell m}^{T/E} \to {\mathbb{P}}[a_{\ell m}^{T/E}] = (-1)^{\ell} a_{\ell m}^{T/E}, a_{\ell m}^{B} \to {\mathbb{P}}[a_{\ell m}^{B}] = -(-1)^{\ell} a_{\ell m}^{B}$, thus, the CMB bispectrum transforms as
\begin{align}
 \Braket{\prod_{i=1}^3 a_{\ell_i m_i}^{X_i} } \to {\mathbb{P}}\left[\Braket{\prod_{i=1}^3 a_{\ell_i m_i}^{X_i} }\right] = (-1)^{\ell_1 + \ell_2 + \ell_3 + n_B} \Braket{\prod_{i=1}^3 a_{\ell_i m_i}^{X_i} } .
\end{align}
where $n_B$ is the number of $B$-modes in the statistic. As such, we find the following properties \citep{Shiraishi:2011st,Shiraishi:2016ads,Philcox:2023xxk,Kamionkowski:2010rb}:
\begin{itemize}
\item Parity-even primordial bispectra source non-zero CMB bispectra with even $\ell_1+\ell_2+\ell_3$ if there are zero or two $B$-modes, and odd $\ell_1+\ell_2+\ell_3$ else.
\item Parity-odd primordial bispectra source non-zero CMB bispectra with odd $\ell_1+\ell_2+\ell_3$ if there are zero or two $B$-modes, and even $\ell_1+\ell_2+\ell_3$ else.
\end{itemize}

\subsection{Tensor-Tensor-Tensor Bispectra}\label{subsec: ttt-bispectra}

\subsubsection{A Squeezed, Scale-Invariant \& Parity-Conserving Model}

\noindent Perhaps the simplest bispectrum model is a parity-even correlator peaking in the squeezed configuration $k_1 \ll k_2 \sim k_3$ (and permutations). According to the Maldacena consistency relation, this is vanishingly small in single-field inflation \cite{Maldacena:2002vr}. In contrast, the presence of sources such as primordial magnetic fields or inflationary massive fields can make the squeezed signal visibly large \cite{Shiraishi:2011dh,Shiraishi:2012rm,Shiraishi:2012sn,Dimastrogiovanni:2018gkl,Dimastrogiovanni:2022afr}. A convenient template for parity-even scale-invariant squeezed bispectra is given by
\begin{align}
  B_{\vk_1 \vk_2 \vk_3}^{\lambda_1 \lambda_2 \lambda_3}
  = \sqrt{2} f_{{\rm NL}}^{ttt, {\rm sq}} \,\times\,B^{\rm sq}(k_1,k_2,k_3)\,\times\,
  e_{ij}^{(-\lambda_1)} (\hk_1) e_{jk}^{(-\lambda_2)} (\hk_2)
  e_{ki}^{(-\lambda_3)} (\hk_3) , \label{eq:h3_sq}
\end{align}
\cite{Shiraishi:2019yux}, where $\lambda_i\in\{\pm2\}$ and $B^{\rm sq}(k_1,k_2,k_3)$ is the well-known local-type bispectrum template:
\begin{align}
B^{\rm sq}(k_1,k_2,k_3)
\equiv \frac{6}{5}(2\pi^2 A_s)^2
\left[\frac{1}{k_1^3 k_2^3} + 2 \, {\rm perms.} 
  \right].
\end{align}
Here, $A_s$ the scalar spectral amplitude and $f_{{\rm NL}}^{ttt, {\rm sq}}$ is an amplitude parameter, normalized according to
\begin{align}
  f_{{\rm NL}}^{ttt, {\rm sq}} = \lim_{\substack{ k_{1,2} \to k \\ k_3 \to 0}} \frac{B_{\vk_1 \vk_2 \vk_3}^{+2 +2 +2}}{B^{\rm sq}(k_1,k_2,k_3)},
\end{align}
choosing the $\vk_2$($\approx-\vk_3$) vector to be maximally aligned with the polarization tensor as in \citep{Meerburg:2016ecv}. 

If some form of magnetogenesis occurs around at the Grand Unified Theory epoch, magnetic fields are expected to be present on ultra-large scales, potentially seeding intergalactic magnetic fields at late times \citep[e.g.,][]{Kulsrud:2007an}. By the time of neutrino decoupling, these can quadratically generate GWs on superhorizon scales due to their inherent anisotropic stress \cite{Paoletti:2008ck,Shaw:2009nf}. Assuming Gaussianity of primordial magnetic fields and nearly scale invariance of their power spectrum, this leads to a model well-approximated by \eqref{eq:h3_sq} with
\begin{align}\label{eq: pmf-scaling}
  f_{{\rm NL}}^{ttt, {\rm sq}} \simeq \left( \frac{B_{1 \rm \, Mpc}}{1 \, \rm nG} \right)^6 ,
\end{align}
where $B_{1 \rm \, Mpc}$ is a magnetic field strength smoothed on $1 \, {\rm Mpc}$ \cite{Shiraishi:2011dh,Shiraishi:2012rm,Shiraishi:2013vha} (with $B_{1\rm\,Mpc}\lesssim \mathcal{O}(1\,\mathrm{nG})$ implied from \textit{Planck} two-point analyses \citep{Pogosian:2018vfr,Planck:2015zrl}). Helical primordial magnetic fields could also arise; these would produce a similar spectrum to \eqref{eq:h3_sq} but with odd parity \citep{Shiraishi:2012sn}.

CMB bispectra generated from \eqref{eq:h3_sq} have large signals for squeezed configurations $\ell_1 \ll \ell_2 \sim \ell_3$ (and permutations), which dominate the overall signal-to-noise \cite{Shiraishi:2012rm,Shiraishi:2013vha,Shiraishi:2019yux} (though one requires $\ell_{2,3}\lesssim 100$ to avoid the decay in the tensor transfer function). Such a model has been analyzed using WMAP and \textit{Planck} temperature data \citep{Shiraishi:2013wua,Planck:2015zrl}; these constraints (consistent with zero at $2\sigma)$ are summarized in Tab.\,\ref{tab:fNL_prev}, and correspond to $B_{1 \rm \, Mpc} \lesssim 3\,\mathrm{nG}$.

\subsubsection{An Equilateral, Scale-Invariant \& Parity-Conserving Model}

\noindent In the presence of deviations from the simplest inflationary model, dominant bispectrum signals can arise also from equilateral configurations with $k_1 \sim k_2 \sim k_3$. Assuming scale-invariance and parity-conservation, a simple equilateral bispectrum template is given by
\begin{align}
  B_{ \vk_1 \vk_2 \vk_3}^{\lambda_1 \lambda_2 \lambda_3}
  = 
\frac{16\sqrt{2}}{27} f_{\rm NL}^{ttt, {\rm eq}}\,\times\,
B^{\rm eq}(k_1,k_2,k_3)\,\times\,
e_{ij}^{(-\lambda_1)} (\hk_1) e_{jk}^{(-\lambda_2)} (\hk_2)
  e_{ki}^{(-\lambda_3)} (\hk_3) , \label{eq:h3_eq}
  \end{align}
where $B^{\rm eq}(k_1,k_2,k_3)$ is the well-known equilateral-type bispectrum template:
\begin{align}
B^{\rm eq}(k_1,k_2,k_3)
&\equiv \frac{18}{5}(2\pi^2 A_s)^2
\left[
    - \left(\frac{1}{k_1^3 k_2^3} + 2 \, {\rm perms.}  \right)
    - \frac{2}{k_1^2 k_2^2 k_3^2}
+ \left( \frac{1}{k_1 k_2^2 k_3^3} + 5 \, {\rm perms.} \right)
\right] , 
\label{eq: b-eq-scalar}
\end{align}
and the amplitude parameter is defined by\footnote{
Note that this definition of $f_{\rm NL}^{ttt, {\rm eq}}$ differs from that found in \cite{Shiraishi:2019yux}. Theirs (which refers to a chiral axion model, and is equivalent to $f_{\rm NL}^{\rm tens}$ of \citep{Planck:2015zfm,Planck:2019kim}) is equivalent to our $f_{\rm NL}^{ttt,\widetilde{F}F}$ in \eqref{eq:h3_*FF}.}
\begin{align}
f_{\rm NL}^{ttt, {\rm eq}} &= \lim_{k_{1,2,3} \to k}
\frac{B_{ \vk_1 \vk_2 \vk_3}^{+2 +2 +2}}{B^{\rm eq}(k_1,k_2,k_3)} . \label{eq:fNLttteq_def}
\end{align}

A physical manifestation of this arises in the generalized Galileon inflation model (as well as multi-field inflationary models, in certain limits \citep{Dimastrogiovanni:2018gkl}), which sets out the most general Lagrangian involving only the metric tensor and a scalar field with second-order field equations (equivalent to the Horndeski form \citep{Horndeski:1974wa}) \citep{Kobayashi:2011nu,Deffayet:2011gz}. This encompasses a variety of modifications to general relativity and can source both squeezed and equilateral tensor-tensor-tensor bispectra (as well as mixed and scalar forms) \cite{Gao:2011vs,Gao:2012ib,Tahara:2017wud}, with the latter taking the form of \eqref{eq:h3_eq} with amplitude parameter
\begin{align}
  f_{\rm NL}^{ttt, \rm eq} \simeq 1.9 \times 10^{-3} r^2 \frac{H\mu}{{\cal G}_T}
\end{align}
where $r$ is the tensor-to-scalar ratio, $H$ is the Hubble parameter and $\mu /{\cal G}_T$ is a function of the scalar field and coupling strength parameters.

CMB bispectra generated from \eqref{eq:h3_eq} exhibit large signals for equilateral configurations $\ell_1 \sim \ell_2 \sim \ell_3$, which dominate the signal-to-noise ratios. To date, there are so far no published observational constraints on such a model.

\subsubsection{Equilateral \& Scale-Dependent Templates Motivated by Weyl Gravity}

\noindent A natural extension to the simplest inflationary action is provided by the higher-derivative interaction $W^3$, involving the Weyl tensor, $W$ \cite{Maldacena:2011nz} (see also \citep{Huang:2013epa}). This generates parity-even tensor-tensor-tensor bispectra which peak in the equilateral limit and is described by the template
\begin{align}
  B_{ \vk_1 \vk_2 \vk_3}^{\lambda_1 \lambda_2 \lambda_3} 
  = 
\frac{32\sqrt{2}}{15} f_{{\rm NL}}^{ttt, W^3(n_{\rm NL})}\,\times\,
 (2\pi^2 A_s)^2 \left( \frac{3}{k_t} \right)^6
\left( \frac{k_t}{3k_*} \right)^{n_{\rm NL}}\,\times\,
e_{ij}^{(-\lambda_1)} (\hk_1) e_{jk}^{(-\lambda_2)} (\hk_2)
e_{ki}^{(-\lambda_3)} (\hk_3)\,\times\,
  \delta^{\rm K}_{\lambda_1, \lambda_2} \delta^{\rm K}_{\lambda_2, \lambda_3} , \label{eq:h3_W3}
\end{align}
with $k_t \equiv k_1 + k_2 + k_3$, where $k_*$ is some pivot scale, which we here fix to $k_* = 0.05\,\mathrm{Mpc}^{-1}$.\footnote{This form is highly correlated ($>95\%$) with that obtained by replacing $(3/k_t)^6$ with the (separable) equilateral template of \eqref{eq: b-eq-scalar}. In practice, we adopt the latter form for computational efficiency (and proceed similarly for the parity-odd case). We further note that this differs from \eqref{eq:h3_eq} even for $n_{\rm NL}=0$ due to the different helicity restrictions.} Note that this is non-zero only if $\lambda_1=\lambda_2=\lambda_3$. Here, we have defined the magnitude
\begin{align}
f_{\rm NL}^{ttt, W^3(n_{\rm NL})} = \lim_{k_{1,2,3} \to k_*}
\frac{B_{ \vk_1 \vk_2 \vk_3}^{+2 +2 +2}}{B^{\rm eq}(k_1,k_2,k_3)} , \label{eq:fNLtttW3_def}
\end{align}
and the spectral tilt, $n_{\rm NL}$, which depend on the coupling of the $W^3$ term. If this is constant in time, a scale-invariant spectrum is realized \cite{Maldacena:2011nz} with $n_{\rm NL} = 0$. In contrast, a time-dependent coupling function will source scale-dependence in the bispectrum. Assuming the coupling function can be written $\Lambda_{W^3}^{-2} (\tau / \tau_*)^n$ for conformal time $\tau$ and energy scale $\Lambda_{W^3}$, the bispectrum takes the form of \eqref{eq:h3_W3} in the de Sitter limit with
\begin{align}
  f_{{\rm NL}}^{ttt, W^3(n_{\rm NL}=-n)} \simeq \frac{5\sqrt{2} \pi ^4 }{ 10368 } \cos \left(\frac{n \pi}{2} \right) \Gamma (6+n) (-3 k_* \tau_*)^{-n} A_s^2 r^4 \left( \frac{M_p}{\Lambda_{W^3}} \right)^2 
\end{align}
and $n_{\rm NL} = -n$, where $M_p$ the reduced Planck mass \cite{Shiraishi:2011st}.

It is also possible to have cubic interactions involving the dual of the Weyl tensor, \textit{i.e.} $\widetilde{W}W^2$.\footnote{In the exact de Sitter limit, $W^3$ and $\widetilde{W}W^2$ are the \textit{only} GW-sourcing modifications allowed in single-field inflation \citep{Maldacena:2011nz}.} This induces a similar bispectrum with opposite parity, which can be expressed as
\begin{align}
  B_{ \vk_1 \vk_2 \vk_3}^{\lambda_1 \lambda_2 \lambda_3} 
  = 
  \frac{32\sqrt{2}}{15} f_{{\rm NL}}^{ttt, \widetilde{W}W^2(n_{\rm NL})}\,\times\,
   (2\pi^2 A_s)^2 \left( \frac{3}{k_t} \right)^6
\left( \frac{k_t}{3k_*} \right)^{n_{\rm NL}}\,\times\,
  e_{ij}^{(-\lambda_1)} (\hk_1) e_{jk}^{(-\lambda_2)} (\hk_2)
  e_{ki}^{(-\lambda_3)} (\hk_3)\,\times\,
\frac{\lambda_1}{2}
  \delta^{\rm K}_{\lambda_1, \lambda_2} \delta^{\rm K}_{\lambda_2, \lambda_3} , \label{eq:h3_*WW2}
  \end{align}
where the amplitude parameter is again normalized via \eqref{eq:fNLtttW3_def}. We note the antisymmetry under $\lambda_1\to -\lambda_1$ transformations, encoding parity asymmetry. As for the $W^3$ case, the bispectrum's shape and amplitude are controlled by the coupling of $\widetilde{W}W^2$ in the action. Assuming a coupling of the form $\Lambda_{\widetilde{W}W^2}^{-2}(\tau / \tau_*)^n$, the bispectrum in the de-Sitter limit is given by \eqref{eq:h3_*WW2} with
\begin{align}
  f_{{\rm NL}}^{ttt, \widetilde{W}W^2(n_{\rm NL}=-n)} \simeq \frac{5\sqrt{2} \pi ^4}{5184} \sin \left( \frac{n \pi}{ 2} \right) \Gamma (6+n) (-3 k_* \tau_*)^{-n} A_s^2 r^4 \left( \frac{M_p}{\Lambda_{\widetilde{W}W^2}} \right)^2
\end{align}
and $n_{\rm NL} = -n$ \cite{Shiraishi:2011st}. For a constant coupling case ($n = 0$), the bispectrum vanishes in the de-Sitter limit; however, a contribution is sourced at first order in slow roll \cite{Soda:2011am}.

If the primordial bispectra~\eqref{eq:h3_W3} and \eqref{eq:h3_*WW2} are not so tilted, the associated CMB bispectra are dominated by equilateral configurations with $\ell_1 \sim \ell_2 \sim \ell_3$ \cite{Shiraishi:2011st,DeLuca:2019jzc}. To date, only two cases have been compared to data (from WMAP temperature data), which are given in Tab.\,\ref{tab:fNL_prev}. Forecasts indicate that polarization data can strongly aid such constraints \citep{DeLuca:2019jzc}, with LiteBIRD $B$-modes poised to reduce the errorbars by orders of magnitude compared to temperature alone.

\subsubsection{An Equilateral \& Scale-Invariant Chiral Template Motivated by Axions}
\noindent The presence of pseudo-scalar / axionic fields during inflation generates a similar bispectrum. Assuming the classic coupling of the axion $\phi$ to a gauge field electromagnetic tensor $F$ via the axial interaction $\phi \widetilde{F}F$, we can source GWs in a quadratic (non-linear) process \cite{Sorbo:2011rz,Barnaby:2011vw,Barnaby:2012xt,Namba:2015gja,Dimastrogiovanni:2016fuu}. Since (a) the gauge field becomes maximally chiral and (b) the gauge quanta production becomes most efficient when its Fourier mode crosses the horizon, the induced tensor-tensor-tensor bispectrum is maximally helical and enhanced for equilateral configurations \cite{Cook:2013xea,Namba:2015gja,Agrawal:2017awz,Agrawal:2018mrg}. Assuming that the axion rolls slowly down a not-so-steep potential for a not-so-small period, the bispectrum is nearly scale-invariant\footnote{
See \cite{Namba:2015gja,Dimastrogiovanni:2016fuu,Shiraishi:2016yun} for cases inducing signals with strong scale dependence.} and therefore well fitted by a slight modification to \eqref{eq:h3_eq} \cite{Shiraishi:2013kxa,Shiraishi:2014ila,Planck:2015zfm,Planck:2019kim,Shiraishi:2019yux}:
\begin{align}
  B_{ \vk_1 \vk_2 \vk_3}^{\lambda_1 \lambda_2 \lambda_3}
  = 
\frac{16\sqrt{2}}{27} f_{\rm NL}^{ttt, \widetilde{F}F}\,\times\,
B^{\rm eq}(k_1,k_2,k_3)\,\times\,
e_{ij}^{(-\lambda_1)} (\hk_1) e_{jk}^{(-\lambda_2)} (\hk_2)
  e_{ki}^{(-\lambda_3)} (\hk_3)\,\times\,\delta^{\rm K}_{\lambda_1, +2} \delta^{\rm K}_{\lambda_2, +2} \delta^{\rm K}_{\lambda_3, +2} , \label{eq:h3_*FF}
\end{align}
where we have chosen $\lambda = +2$ as a surviving GW helicity without loss of generality. This template can be realized as a linear combination of the $W^3(n_{\rm NL}=0)$ and $\widetilde{W}W^2(n_{\rm NL}=0)$ forms;\footnote{Rewriting the Weyl bispectra in terms of the equilateral templates as before, this is given explicitly by $B^{\widetilde{F}F}=0.5291[B^{W^3(n_{\rm NL}=0)}+B^{\widetilde{W}W^2(n_{\rm NL}=0)}]$, with all quantities evaluated at $f_{\rm NL}=1$.} as such, we avoid a joint inference of all three templates in the analyses below.

The particular value of the amplitude parameter, $f_{\rm NL}^{ttt, \widetilde{F}F}$ (which is normalized according to \eqref{eq:fNLttteq_def}) depends on the form of the axial coupling. For a $U(1)$ gauge field,
\begin{align}
f_{\rm NL}^{ttt, \widetilde{F}F} \simeq 6.4 \times 10^{11} A_s^3 \epsilon^3 \frac{e^{6\pi \xi}}{ \xi^9 } ,
\end{align}
where $\epsilon$ is the inflaton slow-roll parameter and $\xi$ is a parameter proportional to the axion rolling speed and the axial coupling constant \cite{Cook:2013xea,Shiraishi:2013kxa,Planck:2015zfm,Planck:2019kim}. For an SU(2) coupling, 
\begin{align}
  f_{\rm NL}^{ttt, \widetilde{F}F} \simeq 2.5 \frac{r^2}{\Omega_A} ,
\end{align}
where $\Omega_A$ is the energy fraction of the gauge field and $r$ is the tensor-to-scalar ratio \cite{Agrawal:2017awz,Agrawal:2018mrg}.

CMB bispectra generated from \eqref{eq:h3_*FF} are dominated by equilateral configurations with $\ell_1 \sim \ell_2 \sim \ell_3$ \cite{Shiraishi:2013kxa,Shiraishi:2019yux}. In contrast to the previous forms, both parity-even and parity-odd spectra are generated with equal magnitudes. These models have been analyzed before: previous constraints from WMAP and {\it Planck} data (involving $T$-, $E$- and $B$-modes) are summarized in Tab.\,\ref{tab:fNL_prev} \cite{Shiraishi:2014ila,Planck:2015zfm,Planck:2019kim,Shiraishi:2019yux,Philcox:2023xxk}. Future polarization experiments such as LiteBIRD are also poised to yield significant improvements from $B$-mode spectra, potentially reaching $\mathcal{O}(1)$ if $r$ is small (and thus cosmic variance is suppressed) \citep{Shiraishi:2013kxa,Shiraishi:2019yux}.

\subsection{Tensor-Tensor-Scalar Bispectra}\label{subsec: tts-model}

\noindent Chern-Simons models of gravity may contain the Weyl quadratic interaction $\widetilde{W}W$. If this has a time-dependent coupling, chiral GWs are produced \cite{Bartolo:2017szm,Bartolo:2018elp}. Within the section of parameter space free from theoretical issues such as ghost instabilities, the tensor-tensor power spectrum, tensor-tensor-tensor bispectrum and tensor-scalar-scalar bispectrum are highly suppressed; in contrast, the tensor-tensor-scalar bispectrum could be large \cite{Bartolo:2017szm,Bartolo:2018elp,Bordin:2020eui} (\resub{and, if the vacuum is taken to be an excited state, peaked in the flattened configuration} \citep{Christodoulidis:2024ric}).%
\footnote{This also sources a scalar four-point function; see \cite{CyrilCS} for a discussion of this signal.} This has odd parity and peaks in the squeezed limit, which involves two short-wavelength tensors and one long-wavelength scalar ($k_1 \sim k_2 \gg k_3$). We find an antisymmetric bispectrum with $\lambda_1=\lambda_2$ (noting that $\lambda_3=0$ since we have a scalar field):
\begin{align} 
B_{ \vk_1 \vk_2 \vk_3}^{\lambda_1 \lambda_2 0}
  &= - \frac{3}{5} f_{\rm NL}^{tts, \widetilde{W}W}\,\times\,
   (2 \pi^2 A_s)^2 
\frac{k_1 + k_2}{k_1^2 k_2^2 k_3^3}
 (\hk_1 \cdot \hk_2)\,\times\,e_{ij}^{(-\lambda_1)}(\hk_1) e_{ij}^{(-\lambda_2)}(\hk_2) \,\times\,\frac{\lambda_1}{2}\delta^{\rm K}_{\lambda_1, \lambda_2} , \label{eq:h2zeta}
\end{align}
\citep{Bartolo:2018elp} where the amplitude parameter is normalized according to
\begin{align}
  f_{\rm NL}^{tts, \widetilde{W}W} 
  \equiv \lim_{ \substack{k_{1,2} \to k \\ k_3 \to 0}} 
  \frac{B_{ \vk_1 \vk_2 \vk_3}^{+2 +2 0 }}{B^{\rm sq}(k_1,k_2,k_3)} .
  \end{align}
If one assumes the coupling of $\widetilde{W}W$ to be a function of the inflaton field, $g_{\widetilde{W}W}(\phi)$, the amplitude can be expressed as
\begin{align}
  f_{\rm NL}^{tts, \widetilde{W}W} \simeq \frac{5\pi^3}{192} A_s r^3 M_p^2 \frac{\partial^2 g_{\widetilde{W}W}}{\partial \phi^2}.
\end{align}
It is also possible to form a (parity-even) tensor-tensor-scalar bispectra from standard Einsteinian gravity \citep{Maldacena:2002vr,Baumann:2020dch}; however, this is highly suppressed by the consistency relation, thus we do not consider it in this work. Furthermore, one could have higher-derivative corrections from $W^2$ terms, as well as from multi-field inflation (for example with a chiral axion -- gauge-field coupling); these are discussed in \citep{Baumann:2020dch,Dimastrogiovanni:2022afr,Dimastrogiovanni:2018xnn}.

CMB bispectra generated from \eqref{eq:h2zeta} are dominated by squeezed configurations $\ell_1 \sim \ell_2 \gg \ell_3$, particularly those involving two $B$-modes (noting the third field is a scalar), as demonstrated in the forecasts of \cite{Bartolo:2018elp}. So far there are no published observational constraints. 

\subsection{Tensor-Scalar-Scalar Bispectra}

\noindent In single-field, slow-roll, and Einsteinian models of inflation, a parity-even tensor-scalar-scalar bispectrum is sourced which peaks at the squeezed limit, corresponding to one long-wavelength tensor and two short-wavelength scalars ($k_1 \ll k_2 \sim k_3$). This takes the following form
\begin{align}
B_{ \vk_1 \vk_2 \vk_3}^{\lambda_1 0 0}
&= -\frac{8 \sqrt{2}}{5} f_{\rm NL}^{tss, \rm sq}\,\times\,(2\pi^2 A_s)^2 
\frac{1}{k_1^3 k_2^2 k_3^2} \left[  -k_t + \frac{k_1 k_2 + k_2 k_3 + k_3 k_1}{k_t} + \frac{k_1 k_2 k_3}{k_t^2}  \right] \,\times\,
e_{ij}^{(-\lambda_1)}(\hk_1) \hat{k}_{2i} \hat{k}_{3j} , \label{eq:hzeta2}
\end{align}
\citep[e.g.,][]{Shiraishi:2010kd,Mata:2012bx,Maldacena:2002vr,Baumann:2020dch,Domenech:2017kno,Meerburg:2016ecv,Shiraishi:2019yux}, where the amplitude parameter is normalized according to\footnote{Our parameter is related to $g_{tss}$ from \citep{Shiraishi:2010kd,Shiraishi:2017yrq}, $f^{h\zeta\zeta}_{\rm NL}$ from \citep{Meerburg:2016ecv}, $f^{\rm tot}_{\rm NL}$ from \citep{Duivenvoorden:2019ses}, and $\lambda_{sst}\epsilon$ from \citep{Domenech:2017kno} by 
$-2\sqrt{2}f_{\rm NL}^{tss,\rm sq}/5 = g_{tss} = \sqrt{r}f^{h\zeta\zeta}_{\rm NL} = f^{\rm tot}_{\rm NL} = -\lambda_{sst} \epsilon$, equal to $r/16$ in the single-field slow-roll limit \citep{Maldacena:2002vr}.}
\begin{align}
  f_{\rm NL}^{tss, \rm sq} 
  \equiv \lim_{ \substack{k_1 \to 0 \\ k_{2,3} \to k}} 
  \frac{B_{ \vk_1 \vk_2 \vk_3}^{+2 0 0 }}{B^{\rm sq}(k_1,k_2,k_3)}.
\end{align}
As in the tensor-tensor-tensor case, the expected value of $f_{\rm NL}^{tss,\rm sq}$ is slow-roll suppressed and practically unmeasurable \cite{Maldacena:2002vr}. However, modifications to the particle or gravity sector can lead to large signals, with notable examples being massive gravity \citep{Domenech:2017kno,Fujita:2018ehq} (which also sources a tensor-tensor-tensor spectrum akin to \eqref{eq:h3_sq} \citep{Fujita:2019tov}) or axion monodromy \citep{Chowdhury:2016yrh}. The amplitude parameter can be shown to take the approximate form
\beq
    f_{\rm NL}^{tss,\rm sq} \sim  0.1r\quad\text{(Einstein)}\qquad f_{\rm NL}^{tss,\rm sq} \sim 0.1r\lambda_{sst}\quad\text{(Massive Gravity)},
\eeq
with the former being clearly unobservable for reasonable values of $r$. The latter depends on the strength of a non-trivial non-linear interaction, $\lambda_{sst}$, which could be large enough to facilitate practical observation \citep{Domenech:2017kno}. 

CMB bispectra generated from \eqref{eq:hzeta2} display large signals for squeezed configurations $\ell_1 \ll \ell_2 \sim \ell_3$, particularly from $TTB$, which does not suffer from cosmic variance at leading order (subject to sufficient delensing) \cite{Shiraishi:2010kd,Domenech:2017kno,Shiraishi:2019yux,Meerburg:2016ecv,Coulton:2019odk}. Whilst forecasts exist for $TTT$ \citep{Shiraishi:2010kd,Domenech:2017kno}, $TTB$ \citep{Meerburg:2016ecv,Domenech:2017kno} and $EEB$ \citep{Domenech:2017kno} and $TTB+TEB+EEB$ \citep{Duivenvoorden:2019ses}, the model has been practically constrained only with WMAP temperature anisotropies \citep{Shiraishi:2017yrq}, leading to the bound given in Tab.\,\ref{tab:fNL_prev}. The addition of polarization information is expected to significantly sharpen constraints, with LiteBIRD and CMB-S4 constraints expected to reach $\mathcal{O}(1)$, with further gains expected with improved delensing and foreground removal techniques \citep{Shiraishi:2019yux,Duivenvoorden:2019ses} (see also \citep{Meerburg:2016ecv,Duivenvoorden:2019ses,Shiraishi:2019yux,Domenech:2017kno,SimonsObservatory:2018koc,Coulton:2019odk}).


\section{Fisher Forecasts \& Quasi-Optimal Binning}\label{sec: fisher}
\noindent With the above primordial models in hand, we now proceed to forecast the constraining power on $f_{\rm NL}^{ttt}$, $f_{\rm NL}^{tts}$ and $f_{\rm NL}^{tss}$ from the latest \textit{Planck} data. We will perform these forecasts both with and without binning, which will facilitate later comparison with the observational data constraints.

\subsection{Binned and Unbinned Bispectra}

\noindent Given a triplet of fields $\{u_1,u_2,u_3\}$ (with $u_i\in\{T,E,B\}$), we define the harmonic-space bispectrum $B_{\ell_1\ell_2\ell_3}^{u_1u_2u_3}$ from the underlying field $a_{\ell m}^u$ via
\beq
    \av{a_{\ell_1m_1}^{u_1}a_{\ell_2m_2}^{u_2}a_{\ell_3m_3}^{u_3}}_c = \tj{\ell_1}{\ell_2}{\ell_3}{m_1}{m_2}{m_3}B_{\ell_1\ell_2\ell_3}^{u_1u_2u_3},
\eeq
such that $B_{\ell_2\ell_1\ell_3}^{u_2u_1u_3}=(-1)^{\ell_1+\ell_2+\ell_3}B_{\ell_1\ell_2\ell_3}^{u_1u_2u_3}$ \textit{et cetera}. This has the inverse relation
\beq
   \sum_{m_1m_2m_3}\tj{\ell_1}{\ell_2}{\ell_3}{m_1}{m_2}{m_3} \av{a_{\ell_1m_1}^{u_1}a_{\ell_2m_2}^{u_2}a_{\ell_3m_3}^{u_3}}_c = B_{\ell_1\ell_2\ell_3}^{u_1u_2u_3}.
\eeq
In practice, the pipeline discussed below estimates a \textit{reduced} bispectrum, $b$, here defined via
\beq
    \av{a_{\ell_1m_1}^{u_1}a_{\ell_2m_2}^{u_2}a_{\ell_3m_3}^{u_3}}_c &=& w_{\ell_1}^{u_1}w_{\ell_2}^{u_2}w_{\ell_3}^{u_3}\sqrt{\frac{(2\ell_1+1)(2\ell_2+1)(2\ell_3+1)}{4\pi}}\left[\frac{1}{3}\tj{\ell_1}{\ell_2}{\ell_3}{-1}{-1}{2}+\text{2 perms.}\right]\tj{\ell_1}{\ell_2}{\ell_3}{m_1}{m_2}{m_3}b_{\ell_1\ell_2\ell_3}^{u_1u_2u_3}\nonumber\\
    &\equiv& \Omega_{\ell_1\ell_2\ell_3}\tj{\ell_1}{\ell_2}{\ell_3}{m_1}{m_2}{m_3}b_{\ell_1\ell_2\ell_3}^{u_1u_2u_3}.
\eeq
The choice of weighting matrix $\Omega_{\ell_1\ell_2\ell_3}$ is such that (a) the bispectrum estimators become separable by way of an angular integral and (b) the weighting is non-vanishing for both even and odd parities (and thus either sign of $\ell_1+\ell_2+\ell_3$). In contrast to previous works \resub{\citep[]{Philcox:2023xxk,Philcox:2023uwe,Philcox:2023psd}}, we include additional \textit{optimality weights}, $w_{\ell}^u$, which weight each $\ell$-mode within a bin, and can lead to less lossy binned estimators. \resub{This is morally similar to the (optimal) approach used in Komatsu-Spergel-Wandelt (KSW) estimators \citep{Komatsu:2003iq}, which apply a set of $\ell$-dependent weights to each Fourier-space field, and sum over a set of auxiliary variables. We adopt a data-driven approach to construct the weights, as discussed in \S\ref{subsec: optimal-binning}.}

Given a set of bins $\vec b\equiv\{b_1,b_2,b_3\}$, fields $\vec u\equiv\{u_1,u_2,u_3\}$, and parities $\chi\in\{\pm1\}$, we define the theoretical binned spectrum, $b^{\vec u}_\chi(\vec b)$, as
\beq\label{eq: binned-def}
    \left[\mathcal{F}_{\rm ideal}b\right]^{\vec u}_\chi(\vec b) = \frac{1}{\Delta^{\vec u}(\vec b)}\sum_{\ell_i}\Theta_{\ell_1\ell_2\ell_3}(\vec b)\Omega_{\ell_1\ell_2\ell_3}\sum_{v_i}S_{\ell_1}^{-1,u_1v_1}S_{\ell_2}^{-1,u_2v_2}S_{\ell_3}^{-1,u_3v_3}B_{\ell_1\ell_2\ell_3}^{v_1v_2v_3}\left[\frac{1+\chi\,p_{\vec u}(-1)^{\ell_1+\ell_2+\ell_3}}{2}\right].
\eeq
where $S_\ell^{uu'}\equiv C_\ell^{uu'}+\delta_{\rm K}^{uu'}N_{\ell}^{u}$ is the signal-plus-noise spectrum (here evaluated at $r=0$, but with lensing), $\Theta_{\ell_1\ell_2\ell_3}(\vec b)$ specifies the binning (zero if the $\ell$-triplet is in the bin and one else), and $p_{\vec u}$ is $1$ if $\vec u$ contains an even number of $B$-modes and $-1$ else.\footnote{We have tacitly assumed that $B_{\ell_1\ell_2\ell_3}^{u_1u_2u_3}$ is replaced with its real or imaginary part (depending on parity); removing this assumption yields additional factors of $(-1)^{\ell_1+\ell_2+\ell_3}$.} Here, the sums are over all $\ell_i$ and $v_i$, and we have defined the normalization
\beq
    \Delta^{\vec u}(\vec b) &=& \begin{cases} 6 & b_1=b_2=b_3\text{ and } u_1 =u_2 =u_3\\
    2 & b_1=b_2\text{ and }u_1=u_2\\
    2 & b_2=b_3\text{ and }u_2=u_3\\
    1 & \text{else,}\end{cases}
\eeq
where the conditions should be read sequentially. In practice, we ensure that the bins contain at least one $\ell_1,\ell_2,\ell_3$ triplet and restrict to $b_2>b_1$ if $u_1=u_2$, and $b_3>b_2$ if $u_2=u_3$, utilizing the following triplets of fields:
\beq
    \vec u\in\{TTT, TTE, TTB, TEE, TEB, TBB, EEE, EEB, EBB, BBB\}.
\eeq
Finally, \eqref{eq: binned-def} involves a normalization matrix, $\F_{\rm ideal}$, which is defined as
\beq\label{eq: ideal-fish}
    \mathcal{F}^{\vec u\vec u'}_{\rm{ideal},\chi\chi'}(\vec b, \vec b') &=& \frac{\delta_{\chi\chi'}^{\rm K}}{\Delta^{\vec u}(\vec b)\Delta^{\vec u'}(\vec b')}\sum_{\ell_i}\Theta_{\ell_1\ell_2\ell_3}(\vec b)\Omega^2_{\ell_1\ell_2\ell_3}\left[\delta^{\rm K}_{(b_1b_2b_3),(b_1'b_2'b_3')}S_{\ell_1}^{-1,u_1u_1'}S_{\ell_2}^{-1,u_2u_2'}S_{\ell_3}^{-1,u_3u_3'}+\text{5 perms.}\right]\\\nonumber
    &&\,\times\,\left[\frac{1+\chi \,p_{\vec u}(-1)^{\ell_1+\ell_2+\ell_3}}{2}\right],
\eeq
where the permutations are over the six orderings of the pairs $\{(b_1',u_1'),(b_2',u_2'),(b_3',u_3')\}$. Under idealized conditions (full-sky, unmasked, Gaussian, with an optimal estimator), the bispectrum covariance is given by the inverse of $\F_{\rm ideal}$, according to the Cram\'er-Rao bound. 

\subsection{Forecast Definitions}
\noindent Without binning, the Fisher matrix obtained from a fraction $f_{\rm sky}$ of the sky is given by \citep[cf.,][]{Duivenvoorden:2019ses}
\beq\label{eq: unbinned-forecast}
    F_{\rm unbin} &=& f_{\rm sky}\sum_{\ell_1\leq \ell_2\leq \ell_3}\sum_{\mathrm{all}\,u_i,u'_i}\frac{1}{\Delta_{\ell_1\ell_2\ell_3}}B_{\ell_1\ell_2\ell_3}^{u_1u_2u_3}\left(S_{\ell_1}^{-1,u_1u_1'}S_{\ell_2}^{-1,u_2u_2'}S_{\ell_3}^{-1,u_3u_3'}\right)\left(B_{\ell_1\ell_2\ell_3}^{u_1'u_2'u_3'}\right)^*
\eeq
where $\Delta_{\ell\ell\ell}=6$, $\Delta_{\ell\ell\ell'}=\Delta_{\ell'\ell\ell}=2$ and $\Delta_{\ell\ell'\ell''}=1$ for $\ell\neq\ell'\neq\ell''$. This involves the theoretical bispectrum template $B_{\ell_1\ell_2\ell_3}^{u_1u_2u_3}$, computed assuming unit $f_{\rm NL}$. Including binning, the Fisher matrix becomes
\beq\label{eq: binned-forecast}
    F_{\rm bin} &=& f_{\rm sky}\sum_{\vec u,\vec u'}\sum_{\vec b,\vec b'}\sum_{\chi\chi'} b^{\vec u}_\chi(\vec b)\,\mathcal{F}^{-1,\vec u\vec u'}_{{\rm ideal},\chi\chi'}(\vec b,\vec b')b^{\vec u'}_{\chi'}(\vec b'),
\eeq
where we sum only over non-degenerate bins (as specified above) and again assume unit $f_{\rm NL}$ to compute the binned numerators. Here, we have assumed that the bispectrum covariance is given by the inverse of $\mathcal{F}_{\rm ideal}$, as before. In all forecasts, we additionally include the beam, $B_\ell^u$, which effectively transforms $S_\ell^{-1,uu'}\to B_\ell^{u}B_\ell^{u'}S_{\ell}^{-1,uu'}$.

\subsection{Quasi-Optimal Binning}\label{subsec: optimal-binning}
\noindent To assess the dependence of our constraints on $\ell$-space binning, we implement the above idealized Fisher forecasts numerically, given a pre-computed table of all $B_{\ell_1\ell_2\ell_3}^{u_1u_2u_3}$ for each template of interest. We first compute the unbinned forecast, $F_{\rm unbin}(\ell_{\rm max}^{\rm long},\ell_{\rm max}^{\rm short})$, where we optionally allow the longest and shortest legs of the triangle to have different $\ell_{\rm max}$, allowing us to restrict to squeezed configurations with $\ell_{\rm max}^{\rm long}\ll \ell_{\rm max}^{\rm short}$. For the three types of bispectra considered above, performing the idealized forecast with a \textit{Planck}-like beam and noise, we find the immediate conclusions:
\begin{itemize}
    \item \textbf{Tensor-Tensor-Tensor}: The $f_{\rm NL}^{ttt}$ information is saturated by $\ell^{\rm long}_{\rm max}=\ell^{\rm short}_{\rm max}=200$, with only a $0.2\%$ gain from increasing to $\ell^{\rm long}_{\rm max}=\ell^{\rm short}_{\rm max}=500$. This is due to the tensor transfer function and allows us to restrict to $\ell^{\rm long}_{\rm max}=\ell^{\rm short}_{\rm max}=200$ in the below.
    \item \textbf{Tensor-Tensor-Scalar}: The $f_{\rm NL}^{tts}$ information is approximately saturated by $\ell_{\rm max}^{\rm long}=100$ and $\ell_{\rm max}^{\rm short}=300$, with $<1\%$ change to SNR imparted by smaller scales. We adopt these values below, which practically restricts to squeezed triangles on small scales (noting that the short legs correspond to tensor modes).
    \item \textbf{Tensor-Scalar-Scalar}: Due to the hard scalar legs, the $f_{\rm NL}^{tss}$ information does not saturate quickly with $\ell_{\rm max}^{\rm short}$. However, we can restrict to squeezed triangles with $\ell_{\rm max}^{\rm long}<120$ with $<1\%$ SNR penalty. We will adopt this below with $\ell_{\rm max}^{\rm short}=2000$ (noting that the Fisher matrix scales as $(\ell_{\rm max}^{\rm short})^2$ in this regime).
\end{itemize}
Armed with the unbinned forecasts, we may compute a \resub{quasi-}optimal binning and weighting strategy for a given bispectrum analysis (noting that some form of dimensionality reduction is necessary for practical implementation of the bispectrum estimators). This is done via two guiding principles: (a) we wish to keep the computation time as small as possible (since bispectrum computation and memory requirements are linear in $N_{\rm bin}$) and (b) we wish for the binned forecast, $F_{\rm bin}$, to match the unbinned forecast, $F_{\rm unbin}$, as closely as possible. To determine the choice of bin edges and optimality weights ($w_\ell^u$) given these restrictions, we utilize the following greedy algorithm:
\begin{enumerate}
    \item Begin with bin-edges $\{2,3\}$ (\textit{i.e.}\ one bin containing $\ell=2$), and compute $F_{\rm bin}$ for each template of interest.
    \item Add an additional bin-edge (e.g.\ $\{2,3\} \to \{2,3,6\}$) and compute the updated Fisher matrix $F_{\rm bin}$ for each template of interest.
    \item Repeat step (2) for all possible additional edges and choose the value which leads to the most optimal Fisher matrices, defined as those for which the minimal value of $F_{\rm bin}/F_{\rm unbin}$ across all templates is largest.
    \item Using the Broyden-Fletcher-Goldfarb-Shanno (BFGS) algorithm, compute the \resub{quasi-}optimal $\ell$- and field-weights, $w_\ell^u$ by maximizing the Fisher matrix, $F_{\rm bin}[w]$ (using the $\ell$-binning obtained above, with gradients computed explicitly).
    \item Repeat steps (2), (3) and (4), gradually building up the bin-edges, and updating the \resub{optimality} weights each time. Terminate when the total number of (three-dimensional) bispectrum bins exceeds $1000$. 
\end{enumerate}
This can be efficiently computed using the above forecasts, which are heavily vectorized, embarrassingly parallelized, and implemented in a combination of Python and Cython. \resub{Notably, the above scheme is not guaranteed to find the optimal set of bins and weights: this is due to the confines of the greedy algorithm and the inherent difficulty of performing optimization in a high-dimensional space (noting that $w_\ell^u$ contains $\mathcal{O}(10^3)$ elements). That said, it gives significant improvements in signal-to-noise in modest computational time and will thus be adopted below.}

Given the above differences in model phenomenology, we will determine three sets of \resub{quasi-}optimal bins, treating tensor-tensor-tensor, tensor-tensor-scalar and tensor-scalar-scalar bispectrum separately. Given that the signal-to-noise in the latter does not saturate at low-$\ell$, we omit step (4) in this case, instead assuming the simplified weight $w_{\ell}^{u} = 1/(2\ell+1)$ for speed. \resub{The output tensor-tensor-tensor and tensor-tensor-scalar weights are plotted in Appendix \ref{app: weights}.}

\begin{figure}
    \centering
    \subfloat[Tensor-Tensor-Tensor]{\includegraphics[width=0.75\textwidth]{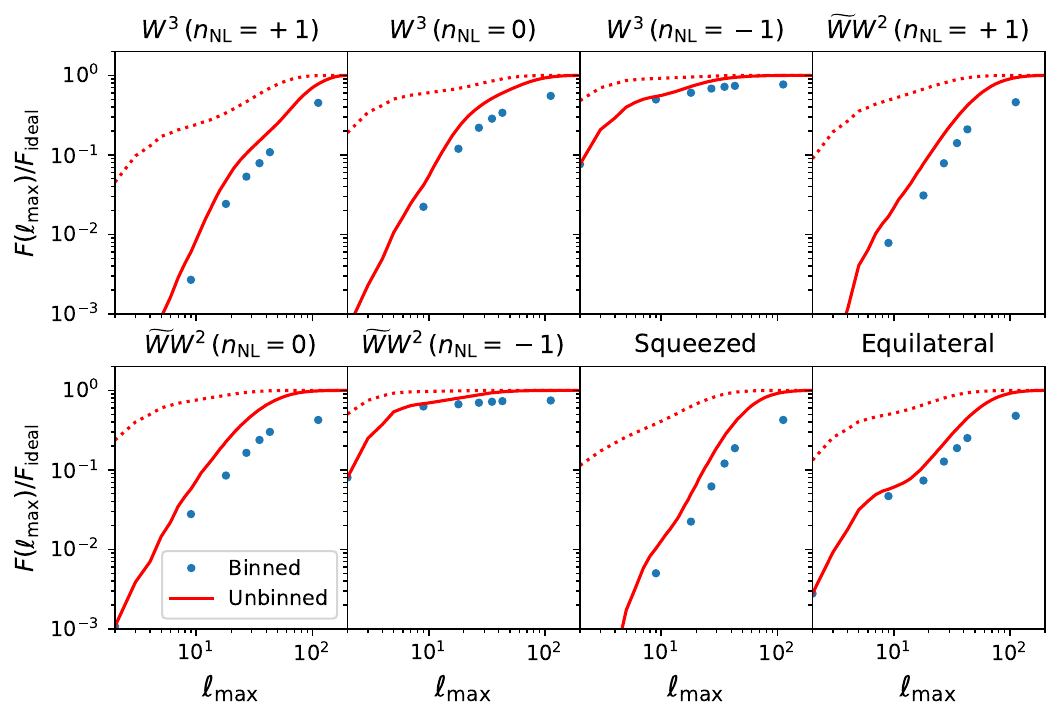}}\\
    \subfloat[Tensor-Tensor-Scalar]{\includegraphics[width=0.35\textwidth]{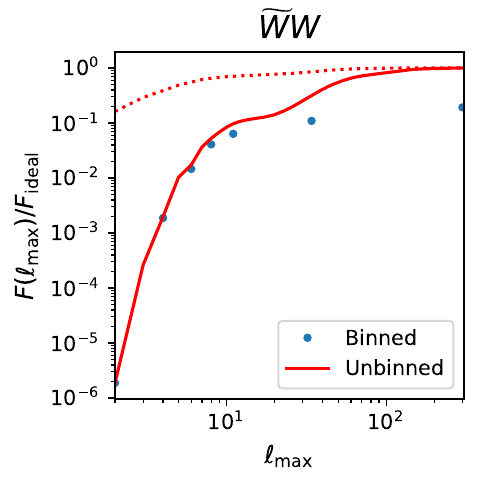}}\qquad
    \subfloat[Tensor-Scalar-Scalar]{\includegraphics[width=0.35\textwidth]{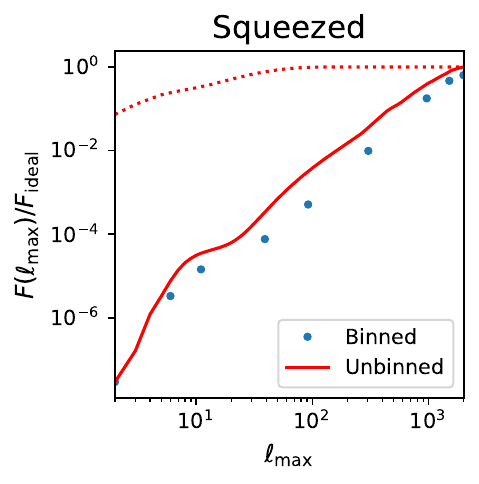}}
    \caption{Fisher analysis forecasting the constraints on $f_{\rm NL}$ parameters for a variety of models (shown in the titles) under idealized conditions, using $T$-, $E$- and $B$-modes with a \textit{Planck}-like beam and noise configuration. We compare the unbinned Fisher results \eqref{eq: unbinned-forecast} with those incorporating (\resub{weighted}) $\ell$-space binning \eqref{eq: binned-forecast}, itself optimized to maximize $F_{\rm bin}/F_{\rm unbin}$. We show the dependence on the maximum scale $\ell_{\rm max}^{\rm long}=\ell_{\rm max}^{\rm short}=\ell_{\rm max}$ included in the analysis, with red dashed lines indicating the constraints if we restrict the longest leg of the triangles to $\ell_{\rm max}^{\rm long}=\ell_{\rm max}$, but allow arbitrarily squeezed triangles (practically setting $\ell_{\rm max}^{\rm short}=200, 300, 2000$ for tensor-tensor-tensor, tensor-tensor-scalar and tensor-scalar-scalar models). In most cases, the signal-to-noise saturates by $\ell_{\rm max}\approx 150$ (due to the tensor transfer functions), except for the tensor-scalar-scalar model, which is dominated by highly squeezed triangles. \resub{The suboptimalities seen at large $\ell_{\rm max}$ reflect the inherent difficulties of compressing (often oscillatory) bispectrum models into a small number of $\ell$-space bins.}}\label{fig: fisher}
\end{figure}

In Fig.\,\ref{fig: fisher}, we show the Fisher errors on the bispectrum amplitudes, $f_{\rm NL}$, obtained from the idealized forecast as a function of the maximum scale, $\ell_{\rm max}^{\rm long}=\ell_{\rm max}^{\rm short}=\ell_{\rm max}$, alongside the binned equivalents. These are computed with the (optimized) bin edges $\{2,   3,  10,  19,  28,  36,  44, 113\}$ (tensor-tensor-tensor), $\{2,   3,   5,   7,   9,  12,  35, 300\}$ (tensor-tensor-scalar) and $\{2,    3,    7,   12,   40,   93,  304,  966, 1511, 2000\}$ (tensor-scalar-scalar). For each of the eight tensor-tensor-tensor models (dropping the degenerate axion case), we find a strong dependence of the constraints on scale for $\ell_{\rm max}\lesssim 100$, particularly for less steep inflationary models such as the Weyl models with $n_{\rm NL}=+1$ and the general squeezed template. For the mixed templates, the scalings are sharper still, and, for the tensor-scalar-scalar model, no saturation is observed until $\ell_{\rm max}^{\rm short}\sim 1000$ (with squeezed templates seen to clearly dominate). Comparing unbinned and binned forecasts at the largest $\ell_{\rm max}$, we find that binning inflates the errorbars by $\approx (10-50)\%$ (tensor-tensor-tensor), $130\%$ (tensor-tensor-scalar), and $25\%$ (tensor-scalar-scalar). \resub{This is larger than for the scalar case \citep{Planck:2019kim}, and is a consequence of restricting the total number of bispectrum bins to $\approx 1000$ for each parity. This is necessary to avoid excessive computational costs for the normalization matrix, and exacerbated compared to the scalar case due to the inclusion of $B$-modes (which increases the dimensionality by $2.5\times$) and the oscillatory tensor transfer functions.}\footnote{The tensor-tensor-scalar bispectra oscillate significantly within $\ell$-bins; \resub{this occurs due to the shape of the underlying tensor bispectra (\S\ref{subsec: tts-model}) and the combination of an enhanced squeezed limit and a parity-odd structure. This sources the higher suboptimality observed and the structure of the weights seen in Appendix \ref{app: weights}.}}

\begin{figure}[t]
    \centering
    \includegraphics[width=0.7\linewidth]{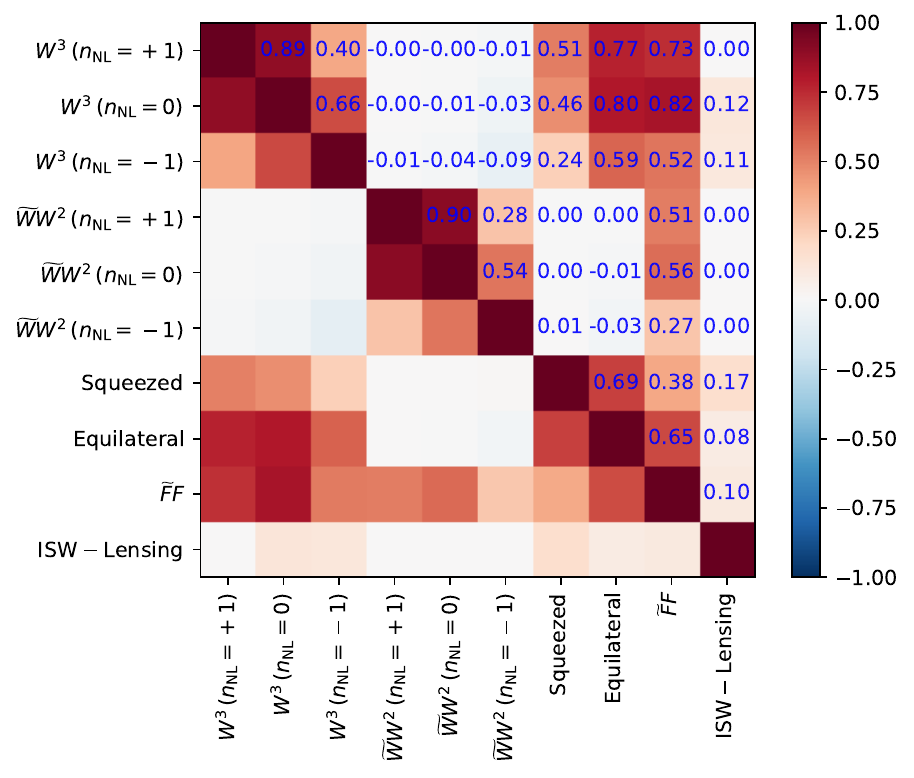}
    \caption{\resub{Correlation matrix between the nine tensor-tensor-tensor bispectrum templates discussed in \S\ref{subsec: ttt-bispectra}, as well as the late-time ISW-lensing signal. We show the empirical correlation, defined as $C_{ab}^{\rm bin}/\sqrt{C^{\rm bin}_{bb}C^{\rm bin}_{aa}}$, where $C^{\rm bin}_{ab}$ is the binned Fisher matrix between templates $a$ and $b$, which accounts for the mask. Several templates exhibit fairly strong correlations (with numerical values labelled in blue), which are largest for the $\tilde{W}W^2(n_{\rm NL}=+1)$ and $\tilde{W}W^2(n_{\rm NL}=0)$ models. As expected, the parity-odd and parity-even sectors are almost uncorrelated (up to small mask-induced leakage), and there is minimal bias from late-time effects.}}
    \label{fig: corr-mat}
\end{figure}

\resub{It is important to assess whether the various bispectrum models discussed above are practically distinguishable. To this end, we plot the theoretical correlation matrix for the binned tensor-tensor-tensor amplitudes in Fig.\,\ref{fig: corr-mat}, easily obtained from the multi-template generalization of \eqref{eq: binned-forecast}.\footnote{We note that the tensor-tensor-scalar and tensor-scalar-scalar amplitudes are roughly independent due to the different transfer functions, parities and scale-cuts.} We observe that the tensor-tensor-tensor Weyl models with different scale dependence are phenomenologically similar, though the parity-even ($W^3$) and parity-odd ($\tilde{W}W^2$) forms are approximately uncorrelated (up to leakage from the non-uniform mask), as expected. Perhaps counterintuitively, the equilateral and squeezed tensor-tensor-tensor models are somewhat degenerate ($\approx 70\%$); this occurs due to the sharp decay of the tensor transfer function with scale, and thus the limited resolution in $\ell$-space (in contrast to scalar studies). That said, most correlations are $\lesssim 70\%$, implying that analyzing them all is a profitable exercise. Finally, we note that degeneracies with the ISW-lensing effect are small, implying limited late-time contamination (in contrast to scalar non-Gaussianity \citep[e.g.,][]{Planck:2019kim}.}

\section{Analysis}\label{sec: analysis}
\noindent We next describe the main datasets and analysis methods used in this work. These are very similar to those used in \citep{Philcox4pt3} (which build on \citep{Philcox:2023xxk}), but we provide a brief recapitulation below for clarity. 

\subsection{Data}
\noindent Our dataset comprises the \textit{Planck} PR4 temperature and polarization maps, \resub{processed using the \textsc{npipe} approach \citep{Planck:2020olo}}. We additionally utilize 600 \textsc{npipe}/FFP10 simulations, 100 of which are used to form \resub{quasi-}optimal estimators, as discussed below.\footnote{Data and simulations are available at \href{https://portal.nersc.gov/project/cmb/planck2020}{portal.nersc.gov/project/cmb/planck2020}.} \resub{To separate primordial and late-time signals, we utilize the \textsc{sevem} component-separation scheme; as a consistency test, we also consider the \textsc{smica} pipeline, implemented as discussed in \citep{Carron:2022eyg,Philcox4pt3} (noting also that residual $B$-mode foregrounds are expected to be small \citep{Coulton:2019bnz}).} Data are filtered using a quasi-optimal scheme, here denoted by the $\Si$ operator, similar to \citep{2015arXiv150200635S,Planck:2015zfm}, \resub{which inpaints small holes in the (smoothly masked) map and applies an $\ell$-space weighting $B_\ell^X[B_{\ell}^{X}B_{\ell}^YC^{XY,\,\rm{theory}}_\ell+N_{\ell}^{XY}]^{-1}$ for noise spectrum $N_{\ell}^{XY}$ and beam $B_\ell^X$ (including the \textsc{npipe} polarization transfer function)}. Given the improved treatment of polarization noise and systematics in \textit{Planck} PR4 and the \textsc{npipe} simulations, we include all scales down to $\ell_{\rm min}=2$ in the analysis (which are of considerable use to squeezed models \citep{Duivenvoorden:2019ses}); removing such scales would be an important consistency test if any model were to be detected.

\subsection{Estimator}
\noindent Given the \resub{quasi-}optimal binning strategies described in \S\ref{subsec: optimal-binning}, reduced bispectra are estimated using the \resub{\textsc{PolySpec}} code\footnote{\resub{Formerly \textsc{PolyBin};} code available at \href{https://github.com/oliverphilcox/PolySpec}{github.com/oliverphilcox/PolySpec}.} \citep{PolyBin} described in \citep{Philcox:2023uwe,Philcox:2023psd,Philcox4pt2} (building on previous binned bispectrum codes \citep{Bucher:2015ura,Bucher:2009nm,Coulton:2019bnz}, utilizing a number of techniques developed in \citep{2011MNRAS.417....2S,2015arXiv150200635S,Philcox4pt1}). This implements the following quasi-optimal bispectrum estimator (applied separately to each of the tensor-tensor-tensor, tensor-tensor-scalar and tensor-scalar-scalar bin configurations):
\beq\label{eq: estimator}
    \left[\F\hat{b}\right]_\chi^{\vec u}(\vec b) &=& \frac{1}{3!}\frac{\partial\av{a^ia^ja^k}}{\partial b_\chi^{\vec u}(\vec b)}\bigg\{[\Si d]_i[\Si d]_j[\Si d]_k-3[\Si d]_i\av{[\Si d]_j[\Si d]_k}_{\rm sim}\bigg\}^*\\\nonumber
    \F_{\chi\chi'}^{\vec u\vec u'}(\vec b,\vec b') &=& \frac{1}{3!}\left[\frac{\partial\av{a^ia^ja^k}^*}{\partial b_\chi^{\vec u}(\vec b)}[\Si\mathsf{P}]_{ii'}[\Si\mathsf{P}]_{jj'}[\Si\mathsf{P}]_{kk'}\frac{\partial\av{a^{i'}a^{j'}a^{k'}}}{\partial b_{\chi'}^{\vec u'}(\vec b')}\right]^*
\eeq
where $i,i',\cdots$ indexes pixels and polarizations \resub{and $\mathsf{P}$ is the pointing matrix, describing the response of the data, $d$, to the primordial signal, $a$}. Essentially, this is a quasi-maximum-likelihood estimator for the binned bispectrum components, which projects three copies of the weighted data ($\Si d$) onto a theoretical template given by the response of the theoretical three-point function $\av{a^ia^ja^k}$ to the coefficient of interest. We additionally include a linear term in the estimator involving an expectation $\av{...}_{\rm sim}$, which is computed using 100 simulations; this leads to a slight reduction in the variance on large scales. 
The normalization matrix, $\F$, accounts for mask-induced leakage between bins, parities, and fields, ensuring that the estimator is unbiased. In the limit of a Gaussian dataset \resub{with $\Si$ equal to $\mathsf{P}^\dagger$ times the inverse covariance of $d$}, this is an optimal estimator and has covariance $\mathcal{F}^{-1}$ \citep{Philcox4pt1}; assuming also a unit mask, unit beam and translation-invariant noise, this reduces to \eqref{eq: ideal-fish}. In practice, all terms can be efficiently computed using spherical harmonic transforms and various numerical tricks, all of which are implemented within \resub{\textsc{PolySpec}}.

For the tensor-tensor-tensor analyses, the \resub{quasi-}optimal binning scheme utilizes seven one-dimensional $\ell$-bins, and, once the fields and parities are taken into account, $2326$ non-trivial bispectrum bins in total. Given that the signal-to-noise saturates at low-$\ell$, we can perform fast analyses using a \textsc{healpix} $N_{\rm side}=256$ \citep{Gorski:2004by}. For the tensor-tensor-scalar analysis, we have seven (six) one-dimensional squeezed (unsqueezed) $\ell$-bins, and $1030$ total bins after restricting to parity-odd configurations. Finally, the tensor-scalar-scalar analysis uses nine (five) one-dimensional squeezed (unsqueezed) $\ell$-bins and $1353$ total bins, now restricting to parity-even configurations. For the latter analysis, we adopt $N_{\rm side}=1024$, since we include significantly smaller scales in the analysis.

Computation of the bispectrum numerator $\F\hat{b}$ required approximately \resub{$0.9$ (tensor-tensor-tensor) and $0.4$ (tensor-tensor-scalar), and $8$ (tensor-scalar-scalar) CPU-hours} per simulation and is dominated by the linear term in \eqref{eq: estimator} (which is omitted for tensor-scalar-scalar models). The normalization matrix, $\F$, was computed using $50$ Monte Carlo iterations (which was found to be more than sufficient for sub-$\sigma$ convergence in the bispectra), \resub{which required around $1$ ($2$) CPU-hours for the tensor-tensor-tensor (tensor-tensor-scalar) model}. For the more-expensive tensor-scalar-scalar computations, we used $10$ Monte Carlo iterations, \resub{each of which required $8$ node-hours.}

\subsection{Theoretical Models \& Likelihood}
\begin{figure}
    \centering
    \includegraphics[width=0.59\textwidth]{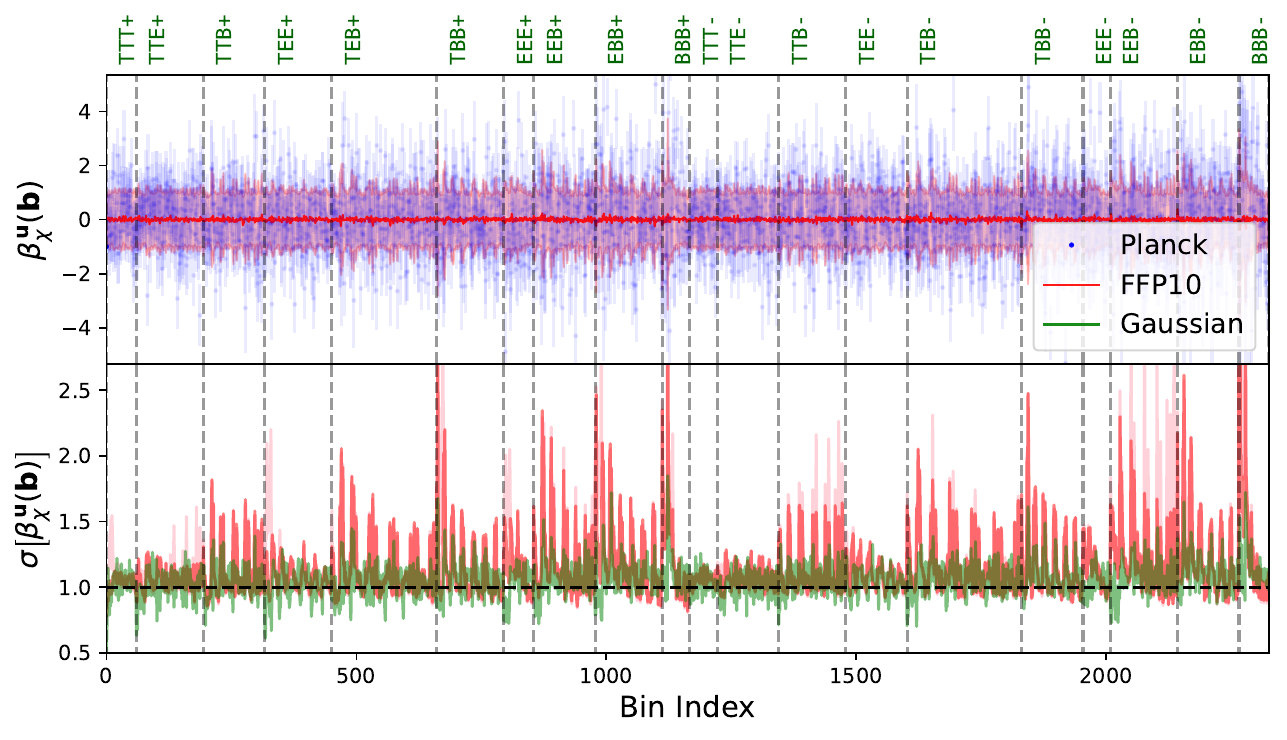}
    \includegraphics[width=0.39\textwidth]{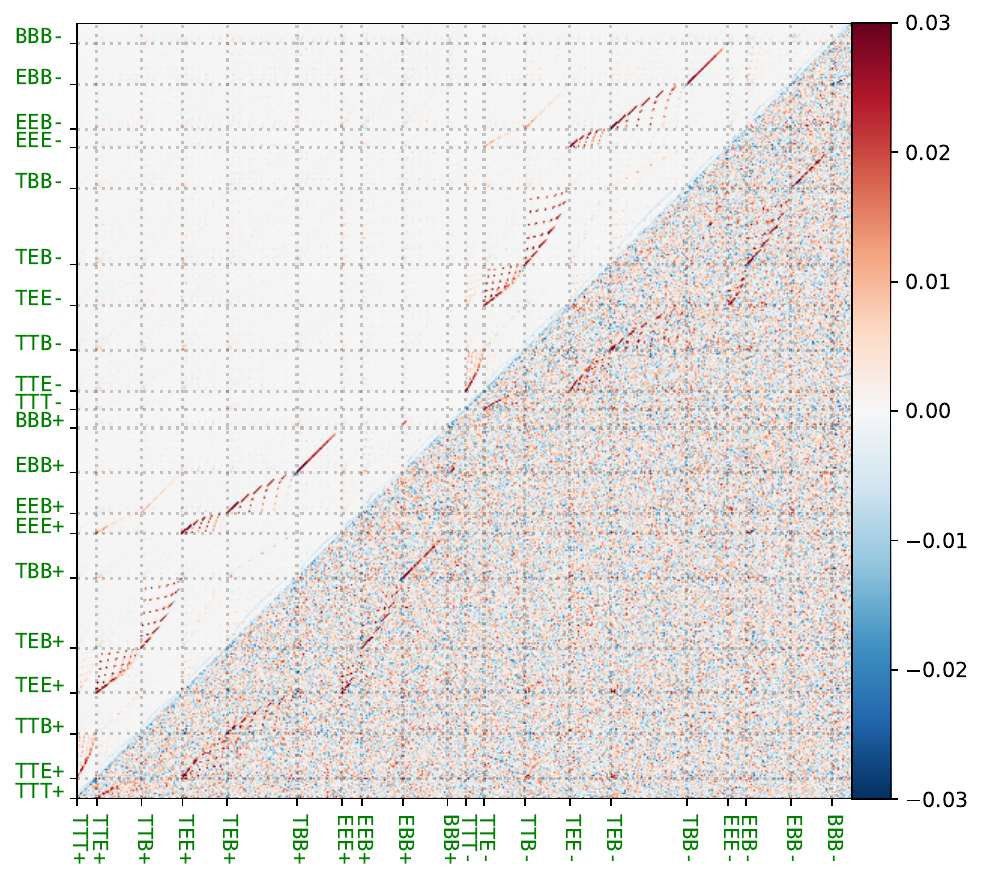}
    \caption{Reduced bispectra measured from \textit{Planck} PR4 data, utilizing a binning scheme optimized for measurement of tensor-tensor-tensor bispectra. \textbf{Left}: normalized bispectrum measurements, $\beta$ (see \eqref{eq: rotator}) from \textit{Planck} (blue) and 500 FFP10/\textsc{npipe} simulations (red) across all combinations of fields and parities, as indicated by the green labels on the top. Smallest bins ($\ell\sim2$) are to the left of each sub-panel whilst the largest ($\ell\sim 200$, for this choice of binning) are towards the right. \resub{The bottom panel shows three sets of errorbars: (1) the theoretical errors from the numerical Fisher matrix, including the mask and bin (black dashed line at unity); (2) synthetic masked Gaussian random fields, including any suboptimalities in the estimator arising from the simplified weighting scheme $\Si$Th ; (3) full FFP10 simulations, including residual foregrounds and spatially-varying noise.} We also show results including only the cubic term in the bispectrum estimator of \eqref{eq: estimator} in light pink. \resub{The errors are close to the theoretical prediction, but there are some deviations from large-scale polarization modes; these are incorporated in the likelihood using the simulation-based variance estimates}. \textbf{Right}: correlation matrix of the reduced bispectra, $b$, from theory (upper left, equal to $\F^{-1}$) and 500 FFP10 simulations (bottom right). \resub{We find that the numerical Fisher matrix well reproduces the structure of the empirical covariance.}}
    \label{fig: beta-plot}
\end{figure}

\noindent The \textit{Planck} reduced bispectra, $\hat{b}$ can be compared to the templates defined above. For this purpose, we compute the theoretical bispectra, $b^{\rm th}$ via \eqref{eq: binned-def}\,\&\,\eqref{eq: ideal-fish} given an input theory model $B_{\ell_1\ell_2\ell_3}^{u_1u_2u_3}$ with unit $f_{\rm NL}$. To compare theory and observation, we adopt the following likelihood, \resub{which combines the theoretical correlation matrix with the empirical simulation-based variances}:
\beq\label{eq: loglik}
    -2\log \mathcal{L}(f_{\rm NL}) &=& \sum_{\vec u,\vec b,\chi}\frac{\left(\hat{\beta}^{\vec u}_\chi(\vec b)-f_{\rm NL}\beta^{\vec u,\rm th}_\chi(\vec b)\right)^2}{\mathrm{var}\left[\beta^{\vec u}_\chi(\vec b)\right]}+\text{const.}
\eeq
defining the rotated bispectra
\beq\label{eq: rotator}
    \beta_\chi^{\vec u}(\vec b) = \left[\mathcal{F}^{\rm T/2}b\right]^{\vec u}_\chi(\vec b),
\eeq
where $\rm{T}/2$ indicates a transposed Cholesky factorization. This involves the normalization matrix $\F$ estimated in \eqref{eq: estimator} (which involves the \textit{Planck} mask and is, in general, more complex than the form given in \eqref{eq: ideal-fish}). As demonstrated in \citep{Philcox:2023psd,Philcox:2023xxk}, rotating according to \eqref{eq: rotator} approximately diagonalizes the bispectra, such that their noise properties can be captured by a variance measured using simulations \resub{(\textit{i.e.}\ the $\mathrm{var}[\beta]$ factor in \eqref{eq: loglik})}. This is confirmed in the right panel of Fig.\,\ref{fig: beta-plot} (using the tensor-tensor-tensor binning scheme); the empirical correlation structure of the bispectra, $b$, is extremely well described by the inverse normalization matrix $\F^{-1}$, informing us that that of $\beta$ is close to diagonal.

In the left panel of Fig.\,\ref{fig: beta-plot}, we plot the rotated and reduced bispectra, $\beta$, obtained from the \textit{Planck} data and FFP10 simulations, across all bins and parities (adopting the \resub{quasi-}optimal tensor-tensor-tensor binning \resub{described in \S\ref{subsec: optimal-binning}}). As expected, the raw data points are distributed around zero with a mostly-constant variance; this is as expected and shows no `by-eye' evidence for a non-Gaussian signal (as was quantified in \citep{Philcox:2023xxk} using a model-agnostic framework). The empirical variance of $\beta$ is not quite equal to the expected value of unity; this indicates \resub{that our estimator is sub-optimal} in some bispectrum configurations. This principally occurs for bins including polarized fields (particularly $B$-modes) and highly squeezed triangles, and is likely caused by mask-induced effects, residual foregrounds, additional variance sourced by weak lensing, \resub{spatially-varying noise}, and likelihood non-Gaussianity.\footnote{If we drop the linear term in the bispectrum estimator \eqref{eq: estimator}, we find that such sub-optimalities increase a fair amount and are present even for $T$-modes, as expected \citep{Philcox:2023uwe}.} To verify this, we analyze also a set of $100$ Gaussian random fields, generated with the same assumed beam and (translation-invariant) noise spectrum. These show much improved variances suggesting that, \resub{if the data are Gaussian and translation-invariant,} our estimators are close to optimal.
\resub{As such,} the above deviations are predominantly sourced to \resub{the other effects contained within the FFP10 simulations, such as residual foregrounds and realistic noise properties}. This will be examined further below, both for this binning scheme as well as those adopted for tensor-tensor-scalar and tensor-scalar-scalar bispectra. \resub{We note that these effects, and any others present in the simulations, are taken into account in the likelihood of \eqref{eq: loglik} through the empirical variance term, \textit{i.e.}\ we do not assume that $\sigma(\beta)=1$ in practice.} 

\subsection{Late-Time Bispectra}
\noindent As CMB photons propagate from the last-scattering-surface to the observer, they are subject to a variety of late-time effects, including (potentially non-linear) CMB lensing and the integrated Sachs-Wolfe (ISW) effect \citep[e.g.,][]{Planck:2015zfm,Planck:2019kim}. Such phenomena can also generate non-Gaussian correlators, which, if sufficiently close in shape to the primordial forms, could lead to false inflationary detections. For the problem at hand, these effects are relatively minor since (a) our primordial bispectra have shapes that are quite distinct from those in the late-Universe (due to the tensor helicity basis functions) and (b) in many cases, our theoretical signal-to-noise saturates at low-$\ell$, where many late-time effects are suppressed. 

In this work, we consider a single late-time contaminant (as in \citep{Planck:2015zfm,Planck:2019kim}): the bispectrum induced by a combination of CMB lensing and the ISW effect. Roughly speaking, this arises since a given field transforms as $a\to a+\phi\star a$ in the presence of a lensing potential $\phi$, which leads to a bispectrum sourced by (a) the correlation of the unlensed field $a$ with a second field and (b) the correlation of $\phi$ with the ISW contribution to a third field  \citep[e.g.,][]{Goldberg:1999xm,Hu:2000ee,Lewis:2011fk,Philcox:2025lxt}. Explicitly, this sources the bispectrum 
\begin{align}
  B_{\ell_1 \ell_2 \ell_3}^{X_1 X_2 X_3}
  &= 
  F_{\ell_3 \ell_1 \ell_2}^{X_3}
  C_{\ell_1}^{X_1 \phi}  
  C_{\ell_2}^{X_2 X_3} + \text{5 perms.}
  \\\nonumber
  B_{\ell_1 \ell_2 \ell_3}^{X_1 X_2 B}
  &= i 
  \left[
    F_{\ell_3 \ell_1 \ell_2}^{B}
    C_{\ell_1}^{X_1 \phi} 
    C_{\ell_2}^{X_2 E} 
    - F_{\ell_3 \ell_2 \ell_1}^{B}
    C_{\ell_2}^{X_2 \phi} 
  C_{\ell_1}^{X_1 E} 
    \right],
  \end{align}
where $X_{1,2,3} \in \{T,E\}$, $C_\ell^{pq}$ is the angular power spectrum of fields $p$ and $q$ and
\beq
  F_{\ell_1 \ell_2 \ell_3}^T &\equiv& \frac{\ell_2(\ell_2 + 1) + \ell_3(\ell_3 + 1) - \ell_1 (\ell_1 + 1)}{2}
 \sqrt{\frac{(2 \ell_1 + 1)(2 \ell_2 + 1)(2 \ell_3 + 1)}{4 \pi}}
  \begin{pmatrix}
  \ell_1 & \ell_2 & \ell_3 \\
  0 & 0 & 0
  \end{pmatrix}
  ~, \\\nonumber
  F_{\ell_1 \ell_2 \ell_3}^E &\equiv& \frac{\ell_2(\ell_2 + 1) + \ell_3(\ell_3 + 1) - \ell_1 (\ell_1 + 1)}{2}
   \sqrt{\frac{(2 \ell_1 + 1)(2 \ell_2 + 1)(2 \ell_3 + 1)}{4 \pi}}
  \begin{pmatrix}
  \ell_1 & \ell_2 & \ell_3 \\
  2 & 0 & -2
  \end{pmatrix}
  \frac{1 + (-1)^{\ell_1 + \ell_2 + \ell_3}}{2} , \\\nonumber
    F_{\ell_1 \ell_2 \ell_3}^B &\equiv& \frac{\ell_2(\ell_2 + 1) + \ell_3(\ell_3 + 1) - \ell_1 (\ell_1 + 1)}{2}
   \sqrt{\frac{(2 \ell_1 + 1)(2 \ell_2 + 1)(2 \ell_3 + 1)}{4 \pi}}
  \begin{pmatrix}
  \ell_1 & \ell_2 & \ell_3 \\
  2 & 0 & -2
  \end{pmatrix}
  \frac{1 - (-1)^{\ell_1 + \ell_2 + \ell_3}}{2} .
\eeq
There are no tree-level contributions to $TBB$, $EBB$ and $BBB$ spectra, since $B$-modes do not contain ISW contributions or primordial signals (assuming $r=0$).

Whilst other secondary contributions are possible (including from the intrinsic non-Gaussianity of $\phi$, stochasticity of the cosmic infrared background, or ISW-tSZ-tSZ cross-correlation, where tSZ is the thermal Sunyaev-Zel'dovich contribution \citep{Lacasa:2013yya,Penin:2013zya,Hill:2018ypf,Coulton:2022wln}), these are expected to be small on large-scales with the latter vanishing for polarization anisotropies (which contribute much of our signal-to-noise). Practically, we can ameliorate any contamination from the above ISW-lensing bispectrum by performing a joint analysis of its amplitude (denoted by $f_{\rm NL}^{\rm lens}$) with the primordial template(s) of interest, optionally imposing a strong prior on the fiducial value $f_{\rm NL}^{\rm lens}=1$. This is done by default in the below, though, in practice, we find it to yield negligible ($<5\%$) degradation on our constraints (and none for our tensor-tensor-scalar model, which is purely parity-odd), even without a restrictive prior on $f_{\rm NL}^{\rm lens}$. \resub{We caution that our binning schemes are not tailored to the ISW-lensing contribution: much stronger constraints on $f_{\rm NL}^{\rm lens}$ can be obtained in a dedicated analysis \citep[e.g.,][]{Philcox:2025lxt,Carron:2022eyg}.}

\section{Constraints on Tensor \& Mixed Non-Gaussianity}\label{sec: results}

\noindent We now present the main results of this work: constraints on tensor and mixed non-Gaussianity from \textit{Planck} PR4 data. In Tab.\,\ref{tab: all-results}, we list the constraints on all the models enumerated in \S\ref{sec: models}, each analyzed independently in a Bayesian scheme \resub{using the \textsc{sevem} component-separation pipeline} (marginalizing over the ISW-lensing contribution, as above). \resub{Analogous results using \textsc{smica} are shown in Tab.\,\ref{tab: all-results-smica}.} Considering first the full-dataset $T+E+B$ results, we find no detections in any case, with a maximum \resub{signal-to-noise of $1.8\sigma$ for \textsc{smica} $W^3(n_{\rm NL}=+1)$}. From the mean of the FFP10/\textsc{npipe} simulations, we find no bias in our pipeline, validating our galactic foreground cuts and treatment of noise. The size of the $f_{\rm NL}$ errorbars differ significantly between models, with, for example, weakest bounds found for the red-tilted Weyl operators with $n_{\rm NL}=-1$ (which show the weakest scalings with $\ell_{\rm max}$, cf.\,Fig.\,\ref{fig: fisher}). \resub{In general, we find good agreement between \textsc{smica} and \textsc{sevem}, with a mean absolute difference of $0.35\sigma$, and a maximum difference of $0.8\sigma$ for $W^3(n_{\rm NL}=-1)$. This indicates that foreground residuals do not strongly bias our results.}

\begin{table}[]
    \centering
    \begin{tabular}{ll||rrr|rrr|rrr}
    \textbf{Model} & & \multicolumn{3}{l|}{\textit{Planck}} & \multicolumn{3}{l|}{FFP10 Simulations} & \multicolumn{3}{l}{Fisher ($f_{\rm sky}=0.76$)}\\
    & & T & T+E & \textbf{T+E+B} & T & T+E & T+E+B & T & T+E & T+E+B\\\hline
\quad\textit{\textbf{Tensor-Tensor-Tensor}} &&&&&&&&&\\
$\mathrm{Squeezed}$ & ($\times 10^{-1}$) & $61 \pm 35$ & $5 \pm 13$ & $\boldsymbol{4 \pm 8}$ & $-0 \pm 33$ & $1 \pm 13$ & $1 \pm 9$ & $\pm 16$ & $\pm 8$ & $\pm 5$\\
$\mathrm{Equilateral}$ & ($\times 10^{-2}$) & $-3 \pm 12$ & $-0 \pm 5$ & $\boldsymbol{-0 \pm 3}$ & $-0 \pm 13$ & $-0 \pm 5$ & $0 \pm 3$ & $\pm 6$ & $\pm 3$ & $\pm 2$\\
$W^3\,(n_{\rm NL}=+1)$ & ($\times 10^{-3}$) & $-66 \pm 34$ & $-2 \pm 7$ & $\boldsymbol{6 \pm 4}$ & $-4 \pm 36$ & $-1 \pm 7$ & $-0 \pm 4$ & $\pm 14$ & $\pm 5$ & $\pm 3$\\
$W^3\,(n_{\rm NL}=0)$ & ($\times 10^{-2}$) & $-8 \pm 14$ & $1 \pm 6$ & $\boldsymbol{4 \pm 4}$ & $-2 \pm 15$ & $0 \pm 6$ & $-0 \pm 4$ & $\pm 8$ & $\pm 4$ & $\pm 2$\\
$W^3\,(n_{\rm NL}=-1)$ & ($\times 10^{0}$) & $6 \pm 42$ & $14 \pm 26$ & $\boldsymbol{3 \pm 15}$ & $-5 \pm 41$ & $-0 \pm 26$ & $-0 \pm 15$ & $\pm 28$ & $\pm 17$ & $\pm 7$\\
$\widetilde{W}W^2\,(n_{\rm NL}=+1)$ & ($\times 10^{-3}$) & $49 \pm 102$ & $-13 \pm 16$ & $\boldsymbol{-4 \pm 7}$ & $-5 \pm 99$ & $-1 \pm 15$ & $-1 \pm 7$ & $\pm 101$ & $\pm 9$ & $\pm 4$\\
$\widetilde{W}W^2\,(n_{\rm NL}=0)$ & ($\times 10^{-2}$) & $47 \pm 63$ & $-5 \pm 11$ & $\boldsymbol{-2 \pm 5}$ & $-11 \pm 62$ & $-0 \pm 10$ & $-1 \pm 6$ & $\pm 55$ & $\pm 6$ & $\pm 3$\\
$\widetilde{W}W^2\,(n_{\rm NL}=-1)$ & ($\times 10^{0}$) & $95 \pm 238$ & $-2 \pm 51$ & $\boldsymbol{5 \pm 17}$ & $-4 \pm 237$ & $-3 \pm 52$ & $-0 \pm 18$ & $\pm 182$ & $\pm 27$ & $\pm 7$\\
$\widetilde{F}F$ & ($\times 10^{-2}$) & $-12 \pm 27$ & $-3 \pm 10$ & $\boldsymbol{3 \pm 6}$ & $-1 \pm 26$ & $-1 \pm 11$ & $-0 \pm 6$ & $\pm 15$ & $\pm 7$ & $\pm 3$\\
\hline
\quad\textit{\textbf{Tensor-Tensor-Scalar}} &&&&&&&&&\\
$\widetilde{W}W$ & ($\times 10^{-2}$) & $-53 \pm 449$ & $70 \pm 68$ & $\boldsymbol{6 \pm 10}$ & $-27 \pm 482$ & $-1 \pm 67$ & $0 \pm 11$ & $\pm 119$ & $\pm 19$ & $\pm 3$\\\hline
\quad\textit{\textbf{Tensor-Scalar-Scalar}} &&&&&&&&&\\
Squeezed & ($\times 10^{0}$) & $-10 \pm 32$ & $-2 \pm 15$ & $\boldsymbol{11 \pm 11}$ & $-0 \pm 31$ & $-2 \pm 14$ & $0 \pm 11$ & $\pm 11$ & $\pm 8$ & $\pm 6$\\
  \end{tabular}
    \caption{Constraints on tensor and mixed tensor-scalar non-Gaussianity $f_{\rm NL}$ amplitudes from \textit{Planck} PR4 binned bispectra, \resub{using the \textsc{sevem} component-separation method}. We show results for eleven templates, as defined in \S\ref{sec: models}, analyzed using $T$-, $E$- and $B$-mode anisotropies. We also show analogous results obtained by analyzing the mean of $500$ FFP10/\textsc{npipe} simulations, as well as idealized Fisher forecasts, which do not include binning, window effects, or foreground-induced non-Gaussianity (with these effects investigated in Fig.\,\ref{fig: breakdown}). The \textit{Planck} $T+E+B$ constraints (shown in bold) are the main results of this work. This table shows results for the analysis of each model in turn; joint analysis of parity-conserving and parity-violating tensor-tensor-tensor models is shown in Fig.\,\ref{fig: corner-plot}. We list analogous constraints using the \textsc{smica} component-separation pipeline in Tab.\,\ref{tab: all-results-smica}.}
    \label{tab: all-results}
\end{table}

\begin{table}[]
    \centering
    \begin{tabular}{ll||rrr|rrr}
    \textbf{Model} & & \multicolumn{3}{l|}{\textit{Planck}} & \multicolumn{3}{l}{FFP10 Simulations}\\
    & & T & T+E & \textbf{T+E+B} & T & T+E & T+E+B\\\hline
\quad\textit{\textbf{Tensor-Tensor-Tensor}} &&&&&&\\
$\mathrm{Squeezed}$ & ($\times 10^{-1}$) & $51 \pm 32$ & $-4 \pm 13$ & $\boldsymbol{7 \pm 9}$ & $-5 \pm 32$ & $0 \pm 13$ & $-0 \pm 8$\\
$\mathrm{Equilateral}$ & ($\times 10^{-2}$) & $-5 \pm 13$ & $-3 \pm 5$ & $\boldsymbol{-0 \pm 3}$ & $-1 \pm 12$ & $-0 \pm 5$ & $0 \pm 3$\\
$W^3\,(n_{\rm NL}=+1)$ & ($\times 10^{-3}$) & $-63 \pm 34$ & $-7 \pm 7$ & $\boldsymbol{8 \pm 4}$ & $-2 \pm 38$ & $-1 \pm 7$ & $1 \pm 4$\\
$W^3\,(n_{\rm NL}=0)$ & ($\times 10^{-2}$) & $-8 \pm 14$ & $-4 \pm 6$ & $\boldsymbol{4 \pm 4}$ & $-1 \pm 15$ & $-1 \pm 6$ & $0 \pm 4$\\
$W^3\,(n_{\rm NL}=-1)$ & ($\times 10^{0}$) & $-3 \pm 41$ & $-7 \pm 27$ & $\boldsymbol{-6 \pm 15}$ & $-7 \pm 39$ & $-6 \pm 26$ & $2 \pm 16$\\
$\widetilde{W}W^2\,(n_{\rm NL}=+1)$ & ($\times 10^{-3}$) & $61 \pm 98$ & $-18 \pm 15$ & $\boldsymbol{-8 \pm 6}$ & $-12 \pm 95$ & $-6 \pm 16$ & $0 \pm 6$\\
$\widetilde{W}W^2\,(n_{\rm NL}=0)$ & ($\times 10^{-2}$) & $42 \pm 63$ & $-9 \pm 11$ & $\boldsymbol{-3 \pm 5}$ & $-9 \pm 60$ & $-4 \pm 11$ & $-1 \pm 5$\\
$\widetilde{W}W^2\,(n_{\rm NL}=-1)$ & ($\times 10^{0}$) & $136 \pm 222$ & $-24 \pm 55$ & $\boldsymbol{5 \pm 20}$ & $-16 \pm 222$ & $-17 \pm 54$ & $-7 \pm 17$\\
$\widetilde{F}F$ & ($\times 10^{-2}$) & $-16 \pm 27$ & $-10 \pm 10$ & $\boldsymbol{3 \pm 6}$ & $-6 \pm 26$ & $-1 \pm 10$ & $-0 \pm 6$\\\hline
\quad\textit{\textbf{Tensor-Tensor-Scalar}} &&&&&&\\
$\widetilde{W}W$ & ($\times 10^{-2}$) & $29 \pm 460$ & $31 \pm 67$ & $\boldsymbol{5 \pm 11}$ & $22 \pm 449$ & $-4 \pm 64$ & $1 \pm 11$\\\hline
\quad\textit{\textbf{Tensor-Scalar-Scalar}} &&&&&&\\
Squeezed & ($\times 10^{0}$) & $-17 \pm 31$ & $9 \pm 15$ & $\boldsymbol{13 \pm 10}$ & $-1 \pm 31$ & $-1 \pm 15$ & $-0 \pm 10$\\
\end{tabular}
    \caption{\resub{As Tab.\,\ref{tab: all-results} but showing results obtained using the \textsc{smica} component-separation pipeline. Monte Carlo errors and simulation results are obtained using $100$ FFP10 realizations. We find only small differences between \textsc{sevem} and \textsc{smica}, indicating minimal bias from residual foregrounds.}}
    \label{tab: all-results-smica}
\end{table}

\resub{We also show} the constraints obtained by restricting our attention to $T$- and $E$-, or just $T$-modes. For parity-conserving bispectrum templates ($W^3$, squeezed and equilateral tensor-tensor-tensor and squeezed tensor-scalar-scalar), we find a modest gain in signal-to-noise from the inclusion of $B$-modes, with $T+E$ capturing most of the information. In contrast, constraints on the parity-violating templates ($\widetilde{W}W^2$ and axion tensor-tensor-tensor and $\widetilde{W}W$ tensor-tensor-scalar) are dominated by the $B$-modes, with a \resub{$(2-7)\times$} improvement factor observed upon their inclusion. This matches our expectations: $B$-modes are particularly useful in parity-violation analyses \citep{Philcox:2023xxk}. In Appendix \ref{app: extra-plots} we consider the contribution of each field-triplet to the overall constraints, which reinforces this conclusion: the largest source of signal-to-noise for the parity-odd models comes from $TTB$ (and sometimes $TEB$) spectra. For parity-even models, the signal-to-noise is distributed more evenly between the ten non-trivial spectra. Furthermore, we consider the dependence of our constraints on $\ell_{\rm max}$, finding that tensor-tensor-tensor and tensor-tensor-scalar constraints saturate by $\ell_{\rm max}\approx 100$, as expected. For the squeezed tensor-scalar-scalar shape, higher $\ell$-modes are found to yield significant gains in constraining power, matching the forecasts of Fig.\,\ref{fig: fisher}.

\begin{figure}
    \centering
    \subfloat[Tensor-Tensor-Tensor]{\includegraphics[width=0.6\textwidth]{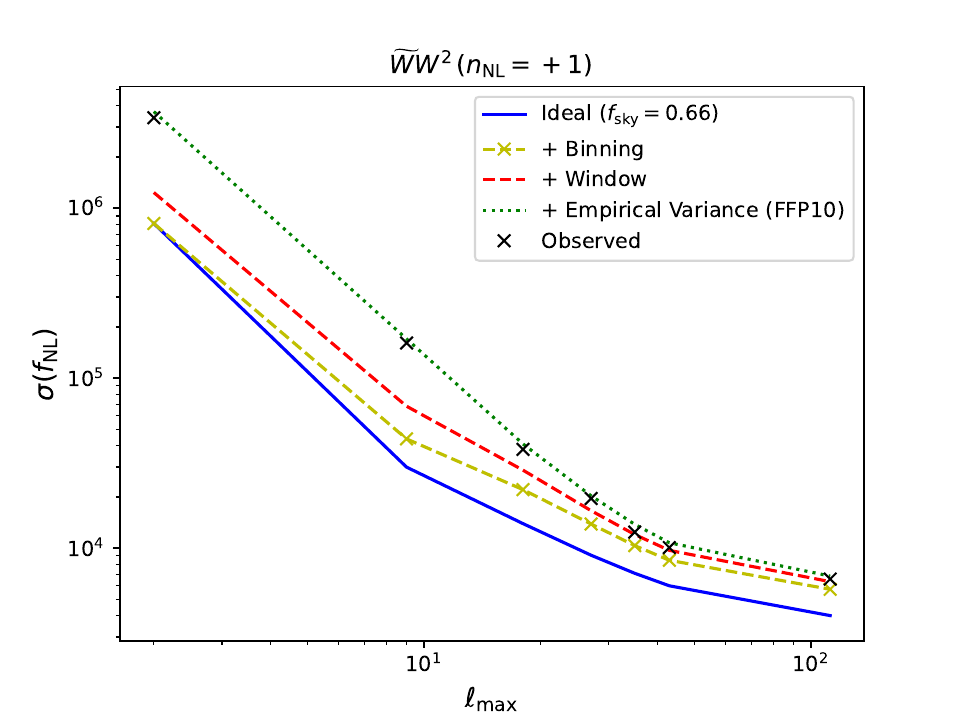}}\\
    \subfloat[Tensor-Tensor-Scalar]{\includegraphics[width=0.48\textwidth]{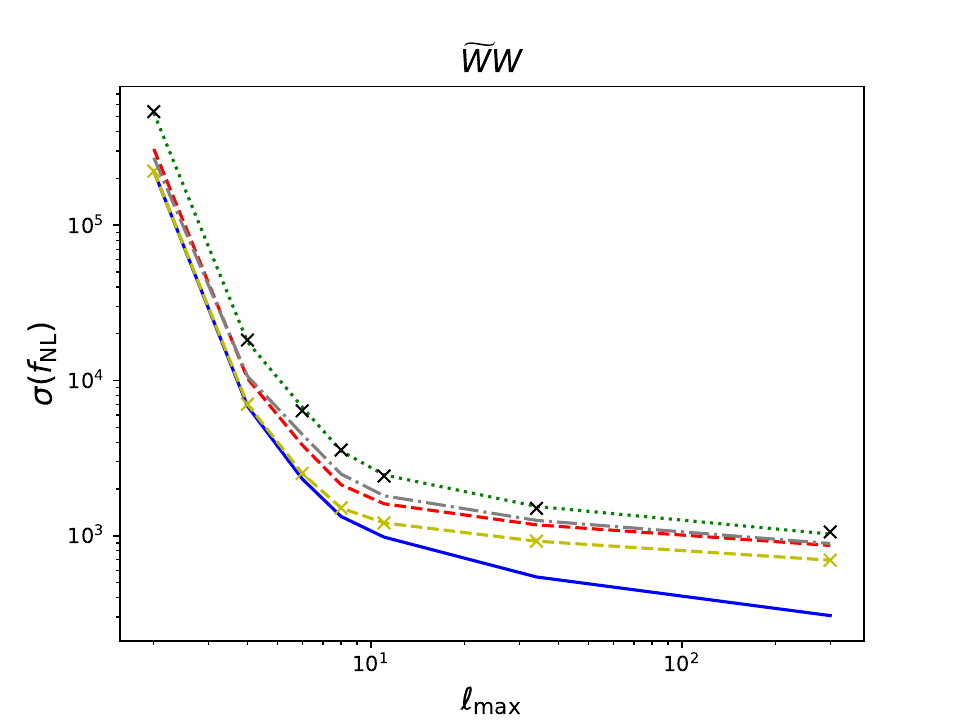}}
    \subfloat[Tensor-Scalar-Scalar]{\includegraphics[width=0.48\textwidth]{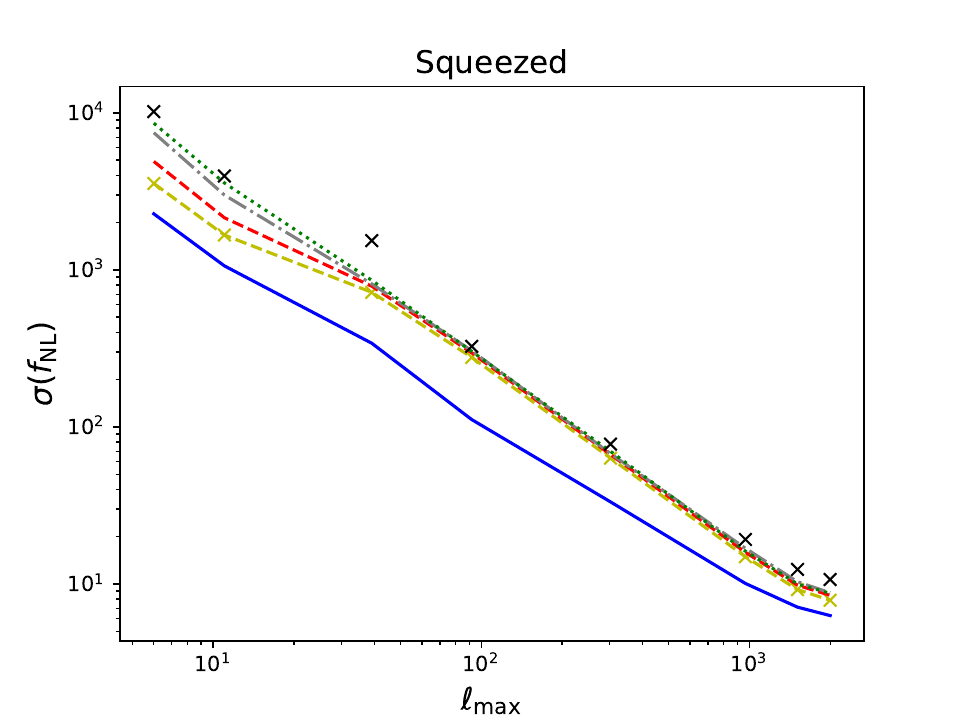}}
    \caption{Forecasted constraints on three bispectrum templates (using a single exemplar tensor-tensor-tensor model), assessing the impact of a number of systematic effects via Fisher analyses with different input covariance matrices. The various forecasts are: without binning or a mask (blue), including binning but with no mask (yellow), including binning and a mask, assuming that the bispectrum estimator is optimal (red), allowing for a suboptimal estimator with a variance measured from Gaussian random fields (gray), including any additional sources of variance from the FFP10 simulations (green). The difference between the idealized and true errors arises from a combination of many of these effects, particularly the non-trivial mask geometry and coarse binning scheme (with the latter dominating for the tensor-tensor-scalar model).}
    \label{fig: breakdown}
\end{figure}

It is instructive to compare our constraints to those expected from Fisher forecasts (performed as discussed in \S\ref{sec: fisher}). In general, the \textit{Planck} errorbars are wider than the idealized forecasts by $\approx (50-100)\%$ (cf.\,Tab\,\ref{tab: all-results}), with the largest deviations seen for the parity-breaking templates, particularly $\widetilde{W}W$ (with a $3\times$ loss of information). This difference reduces if we exclude $B$-modes and is smallest if we consider only $T$-modes, suggesting that the polarization data is to blame. To elucidate this, we perform a number of forecasts with varying degrees of realism, the results of which are shown in Fig.\,\ref{fig: breakdown} for three representative template analyses. 

Firstly, we find significant information loss from binning; this is a natural consequence of our coarse $\ell$-bins and leads to a $(15-50)\%$ inflation in $\sigma(f_{\rm NL})$, as discussed above. For the tensor-tensor-scalar template, this is more significant, and leads to a loss of signal-to-noise by almost $130\%$. As discussed above, this is expected to arise due to the oscillations within each $\ell$-bin that are averaged over. Notably, we also find significant $(10-40)\%$ impacts (across all templates considered) from the spatially varying mask, which has strong power at low-$\ell$, and cannot be simply captured by an $f_{\rm sky}$ parameter in the forecast. Suboptimality of the \resub{\textsc{PolySpec}} bispectrum estimators (traced by a forecast with Gaussian simulations) causes a $(5-30)\%$ effect -- this is a direct result of approximations in our $\Si$ weighting function (particularly for polarization) and could potentially be rectified by more nuanced Wiener filtering schemes \citep[e.g.,][]{Munchmeyer:2019kng}. Finally, we find that forecasts using FFP10 simulations suffer a further $(1-80)\%$ inflation of $\sigma(f_{\rm NL})$ with respect to those of Gaussian random fields (with large effects found for models with significant large-scale power, \textit{i.e.}\ those with $n_{\rm NL}=-1$, \resub{but negligible for the tensor-scalar-scalar model}); this is attributed to residual foreground contamination, spatially varying noise properties and intrinsic non-Gaussianity of the maps \resub{and matches the increased variances seen in Fig.\,\ref{fig: beta-plot}}. Overall, we find that the suboptimality is sourced by a variety of effects, the majority of which cannot be easily fixed by simple modifications to our analysis pipeline.

\begin{figure}
    \centering
    \includegraphics[width=0.9\textwidth]{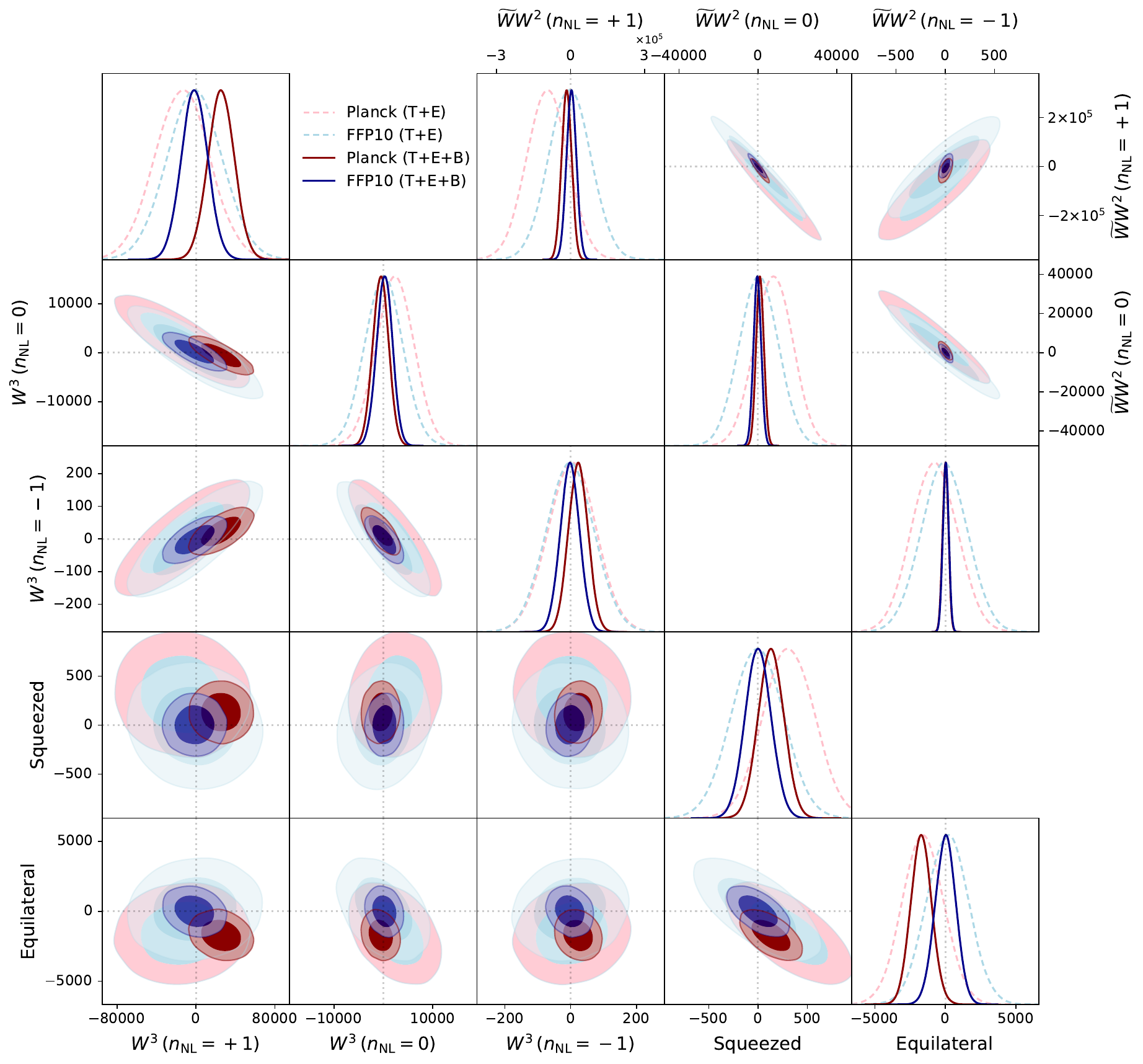}
    \caption{Joint constraints on the tensor-tensor-tensor non-Gaussianity parameters, $f_{\rm NL}^{ttt}$, from the eight models given in \S\ref{sec: models} from \textit{Planck} PR4 data (red) and the mean of 500 FFP10 simulations (blue), \resub{all processed with the \textsc{sevem} component-separation pipeline}. We separately analyze the parity-conserving (bottom left) and parity-breaking (top right) templates, which are uncorrelated (up to masking and systematic effects) and show results from a $(T+E)$-only analysis in light pink and blue. We observe significant correlations between templates, particularly the various Weyl operators, and a slight preference for non-zero $W^3(n_{\rm NL}=0)$ and equilateral, models, though no preference is found in the (more-constraining) single template analyses. We do not include the axion model in this plot since it is degenerate with a combination of $W^3(n_{\rm NL}=0)$ and $\widetilde{W}W^2(n_{\rm NL}=0)$ models.}
    \label{fig: corner-plot}
\end{figure}

Finally, we perform joint analyses of the tensor-tensor-tensor bispectrum templates. This allows an assessment of the correlation between various models and facilitates their distinguishment. For this purpose, we split the models into two groups: parity-conserving and parity-breaking, noting that the bispectra sourced by the two are uncorrelated up to geometry and systematic effects (ignoring the $\widetilde{F}F$ axion model, given its complete degeneracy with the Weyl models). The resulting constraints are shown in Fig.\,\ref{fig: corner-plot}. In almost all cases, the correlations between templates imply that the joint $f_{\rm NL}^{ttt}$ bounds are considerably weaker than those of the single-model analyses reported in Tab.\,\ref{tab: all-results}. We find significant correlation between many of the templates, particularly those of the two sets of Weyl operators with different time-dependencies. Notably, we also find a mild preference for non-zero $W^3(n_{\rm NL
}=+1)$ and equilateral tensor templates which have the marginalized constraints: \resub{$f_{\rm NL}^{ttt} =  (2.6\pm1.3)\times10^4$ and $(-17\pm7)\times 10^2$ respectively, \textit{i.e.}\ $1.9\sigma$ and $(-)2.3\sigma$. When analyzing the \textsc{smica} data, we find comparable results: $f_{\rm NL}^{ttt} =  (2.2\pm1.3)\times10^4$ and $(-20\pm7)\times 10^2$, with the latter reaching $2.7\sigma$}. We do not attribute any detection to these however, given that \resub{(a) the detection significances are small}, (b) the shapes are significantly correlated with other templates, (c) we do not account for look-elsewhere effects, and (d) we find no preference in the individual analyses in Tab.\,\ref{tab: all-results}\,\&\,\ref{tab: all-results-smica}.

\section{Summary \& Discussion}\label{sec: conclusions}
\noindent Unlike scalar fluctuations, little is known about the primordial tensor degrees of freedom. Despite a number of targeted searches using a variety of probes, inflationary gravitational waves remain undetected. A measurement of their power spectrum could yield great insight into the primordial universe; moreover, characterizing their statistical properties would give a new window into physics at the highest energies. In contrast to the scalar sector, the tensor sector is not known to be close-to Gaussian, and there exist many physical mechanisms that could source detectable tensor (or mixed tensor-scalar) non-Gaussianity without a first detection in the two-point function. 

A brief literature search will reveal a wide array of theoretical work devoted to predicting and characterizing such signatures \citep[e.g.,][]{Dimastrogiovanni:2018gkl,Dimastrogiovanni:2022afr,Dimastrogiovanni:2018uqy,Raveendran:2016wjz,Lee:2016vti,Watanabe:2010fh,Barnaby:2012xt,Lue:1998mq,Sorbo:2011rz,Barnaby:2011vw,Barnaby:2012xt,Cook:2013xea,Namba:2015gja,Dimastrogiovanni:2016fuu,Maleknejad:2012fw,Komatsu:2022nvu,Thorne:2017jft,Agrawal:2017awz,Shiraishi:2011st,Shiraishi:2013kxa,Niu:2022quw,Hiramatsu:2020jes,Chowdhury:2016yrh,Gao:2011vs,Gao:2012ib,DeLuca:2019jzc,Maldacena:2011nz,Alexander:2004wk,Soda:2011am,Bartolo:2017szm,Bartolo:2018elp,Mylova:2019jrj,Bartolo:2020gsh,Huang:2013epa,Ozsoy:2019slf,Namjoo:2012aa,Martin:2012pe,Shiraishi:2012sn,Gong:2023kpe,Kanno:2022mkx,Shiraishi:2016yun,Berezhiani:2014kga,Akama:2020jko,Naskar:2020vkd,Peng:2024eok,Maldacena:2002vr,Duivenvoorden:2019ses,Meerburg:2016ecv,Akama:2020jko,Adshead:2009bz,Naskar:2019shl,Cabass:2021fnw,Cabass:2022jda,Bordin:2020eui,Cabass:2021iii,Pajer:2020wxk,Baumann:2020dch,Naskar:2019shl,Naskar:2018rmu,Domenech:2017kno,Fujita:2018ehq,Fujita:2019tov,Gao:2011vs,Gao:2012ib,Shiraishi:2011dh}. Much less effort has been devoted to forecasting current or future constraints on such parameters (though see \citep[e.g.,][]{Meerburg:2016ecv,Duivenvoorden:2019ses,Shiraishi:2019yux,Shiraishi:2012rm,Shiraishi:2013vha,DeLuca:2019jzc,Shiraishi:2011st,Namba:2015gja,Shiraishi:2016yun,Bartolo:2018elp,Shiraishi:2010kd,Domenech:2017kno,SimonsObservatory:2018koc,Shiraishi:2013kxa,Shiraishi:2012sn,LiteBIRD:2024twk,Tahara:2017wud} for notable exceptions). Even fewer papers have actually attempted to constrain such models from data; to our knowledge, these are limited to just five models analyzed with WMAP or \textit{Planck} temperature anisotropies \citep{Shiraishi:2013wua,Planck:2015zrl,DeLuca:2019jzc,Shiraishi:2014ila,Shiraishi:2017yrq,Planck:2019kim,Planck:2015zfm} and just one model with polarization data \citep{Planck:2019kim,Planck:2015zfm,Philcox:2023xxk}. In this work, our goal has been to rectify this situation and place constraints on a wide variety of tensor and mixed tensor-scalar bispectrum templates using the latest observational large-scale temperature and polarization data. 

To achieve this, we have measured $T$-, $E$- and $B$-mode bispectra from \textit{Planck} PR4 data using modern (quasi-optimal) binned bispectrum estimators \citep{Philcox:2023uwe,Philcox:2023psd,PolyBin}. These have then been directly compared to analytic templates which span a variety of phenomenological regimes (equilateral/local, parity-even/parity-odd and tensor-tensor-tensor/tensor-tensor-scalar/tensor-scalar-scalar), leading to constraints on eleven $f_{\rm NL}$ parameters as a function of scale and field configuration (marginalizing over secondary contamination). All our analyses have returned null results; we find no evidence for any tensor non-Gaussianity in the \textit{Planck} dataset. However, we have placed strong constraints on a variety of models, which may be recast into physical parameters as discussed in \S\ref{sec: models}.

It is interesting to place our constraints in context of past bounds. Comparing Tab.\,\ref{tab:fNL_prev}\,\&\,\ref{tab: all-results}, we find significant ($1.2-7\times$) reduction in $\sigma(f_{\rm NL})$ for all templates analyzed, with six models analyzed for the first time. Much of this improvement is driven by the addition of polarization information, with $B$-modes seen to contain considerable constraining power. When restricting to just temperature anisotropies, the constraints from our pipeline are generally somewhat weaker than those of previous analyses (up to $2\times$ in some cases). This is not surprising; our binning strategies are optimized only for the full $T+E+B$-mode analysis, thus we suffer from significant information loss when restricting to only $T$-modes (and drop the majority of the bins in our data-vector). Comparing the full dataset $\widetilde{F}F$ constraints from Tab.\,\ref{tab: all-results} to those from \citep{Philcox:2023xxk} (which used the same analysis methods and dataset), we find $\approx 40\%$ improvements. This is due to our newly developed optimized binning and weighting strategies (\S\ref{sec: fisher}), which significantly reduces the losses inherent to bin-averaging. 

Looking to the future, the constraints on many of the models considered herein are expected to tighten dramatically. In particular, \citep{Shiraishi:2013vha,Shiraishi:2019yux,LiteBIRD:2024twk} forecasts a factor of $\approx 3$ improvement on $\sigma(f_{\rm NL}^{ttt,\rm sq})$ from a futuristic $T+E$-mode analysis, or up to $100\times$ improvement when $B$-modes are included. This could tighten the bounds on the amplitude of primordial magnetic fields by $\approx (2-3)\times$ (noting the prohibitive sextic scaling in \eqref{eq: pmf-scaling}). The axion $\widetilde{F}F$ tensor constraints are also predicted to tighten significantly: for a LiteBIRD-like experiment with $r \lesssim 0.03$, one expects $\sigma(f_{\rm NL}^{ttt,\widetilde{F}F}) = \mathcal{O}(1)$ (which can be compared to the current $\mathcal{O}(500)$ constraint) \citep{Shiraishi:2019yux,Shiraishi:2013kxa}. Templates that peak in squeezed configurations are subject to particularly large improvements: \citep{Bartolo:2018elp} predict $\sigma(f_{\rm NL}^{tts,\widetilde{W}W}) =\mathcal{O}(10)$ from a
futuristic experiment using $TBB$ and $EBB$ spectra and
\citep{Shiraishi:2019yux,Domenech:2017kno,Meerburg:2016ecv,Duivenvoorden:2019ses} forecast $\sigma(f_{\rm NL}^{tss,\rm sq}) = \mathcal{O}(0.1 - 1)$ from LiteBIRD, CMB-S4 or upcoming experiments (depending on the efficacy of foreground removal and delensing). 
$B$-modes play a key role in these forecasts; unlike $T$- and $E$-modes they will not be cosmic-variance-limited in the near future (unless $r$ is large or we do not delens sufficiently). This explains why bounds on tensor non-Gaussianity are expected to tighten far faster than those on scalar non-Gaussianity with next generation surveys.

Finally, let us reflect on the limitations of our approach, as well as hurdles that will need to be overcome to realize the above forecasts in practice. As is evident from Tab.\,\ref{tab: all-results} and Fig.\,\ref{fig: breakdown}, the \textit{Planck} $f_{\rm NL}$ constraints obtained herein are considerably weaker than the Fisher forecast predictions (by a factor of $1.5-3$). This is due to a number of effects: information is lost due to the coarse $\ell$-bins adopted for computational efficiency, the mask has a non-trivial structure at low-$\ell$, residual foregrounds may exist in the \textit{Planck} data (and FFP10/\textsc{npipe} simulations), our estimators are not quite optimal, and there may be additional sources of covariance beyond the fiducial power spectra, \resub{but present in the simulations}. As shown in Fig.\,\ref{fig: fisher}, the binning has a moderately large effect -- although its impacts have been somewhat mollified by our data-driven weighting schemes, we still lose information, particularly for the $\widetilde{W}W$ template. An alternative approach (used in the \textit{Planck} tensor analyses \citep[e.g.,][]{Planck:2015zrl,Planck:2015zfm,Planck:2019kim}) would be to abandon the binned estimators and instead use modal decompositions \cite{Shiraishi:2019exr,Shiraishi:2014roa,2009PhRvD..80d3510F}. This was shown to perform well in previous analyses, though of course is subject to its own limitations. The optimal approach would be to perform direct template analyses as in \citep{Duivenvoorden:2019ses}; however, this is often highly computationally demanding and must be performed separately for each template. Suboptimalities in the estimator are a smaller effect (though particularly relevant for squeezed templates and future inhomogeneous surveys); to reduce these, one can adopt improved (possibly non-linear) weighting schemes, described, for example in \citep{Munchmeyer:2019kng}. Finally, there may be additional covariance arising from residual foregrounds and CMB lensing. Whilst the former is difficult to remove without better cleaning methods, the latter may be aided by delensing -- indeed, \citep{Shiraishi:2016yun,Shiraishi:2019yux,Domenech:2017kno,Duivenvoorden:2019ses,Meerburg:2016ecv,Coulton:2019odk} show that this will be crucial if we wish to obtain tight constraints from $B$-mode data in future low-noise experiments. 

By combining the technical developments discussed above with future high-resolution data, we will soon be able to obtain high-precision constraints on many models of tensor non-Gaussianity. Besides yielding dramatic improvements over the results presented in this work, this will allow a range of new, and hopefully exciting, insights into the primordial Universe.

\vskip 8pt
\acknowledgments
{\small
\begingroup
\hypersetup{hidelinks}
\noindent 
\resub{We thank Adri Duivenvoorden \& Maria Mylova for comments. We are additionally grateful to the anonymous referees for insightful feedback.} OHEP is a Junior Fellow of the Simons Society of Fellows and thanks \href{https://i.redd.it/81bsw0ph66x41.jpg}{Llewellyn the Llama} for his Cambrian consultations. MS is supported by JSPS KAKENHI Grant Nos. JP20H05859 and JP23K03390. MS also acknowledges the Center for Computational Astrophysics, National Astronomical Observatory of Japan, for providing the computing resources of the Cray XC50.
\endgroup
}

\appendix

\section{Dependence of Constraints on Scale-Cuts and Fields}\label{app: extra-plots}

\begin{figure}[!t]
    \centering
    \subfloat[Tensor-Tensor-Tensor]{\includegraphics[width=0.85\textwidth]{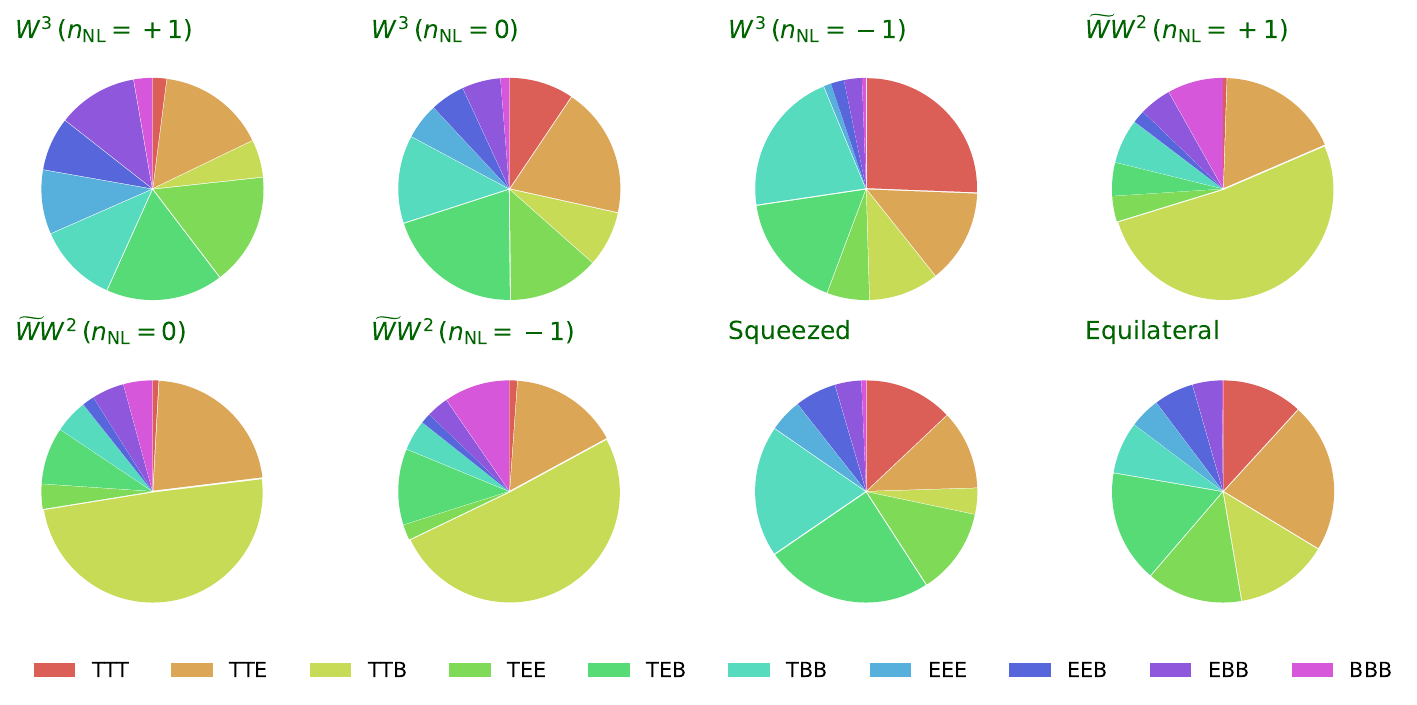}}\\
    \subfloat[Tensor-Tensor-Scalar]{\includegraphics[width=0.3\textwidth]{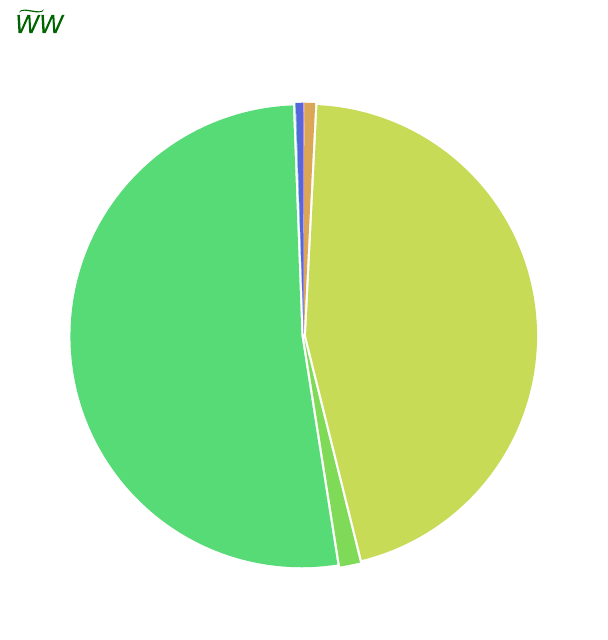}}\qquad
    \subfloat[Tensor-Scalar-Scalar]{\includegraphics[width=0.3\textwidth]{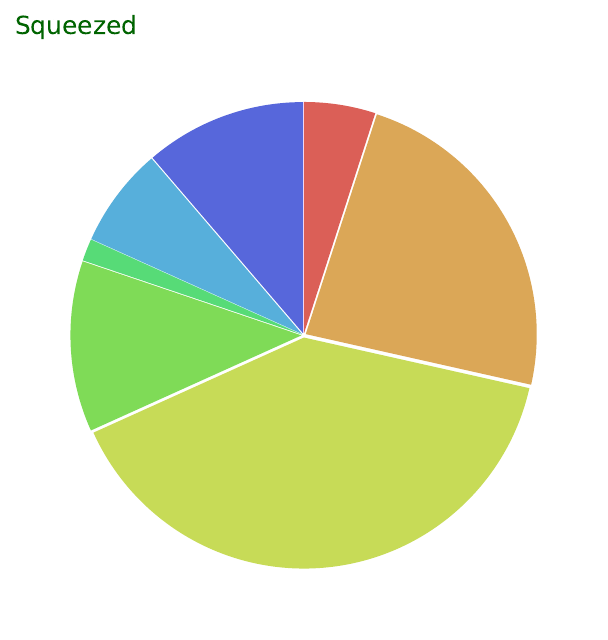}}
    \caption{Fractional contributions to $\sigma^{-2}(f_{\rm NL})$ from a given triplet of fields for each of the ten templates considered in this work (dropping the axion model, which is a sum of $\widetilde{W}W^2(n_{\rm NL}=0)$ and $W^3(n_{\rm NL}=0)$). This is obtained by performing MCMC analyses for each of the ten triplets in turn and for each model separately, \resub{assuming the \textsc{sevem} component-separation pipeline}. The dominant triplets depend heavily on the models in question; for many of the parity-even models, there are contributions from a range of fields, though the parity-odd models ($\widetilde{W}W^2$ and $\widetilde{W}W$) are dominated primarily by spectra containing a single $B$-mode.}
    \label{fig: pie-plots}
\end{figure}

\noindent In Fig.\,\ref{fig: pie-plots}, we consider the contribution of each triplet of fields (comprising $T$, $E$ and $B$) to the constraints on non-Gaussianity models, \resub{assuming the \textsc{sevem} component-separation pipeline}. For the parity-conserving templates, the signal-to-noise is similarly distributed across almost all the configurations (except $BBB$), highlighting the importance of including all fields in the tensor analysis. For parity-breaking templates, the constraints are dominated by $TTB$ and, for $\widetilde{W}W$, $TEB$, with almost vanishing contributions from $TTT$ and $EEE$. For the tensor-scalar-scalar model, we find $TTB$ to dominate; this matches the forecast of \citep{Meerburg:2016ecv,Domenech:2017kno}. Constraints sourced by $B$-modes are limited primarily by the polarization noise (rather than cosmic variance, assuming small $r$); as such, they are expected to tighten considerably with future experiments such as LiteBIRD \citep{Shiraishi:2019yux,LiteBIRD:2022cnt}.

\begin{figure}
    \centering
    \subfloat[Tensor-Tensor-Tensor]{\includegraphics[width=0.85\textwidth]{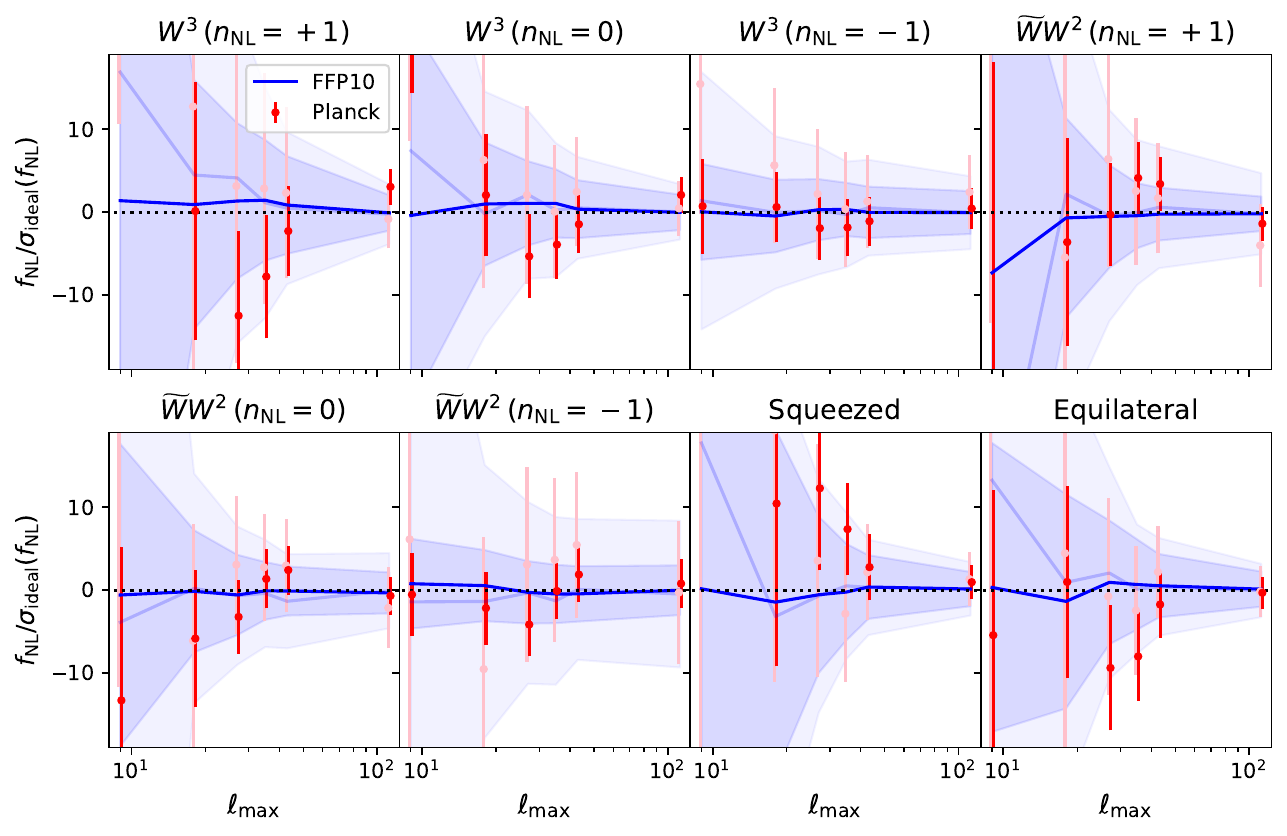}}\\
    \subfloat[Tensor-Tensor-Scalar]{\includegraphics[width=0.35\textwidth]{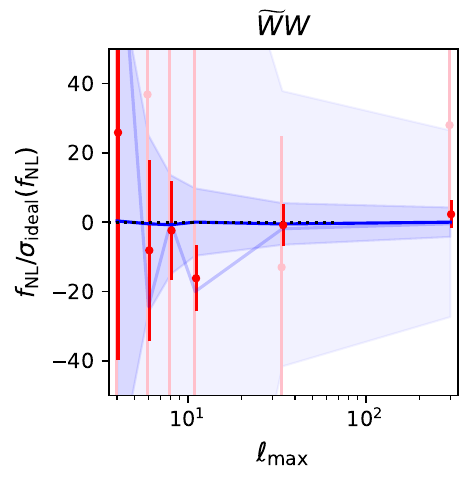}}\qquad
    \subfloat[Tensor-Scalar-Scalar]{\includegraphics[width=0.35\textwidth]{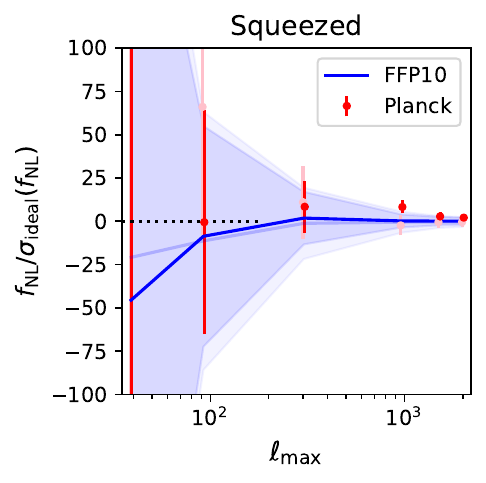}}
    \caption{Dependence of the $f_{\rm NL}$ constraints on the maximum scale included in the analysis, $\ell_{\rm max}$, considering each of the separately. We plot results relative to the unbinned Fisher matrix constraint given in Tab.\,\ref{tab: all-results}. Red data points show the results from \textit{Planck} \resub{(using \textsc{sevem})}, whilst dark blue bands give those from FFP10. The light blue regions and pink errorbars show results from an analysis using only $T$- and $E$-modes. We find sharp dependence on $\ell_{\rm max}$ for squeezed models and equilateral models with $n_{\rm NL}>0$ (blue-tilted), as well as significant improvements on parity-violating model constraints when $B$-modes are added. Whilst the errorbars dance around with scale, we find no clear detection of any model.}
    \label{fig: lmax}
\end{figure}

Fig.\,\ref{fig: lmax} assesses the dependence of the $f_{\rm NL}$ constraints on the maximum scale in the analysis. For tensor-tensor-tensor models, we find strong dependence at low $\ell_{\rm max}$ (particularly for steep models with $n_{\rm NL}>0$, matching Fig.\,\ref{fig: fisher}), but a quick saturation at higher $\ell$, with stable results obtained by $\ell\approx 80$. This matches expectations and arises due to the tensor transfer functions (which do not have integrated Sachs-Wolfe contributions). For the mixed bispectrum, the scaling is stronger due to the presence of scalar transfer functions; in particular, the tensor-scalar-scalar forecast saturates only at $\ell_{\rm max}\gtrsim 1000$. We additionally observe that the $T+E$ constraints are much weaker than $T+E+B$ in almost all cases (particularly for parity-breaking models). Some deviation of the constraints from zero is seen in the $T+E$ data, though this is not statistically significant.

\section{Quasi-Optimal Weights}\label{app: weights}
\begin{figure}[t]
    \centering
    \includegraphics[width=0.9\linewidth]{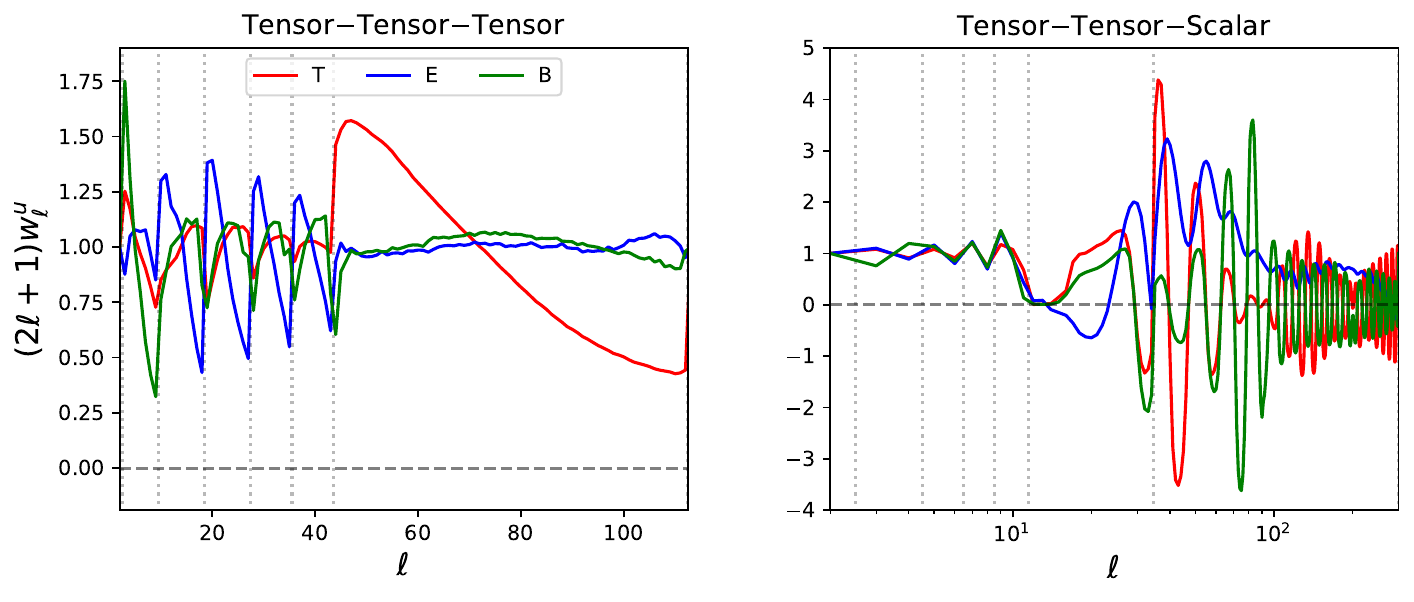}
    \caption{\resub{Quasi-optimal weights used in the tensor-tensor-tensor (left) and tensor-tensor-scalar (right) analyses, computed as described in \S\ref{subsec: optimal-binning}. In each case, we show the weightings as a function of $\ell$, which are arbitrarily normalized within each $\ell$-bin (indicated by vertical dashed lines). The red, blue and green curves show the weights applied to $T$-, $E$-, and $B$-modes respectively. Whilst the tensor-tensor-tensor weights are relatively flat within each bin, the tensor-tensor-scalar weights oscillate at high $\ell$.}}
    \label{fig: weights}
\end{figure}

\noindent \resub{In Fig.\,\ref{fig: weights}, we plot the quasi-optimal weights utilized in the tensor-tensor-tensor and tensor-tensor-scalar analyses. These are computed by numerically maximizing the Fisher information as described in \S\ref{subsec: optimal-binning}, and depend sensitively on both the fiducial power spectrum (including the beam and noise) and the theoretical template(s) of interest. For the tensor-tensor-tensor weights (which are designed to jointly optimize a suite of models), we find fairly small variations within each bin, with a general tendency to upweight low-$\ell$ modes (the left side of each bin), particularly for $E$-modes and at low $\ell$, where squeezed templates are most sensitive. For $\ell\gtrsim 50$, the optimal weights for $E$- and $B$-modes are roughly consistent with the input forms ($w_\ell^u\propto 1/(2\ell+1)$); this occurs due to the sharp fall-off in the tensor transfer function and the increasing polarization noise on small-scales. For the tensor-tensor-scalar bispectrum, we find large variation in the optimal weights for $\ell\gtrsim 30$, particularly for $T$- and $B$-modes. This occurs due to the oscillatory nature of the underlying parity-odd bispectrum template and the accompanying Wigner $3j$ factor -- in essence, we are trying to reproduce the complex weighting schemes present in KSW-type estimators \citep[e.g.,][]{Duivenvoorden:2019ses}. The information loss induced by projecting these oscillatory templates onto broad bins represents a general difficulty for binned bispectrum analyses.}

\bibliographystyle{apsrev4-1}
\bibliography{refs}

\begin{thebibliography}{132}%
\makeatletter
\providecommand \@ifxundefined [1]{%
 \@ifx{#1\undefined}
}%
\providecommand \@ifnum [1]{%
 \ifnum #1\expandafter \@firstoftwo
 \else \expandafter \@secondoftwo
 \fi
}%
\providecommand \@ifx [1]{%
 \ifx #1\expandafter \@firstoftwo
 \else \expandafter \@secondoftwo
 \fi
}%
\providecommand \natexlab [1]{#1}%
\providecommand \enquote  [1]{``#1''}%
\providecommand \bibnamefont  [1]{#1}%
\providecommand \bibfnamefont [1]{#1}%
\providecommand \citenamefont [1]{#1}%
\providecommand \href@noop [0]{\@secondoftwo}%
\providecommand \href [0]{\begingroup \@sanitize@url \@href}%
\providecommand \@href[1]{\@@startlink{#1}\@@href}%
\providecommand \@@href[1]{\endgroup#1\@@endlink}%
\providecommand \@sanitize@url [0]{\catcode `\\12\catcode `\$12\catcode `\&12\catcode `\#12\catcode `\^12\catcode `\_12\catcode `\%12\relax}%
\providecommand \@@startlink[1]{}%
\providecommand \@@endlink[0]{}%
\providecommand \url  [0]{\begingroup\@sanitize@url \@url }%
\providecommand \@url [1]{\endgroup\@href {#1}{\urlprefix }}%
\providecommand \urlprefix  [0]{URL }%
\providecommand \Eprint [0]{\href }%
\providecommand \doibase [0]{http://dx.doi.org/}%
\providecommand \selectlanguage [0]{\@gobble}%
\providecommand \bibinfo  [0]{\@secondoftwo}%
\providecommand \bibfield  [0]{\@secondoftwo}%
\providecommand \translation [1]{[#1]}%
\providecommand \BibitemOpen [0]{}%
\providecommand \bibitemStop [0]{}%
\providecommand \bibitemNoStop [0]{.\EOS\space}%
\providecommand \EOS [0]{\spacefactor3000\relax}%
\providecommand \BibitemShut  [1]{\csname bibitem#1\endcsname}%
\let\auto@bib@innerbib\@empty
\bibitem [{\citenamefont {{Planck Collaboration}}\ \emph {et~al.}(2020)\citenamefont {{Planck Collaboration}}, \citenamefont {{Aghanim}}, \citenamefont {{Akrami}}, \citenamefont {{Ashdown}}, \citenamefont {{Aumont}}, \citenamefont {{Baccigalupi}}, \citenamefont {{Ballardini}}, \citenamefont {{Banday}}, \citenamefont {{Barreiro}}, \citenamefont {{Bartolo}} \emph {et~al.}}]{2020A&A...641A...6P}%
  \BibitemOpen
  \bibfield  {author} {\bibinfo {author} {\bibnamefont {{Planck Collaboration}}}, \bibinfo {author} {\bibfnamefont {N.}~\bibnamefont {{Aghanim}}}, \bibinfo {author} {\bibfnamefont {Y.}~\bibnamefont {{Akrami}}}, \bibinfo {author} {\bibfnamefont {M.}~\bibnamefont {{Ashdown}}}, \bibinfo {author} {\bibfnamefont {J.}~\bibnamefont {{Aumont}}}, \bibinfo {author} {\bibfnamefont {C.}~\bibnamefont {{Baccigalupi}}}, \bibinfo {author} {\bibfnamefont {M.}~\bibnamefont {{Ballardini}}}, \bibinfo {author} {\bibfnamefont {A.~J.}\ \bibnamefont {{Banday}}}, \bibinfo {author} {\bibfnamefont {R.~B.}\ \bibnamefont {{Barreiro}}}, \bibinfo {author} {\bibfnamefont {N.}~\bibnamefont {{Bartolo}}},  \emph {et~al.},\ }\href {\doibase 10.1051/0004-6361/201833910} {\bibfield  {journal} {\bibinfo  {journal} {\aap}\ }\textbf {\bibinfo {volume} {641}},\ \bibinfo {eid} {A6} (\bibinfo {year} {2020})},\ \Eprint {http://arxiv.org/abs/1807.06209} {arXiv:1807.06209 [astro-ph.CO]} \BibitemShut {NoStop}%
\bibitem [{\citenamefont {Akrami}\ \emph {et~al.}(2020{\natexlab{a}})\citenamefont {Akrami} \emph {et~al.}}]{Planck:2018jri}%
  \BibitemOpen
  \bibfield  {author} {\bibinfo {author} {\bibfnamefont {Y.}~\bibnamefont {Akrami}} \emph {et~al.} (\bibinfo {collaboration} {Planck}),\ }\href {\doibase 10.1051/0004-6361/201833887} {\bibfield  {journal} {\bibinfo  {journal} {Astron. Astrophys.}\ }\textbf {\bibinfo {volume} {641}},\ \bibinfo {pages} {A10} (\bibinfo {year} {2020}{\natexlab{a}})},\ \Eprint {http://arxiv.org/abs/1807.06211} {arXiv:1807.06211 [astro-ph.CO]} \BibitemShut {NoStop}%
\bibitem [{\citenamefont {Ade}\ \emph {et~al.}(2021)\citenamefont {Ade} \emph {et~al.}}]{BICEP:2021xfz}%
  \BibitemOpen
  \bibfield  {author} {\bibinfo {author} {\bibfnamefont {P.~A.~R.}\ \bibnamefont {Ade}} \emph {et~al.} (\bibinfo {collaboration} {BICEP, Keck}),\ }\href {\doibase 10.1103/PhysRevLett.127.151301} {\bibfield  {journal} {\bibinfo  {journal} {Phys. Rev. Lett.}\ }\textbf {\bibinfo {volume} {127}},\ \bibinfo {pages} {151301} (\bibinfo {year} {2021})},\ \Eprint {http://arxiv.org/abs/2110.00483} {arXiv:2110.00483 [astro-ph.CO]} \BibitemShut {NoStop}%
\bibitem [{\citenamefont {Meerburg}(2016)}]{Meerburg:2016nhs}%
  \BibitemOpen
  \bibfield  {author} {\bibinfo {author} {\bibfnamefont {P.~D.}\ \bibnamefont {Meerburg}}\ }(\bibinfo {year} {2016})\ \Eprint {http://arxiv.org/abs/1605.04431} {arXiv:1605.04431 [astro-ph.CO]} \BibitemShut {NoStop}%
\bibitem [{\citenamefont {Beutler}\ and\ \citenamefont {McDonald}(2021)}]{Beutler:2021eqq}%
  \BibitemOpen
  \bibfield  {author} {\bibinfo {author} {\bibfnamefont {F.}~\bibnamefont {Beutler}}\ and\ \bibinfo {author} {\bibfnamefont {P.}~\bibnamefont {McDonald}},\ }\href {\doibase 10.1088/1475-7516/2021/11/031} {\bibfield  {journal} {\bibinfo  {journal} {JCAP}\ }\textbf {\bibinfo {volume} {11}},\ \bibinfo {pages} {031} (\bibinfo {year} {2021})},\ \Eprint {http://arxiv.org/abs/2106.06324} {arXiv:2106.06324 [astro-ph.CO]} \BibitemShut {NoStop}%
\bibitem [{\citenamefont {Guzzetti}\ \emph {et~al.}(2016)\citenamefont {Guzzetti}, \citenamefont {Bartolo}, \citenamefont {Liguori},\ and\ \citenamefont {Matarrese}}]{Guzzetti:2016mkm}%
  \BibitemOpen
  \bibfield  {author} {\bibinfo {author} {\bibfnamefont {M.~C.}\ \bibnamefont {Guzzetti}}, \bibinfo {author} {\bibfnamefont {N.}~\bibnamefont {Bartolo}}, \bibinfo {author} {\bibfnamefont {M.}~\bibnamefont {Liguori}}, \ and\ \bibinfo {author} {\bibfnamefont {S.}~\bibnamefont {Matarrese}},\ }\href {\doibase 10.1393/ncr/i2016-10127-1} {\bibfield  {journal} {\bibinfo  {journal} {Riv. Nuovo Cim.}\ }\textbf {\bibinfo {volume} {39}},\ \bibinfo {pages} {399} (\bibinfo {year} {2016})},\ \Eprint {http://arxiv.org/abs/1605.01615} {arXiv:1605.01615 [astro-ph.CO]} \BibitemShut {NoStop}%
\bibitem [{\citenamefont {Maldacena}(2003)}]{Maldacena:2002vr}%
  \BibitemOpen
  \bibfield  {author} {\bibinfo {author} {\bibfnamefont {J.~M.}\ \bibnamefont {Maldacena}},\ }\href {\doibase 10.1088/1126-6708/2003/05/013} {\bibfield  {journal} {\bibinfo  {journal} {JHEP}\ }\textbf {\bibinfo {volume} {05}},\ \bibinfo {pages} {013} (\bibinfo {year} {2003})},\ \Eprint {http://arxiv.org/abs/astro-ph/0210603} {arXiv:astro-ph/0210603} \BibitemShut {NoStop}%
\bibitem [{\citenamefont {Maldacena}\ and\ \citenamefont {Pimentel}(2011)}]{Maldacena:2011nz}%
  \BibitemOpen
  \bibfield  {author} {\bibinfo {author} {\bibfnamefont {J.~M.}\ \bibnamefont {Maldacena}}\ and\ \bibinfo {author} {\bibfnamefont {G.~L.}\ \bibnamefont {Pimentel}},\ }\href {\doibase 10.1007/JHEP09(2011)045} {\bibfield  {journal} {\bibinfo  {journal} {JHEP}\ }\textbf {\bibinfo {volume} {09}},\ \bibinfo {pages} {045} (\bibinfo {year} {2011})},\ \Eprint {http://arxiv.org/abs/1104.2846} {arXiv:1104.2846 [hep-th]} \BibitemShut {NoStop}%
\bibitem [{\citenamefont {Dimastrogiovanni}\ \emph {et~al.}(2019)\citenamefont {Dimastrogiovanni}, \citenamefont {Fasiello}, \citenamefont {Tasinato},\ and\ \citenamefont {Wands}}]{Dimastrogiovanni:2018gkl}%
  \BibitemOpen
  \bibfield  {author} {\bibinfo {author} {\bibfnamefont {E.}~\bibnamefont {Dimastrogiovanni}}, \bibinfo {author} {\bibfnamefont {M.}~\bibnamefont {Fasiello}}, \bibinfo {author} {\bibfnamefont {G.}~\bibnamefont {Tasinato}}, \ and\ \bibinfo {author} {\bibfnamefont {D.}~\bibnamefont {Wands}},\ }\href {\doibase 10.1088/1475-7516/2019/02/008} {\bibfield  {journal} {\bibinfo  {journal} {JCAP}\ }\textbf {\bibinfo {volume} {02}},\ \bibinfo {pages} {008} (\bibinfo {year} {2019})},\ \Eprint {http://arxiv.org/abs/1810.08866} {arXiv:1810.08866 [astro-ph.CO]} \BibitemShut {NoStop}%
\bibitem [{\citenamefont {Dimastrogiovanni}\ \emph {et~al.}(2022)\citenamefont {Dimastrogiovanni}, \citenamefont {Fasiello},\ and\ \citenamefont {Pinol}}]{Dimastrogiovanni:2022afr}%
  \BibitemOpen
  \bibfield  {author} {\bibinfo {author} {\bibfnamefont {E.}~\bibnamefont {Dimastrogiovanni}}, \bibinfo {author} {\bibfnamefont {M.}~\bibnamefont {Fasiello}}, \ and\ \bibinfo {author} {\bibfnamefont {L.}~\bibnamefont {Pinol}},\ }\href {\doibase 10.1088/1475-7516/2022/09/031} {\bibfield  {journal} {\bibinfo  {journal} {JCAP}\ }\textbf {\bibinfo {volume} {09}},\ \bibinfo {pages} {031} (\bibinfo {year} {2022})},\ \Eprint {http://arxiv.org/abs/2203.17192} {arXiv:2203.17192 [astro-ph.CO]} \BibitemShut {NoStop}%
\bibitem [{\citenamefont {Dimastrogiovanni}\ \emph {et~al.}(2018{\natexlab{a}})\citenamefont {Dimastrogiovanni}, \citenamefont {Fasiello},\ and\ \citenamefont {Tasinato}}]{Dimastrogiovanni:2018uqy}%
  \BibitemOpen
  \bibfield  {author} {\bibinfo {author} {\bibfnamefont {E.}~\bibnamefont {Dimastrogiovanni}}, \bibinfo {author} {\bibfnamefont {M.}~\bibnamefont {Fasiello}}, \ and\ \bibinfo {author} {\bibfnamefont {G.}~\bibnamefont {Tasinato}},\ }\href {\doibase 10.1088/1475-7516/2018/08/016} {\bibfield  {journal} {\bibinfo  {journal} {JCAP}\ }\textbf {\bibinfo {volume} {08}},\ \bibinfo {pages} {016} (\bibinfo {year} {2018}{\natexlab{a}})},\ \Eprint {http://arxiv.org/abs/1806.00850} {arXiv:1806.00850 [astro-ph.CO]} \BibitemShut {NoStop}%
\bibitem [{\citenamefont {Raveendran}\ and\ \citenamefont {Sriramkumar}(2017)}]{Raveendran:2016wjz}%
  \BibitemOpen
  \bibfield  {author} {\bibinfo {author} {\bibfnamefont {R.~N.}\ \bibnamefont {Raveendran}}\ and\ \bibinfo {author} {\bibfnamefont {L.}~\bibnamefont {Sriramkumar}},\ }\href {\doibase 10.1088/1475-7516/2017/07/035} {\bibfield  {journal} {\bibinfo  {journal} {JCAP}\ }\textbf {\bibinfo {volume} {07}},\ \bibinfo {pages} {035} (\bibinfo {year} {2017})},\ \Eprint {http://arxiv.org/abs/1611.00473} {arXiv:1611.00473 [astro-ph.CO]} \BibitemShut {NoStop}%
\bibitem [{\citenamefont {Lee}\ \emph {et~al.}(2016)\citenamefont {Lee}, \citenamefont {Baumann},\ and\ \citenamefont {Pimentel}}]{Lee:2016vti}%
  \BibitemOpen
  \bibfield  {author} {\bibinfo {author} {\bibfnamefont {H.}~\bibnamefont {Lee}}, \bibinfo {author} {\bibfnamefont {D.}~\bibnamefont {Baumann}}, \ and\ \bibinfo {author} {\bibfnamefont {G.~L.}\ \bibnamefont {Pimentel}},\ }\href {\doibase 10.1007/JHEP12(2016)040} {\bibfield  {journal} {\bibinfo  {journal} {JHEP}\ }\textbf {\bibinfo {volume} {12}},\ \bibinfo {pages} {040} (\bibinfo {year} {2016})},\ \Eprint {http://arxiv.org/abs/1607.03735} {arXiv:1607.03735 [hep-th]} \BibitemShut {NoStop}%
\bibitem [{\citenamefont {Watanabe}\ \emph {et~al.}(2010)\citenamefont {Watanabe}, \citenamefont {Kanno},\ and\ \citenamefont {Soda}}]{Watanabe:2010fh}%
  \BibitemOpen
  \bibfield  {author} {\bibinfo {author} {\bibfnamefont {M.-a.}\ \bibnamefont {Watanabe}}, \bibinfo {author} {\bibfnamefont {S.}~\bibnamefont {Kanno}}, \ and\ \bibinfo {author} {\bibfnamefont {J.}~\bibnamefont {Soda}},\ }\href {\doibase 10.1143/PTP.123.1041} {\bibfield  {journal} {\bibinfo  {journal} {Prog. Theor. Phys.}\ }\textbf {\bibinfo {volume} {123}},\ \bibinfo {pages} {1041} (\bibinfo {year} {2010})},\ \Eprint {http://arxiv.org/abs/1003.0056} {arXiv:1003.0056 [astro-ph.CO]} \BibitemShut {NoStop}%
\bibitem [{\citenamefont {Barnaby}\ \emph {et~al.}(2012)\citenamefont {Barnaby}, \citenamefont {Moxon}, \citenamefont {Namba}, \citenamefont {Peloso}, \citenamefont {Shiu},\ and\ \citenamefont {Zhou}}]{Barnaby:2012xt}%
  \BibitemOpen
  \bibfield  {author} {\bibinfo {author} {\bibfnamefont {N.}~\bibnamefont {Barnaby}}, \bibinfo {author} {\bibfnamefont {J.}~\bibnamefont {Moxon}}, \bibinfo {author} {\bibfnamefont {R.}~\bibnamefont {Namba}}, \bibinfo {author} {\bibfnamefont {M.}~\bibnamefont {Peloso}}, \bibinfo {author} {\bibfnamefont {G.}~\bibnamefont {Shiu}}, \ and\ \bibinfo {author} {\bibfnamefont {P.}~\bibnamefont {Zhou}},\ }\href {\doibase 10.1103/PhysRevD.86.103508} {\bibfield  {journal} {\bibinfo  {journal} {Phys. Rev. D}\ }\textbf {\bibinfo {volume} {86}},\ \bibinfo {pages} {103508} (\bibinfo {year} {2012})},\ \Eprint {http://arxiv.org/abs/1206.6117} {arXiv:1206.6117 [astro-ph.CO]} \BibitemShut {NoStop}%
\bibitem [{\citenamefont {Lue}\ \emph {et~al.}(1999)\citenamefont {Lue}, \citenamefont {Wang},\ and\ \citenamefont {Kamionkowski}}]{Lue:1998mq}%
  \BibitemOpen
  \bibfield  {author} {\bibinfo {author} {\bibfnamefont {A.}~\bibnamefont {Lue}}, \bibinfo {author} {\bibfnamefont {L.-M.}\ \bibnamefont {Wang}}, \ and\ \bibinfo {author} {\bibfnamefont {M.}~\bibnamefont {Kamionkowski}},\ }\href {\doibase 10.1103/PhysRevLett.83.1506} {\bibfield  {journal} {\bibinfo  {journal} {Phys. Rev. Lett.}\ }\textbf {\bibinfo {volume} {83}},\ \bibinfo {pages} {1506} (\bibinfo {year} {1999})},\ \Eprint {http://arxiv.org/abs/astro-ph/9812088} {arXiv:astro-ph/9812088} \BibitemShut {NoStop}%
\bibitem [{\citenamefont {Sorbo}(2011)}]{Sorbo:2011rz}%
  \BibitemOpen
  \bibfield  {author} {\bibinfo {author} {\bibfnamefont {L.}~\bibnamefont {Sorbo}},\ }\href {\doibase 10.1088/1475-7516/2011/06/003} {\bibfield  {journal} {\bibinfo  {journal} {JCAP}\ }\textbf {\bibinfo {volume} {06}},\ \bibinfo {pages} {003} (\bibinfo {year} {2011})},\ \Eprint {http://arxiv.org/abs/1101.1525} {arXiv:1101.1525 [astro-ph.CO]} \BibitemShut {NoStop}%
\bibitem [{\citenamefont {Barnaby}\ \emph {et~al.}(2011)\citenamefont {Barnaby}, \citenamefont {Namba},\ and\ \citenamefont {Peloso}}]{Barnaby:2011vw}%
  \BibitemOpen
  \bibfield  {author} {\bibinfo {author} {\bibfnamefont {N.}~\bibnamefont {Barnaby}}, \bibinfo {author} {\bibfnamefont {R.}~\bibnamefont {Namba}}, \ and\ \bibinfo {author} {\bibfnamefont {M.}~\bibnamefont {Peloso}},\ }\href {\doibase 10.1088/1475-7516/2011/04/009} {\bibfield  {journal} {\bibinfo  {journal} {JCAP}\ }\textbf {\bibinfo {volume} {04}},\ \bibinfo {pages} {009} (\bibinfo {year} {2011})},\ \Eprint {http://arxiv.org/abs/1102.4333} {arXiv:1102.4333 [astro-ph.CO]} \BibitemShut {NoStop}%
\bibitem [{\citenamefont {Cook}\ and\ \citenamefont {Sorbo}(2013)}]{Cook:2013xea}%
  \BibitemOpen
  \bibfield  {author} {\bibinfo {author} {\bibfnamefont {J.~L.}\ \bibnamefont {Cook}}\ and\ \bibinfo {author} {\bibfnamefont {L.}~\bibnamefont {Sorbo}},\ }\href {\doibase 10.1088/1475-7516/2013/11/047} {\bibfield  {journal} {\bibinfo  {journal} {JCAP}\ }\textbf {\bibinfo {volume} {11}},\ \bibinfo {pages} {047} (\bibinfo {year} {2013})},\ \Eprint {http://arxiv.org/abs/1307.7077} {arXiv:1307.7077 [astro-ph.CO]} \BibitemShut {NoStop}%
\bibitem [{\citenamefont {Namba}\ \emph {et~al.}(2016)\citenamefont {Namba}, \citenamefont {Peloso}, \citenamefont {Shiraishi}, \citenamefont {Sorbo},\ and\ \citenamefont {Unal}}]{Namba:2015gja}%
  \BibitemOpen
  \bibfield  {author} {\bibinfo {author} {\bibfnamefont {R.}~\bibnamefont {Namba}}, \bibinfo {author} {\bibfnamefont {M.}~\bibnamefont {Peloso}}, \bibinfo {author} {\bibfnamefont {M.}~\bibnamefont {Shiraishi}}, \bibinfo {author} {\bibfnamefont {L.}~\bibnamefont {Sorbo}}, \ and\ \bibinfo {author} {\bibfnamefont {C.}~\bibnamefont {Unal}},\ }\href {\doibase 10.1088/1475-7516/2016/01/041} {\bibfield  {journal} {\bibinfo  {journal} {JCAP}\ }\textbf {\bibinfo {volume} {01}},\ \bibinfo {pages} {041} (\bibinfo {year} {2016})},\ \Eprint {http://arxiv.org/abs/1509.07521} {arXiv:1509.07521 [astro-ph.CO]} \BibitemShut {NoStop}%
\bibitem [{\citenamefont {Dimastrogiovanni}\ \emph {et~al.}(2017)\citenamefont {Dimastrogiovanni}, \citenamefont {Fasiello},\ and\ \citenamefont {Fujita}}]{Dimastrogiovanni:2016fuu}%
  \BibitemOpen
  \bibfield  {author} {\bibinfo {author} {\bibfnamefont {E.}~\bibnamefont {Dimastrogiovanni}}, \bibinfo {author} {\bibfnamefont {M.}~\bibnamefont {Fasiello}}, \ and\ \bibinfo {author} {\bibfnamefont {T.}~\bibnamefont {Fujita}},\ }\href {\doibase 10.1088/1475-7516/2017/01/019} {\bibfield  {journal} {\bibinfo  {journal} {JCAP}\ }\textbf {\bibinfo {volume} {01}},\ \bibinfo {pages} {019} (\bibinfo {year} {2017})},\ \Eprint {http://arxiv.org/abs/1608.04216} {arXiv:1608.04216 [astro-ph.CO]} \BibitemShut {NoStop}%
\bibitem [{\citenamefont {Agrawal}\ \emph {et~al.}(2018{\natexlab{a}})\citenamefont {Agrawal}, \citenamefont {Fujita},\ and\ \citenamefont {Komatsu}}]{Agrawal:2017awz}%
  \BibitemOpen
  \bibfield  {author} {\bibinfo {author} {\bibfnamefont {A.}~\bibnamefont {Agrawal}}, \bibinfo {author} {\bibfnamefont {T.}~\bibnamefont {Fujita}}, \ and\ \bibinfo {author} {\bibfnamefont {E.}~\bibnamefont {Komatsu}},\ }\href {\doibase 10.1103/PhysRevD.97.103526} {\bibfield  {journal} {\bibinfo  {journal} {Phys. Rev. D}\ }\textbf {\bibinfo {volume} {97}},\ \bibinfo {pages} {103526} (\bibinfo {year} {2018}{\natexlab{a}})},\ \Eprint {http://arxiv.org/abs/1707.03023} {arXiv:1707.03023 [astro-ph.CO]} \BibitemShut {NoStop}%
\bibitem [{\citenamefont {Maleknejad}\ \emph {et~al.}(2013)\citenamefont {Maleknejad}, \citenamefont {Sheikh-Jabbari},\ and\ \citenamefont {Soda}}]{Maleknejad:2012fw}%
  \BibitemOpen
  \bibfield  {author} {\bibinfo {author} {\bibfnamefont {A.}~\bibnamefont {Maleknejad}}, \bibinfo {author} {\bibfnamefont {M.~M.}\ \bibnamefont {Sheikh-Jabbari}}, \ and\ \bibinfo {author} {\bibfnamefont {J.}~\bibnamefont {Soda}},\ }\href {\doibase 10.1016/j.physrep.2013.03.003} {\bibfield  {journal} {\bibinfo  {journal} {Phys. Rept.}\ }\textbf {\bibinfo {volume} {528}},\ \bibinfo {pages} {161} (\bibinfo {year} {2013})},\ \Eprint {http://arxiv.org/abs/1212.2921} {arXiv:1212.2921 [hep-th]} \BibitemShut {NoStop}%
\bibitem [{\citenamefont {Komatsu}(2022)}]{Komatsu:2022nvu}%
  \BibitemOpen
  \bibfield  {author} {\bibinfo {author} {\bibfnamefont {E.}~\bibnamefont {Komatsu}},\ }\href {\doibase 10.1038/s42254-022-00452-4} {\bibfield  {journal} {\bibinfo  {journal} {Nature Rev. Phys.}\ }\textbf {\bibinfo {volume} {4}},\ \bibinfo {pages} {452} (\bibinfo {year} {2022})},\ \Eprint {http://arxiv.org/abs/2202.13919} {arXiv:2202.13919 [astro-ph.CO]} \BibitemShut {NoStop}%
\bibitem [{\citenamefont {Thorne}\ \emph {et~al.}(2018)\citenamefont {Thorne}, \citenamefont {Fujita}, \citenamefont {Hazumi}, \citenamefont {Katayama}, \citenamefont {Komatsu},\ and\ \citenamefont {Shiraishi}}]{Thorne:2017jft}%
  \BibitemOpen
  \bibfield  {author} {\bibinfo {author} {\bibfnamefont {B.}~\bibnamefont {Thorne}}, \bibinfo {author} {\bibfnamefont {T.}~\bibnamefont {Fujita}}, \bibinfo {author} {\bibfnamefont {M.}~\bibnamefont {Hazumi}}, \bibinfo {author} {\bibfnamefont {N.}~\bibnamefont {Katayama}}, \bibinfo {author} {\bibfnamefont {E.}~\bibnamefont {Komatsu}}, \ and\ \bibinfo {author} {\bibfnamefont {M.}~\bibnamefont {Shiraishi}},\ }\href {\doibase 10.1103/PhysRevD.97.043506} {\bibfield  {journal} {\bibinfo  {journal} {Phys. Rev. D}\ }\textbf {\bibinfo {volume} {97}},\ \bibinfo {pages} {043506} (\bibinfo {year} {2018})},\ \Eprint {http://arxiv.org/abs/1707.03240} {arXiv:1707.03240 [astro-ph.CO]} \BibitemShut {NoStop}%
\bibitem [{\citenamefont {Shiraishi}\ \emph {et~al.}(2013)\citenamefont {Shiraishi}, \citenamefont {Ricciardone},\ and\ \citenamefont {Saga}}]{Shiraishi:2013kxa}%
  \BibitemOpen
  \bibfield  {author} {\bibinfo {author} {\bibfnamefont {M.}~\bibnamefont {Shiraishi}}, \bibinfo {author} {\bibfnamefont {A.}~\bibnamefont {Ricciardone}}, \ and\ \bibinfo {author} {\bibfnamefont {S.}~\bibnamefont {Saga}},\ }\href {\doibase 10.1088/1475-7516/2013/11/051} {\bibfield  {journal} {\bibinfo  {journal} {JCAP}\ }\textbf {\bibinfo {volume} {11}},\ \bibinfo {pages} {051} (\bibinfo {year} {2013})},\ \Eprint {http://arxiv.org/abs/1308.6769} {arXiv:1308.6769 [astro-ph.CO]} \BibitemShut {NoStop}%
\bibitem [{\citenamefont {Shiraishi}\ \emph {et~al.}(2016)\citenamefont {Shiraishi}, \citenamefont {Hikage}, \citenamefont {Namba}, \citenamefont {Namikawa},\ and\ \citenamefont {Hazumi}}]{Shiraishi:2016yun}%
  \BibitemOpen
  \bibfield  {author} {\bibinfo {author} {\bibfnamefont {M.}~\bibnamefont {Shiraishi}}, \bibinfo {author} {\bibfnamefont {C.}~\bibnamefont {Hikage}}, \bibinfo {author} {\bibfnamefont {R.}~\bibnamefont {Namba}}, \bibinfo {author} {\bibfnamefont {T.}~\bibnamefont {Namikawa}}, \ and\ \bibinfo {author} {\bibfnamefont {M.}~\bibnamefont {Hazumi}},\ }\href {\doibase 10.1103/PhysRevD.94.043506} {\bibfield  {journal} {\bibinfo  {journal} {Phys. Rev. D}\ }\textbf {\bibinfo {volume} {94}},\ \bibinfo {pages} {043506} (\bibinfo {year} {2016})},\ \Eprint {http://arxiv.org/abs/1606.06082} {arXiv:1606.06082 [astro-ph.CO]} \BibitemShut {NoStop}%
\bibitem [{\citenamefont {Niu}\ \emph {et~al.}(2023)\citenamefont {Niu}, \citenamefont {Rahat}, \citenamefont {Srinivasan},\ and\ \citenamefont {Xue}}]{Niu:2022quw}%
  \BibitemOpen
  \bibfield  {author} {\bibinfo {author} {\bibfnamefont {X.}~\bibnamefont {Niu}}, \bibinfo {author} {\bibfnamefont {M.~H.}\ \bibnamefont {Rahat}}, \bibinfo {author} {\bibfnamefont {K.}~\bibnamefont {Srinivasan}}, \ and\ \bibinfo {author} {\bibfnamefont {W.}~\bibnamefont {Xue}},\ }\href {\doibase 10.1088/1475-7516/2023/02/013} {\bibfield  {journal} {\bibinfo  {journal} {JCAP}\ }\textbf {\bibinfo {volume} {02}},\ \bibinfo {pages} {013} (\bibinfo {year} {2023})},\ \Eprint {http://arxiv.org/abs/2211.14331} {arXiv:2211.14331 [hep-ph]} \BibitemShut {NoStop}%
\bibitem [{\citenamefont {Hiramatsu}\ \emph {et~al.}(2021)\citenamefont {Hiramatsu}, \citenamefont {Murai}, \citenamefont {Obata},\ and\ \citenamefont {Yokoyama}}]{Hiramatsu:2020jes}%
  \BibitemOpen
  \bibfield  {author} {\bibinfo {author} {\bibfnamefont {T.}~\bibnamefont {Hiramatsu}}, \bibinfo {author} {\bibfnamefont {K.}~\bibnamefont {Murai}}, \bibinfo {author} {\bibfnamefont {I.}~\bibnamefont {Obata}}, \ and\ \bibinfo {author} {\bibfnamefont {S.}~\bibnamefont {Yokoyama}},\ }\href {\doibase 10.1088/1475-7516/2021/03/047} {\bibfield  {journal} {\bibinfo  {journal} {JCAP}\ }\textbf {\bibinfo {volume} {03}},\ \bibinfo {pages} {047} (\bibinfo {year} {2021})},\ \Eprint {http://arxiv.org/abs/2008.03233} {arXiv:2008.03233 [astro-ph.CO]} \BibitemShut {NoStop}%
\bibitem [{\citenamefont {Chowdhury}\ \emph {et~al.}(2016)\citenamefont {Chowdhury}, \citenamefont {Sreenath},\ and\ \citenamefont {Sriramkumar}}]{Chowdhury:2016yrh}%
  \BibitemOpen
  \bibfield  {author} {\bibinfo {author} {\bibfnamefont {D.}~\bibnamefont {Chowdhury}}, \bibinfo {author} {\bibfnamefont {V.}~\bibnamefont {Sreenath}}, \ and\ \bibinfo {author} {\bibfnamefont {L.}~\bibnamefont {Sriramkumar}},\ }\href {\doibase 10.1088/1475-7516/2016/11/041} {\bibfield  {journal} {\bibinfo  {journal} {JCAP}\ }\textbf {\bibinfo {volume} {11}},\ \bibinfo {pages} {041} (\bibinfo {year} {2016})},\ \Eprint {http://arxiv.org/abs/1605.05292} {arXiv:1605.05292 [astro-ph.CO]} \BibitemShut {NoStop}%
\bibitem [{\citenamefont {Fujita}\ \emph {et~al.}(2019{\natexlab{a}})\citenamefont {Fujita}, \citenamefont {Namba},\ and\ \citenamefont {Obata}}]{Fujita:2018vmv}%
  \BibitemOpen
  \bibfield  {author} {\bibinfo {author} {\bibfnamefont {T.}~\bibnamefont {Fujita}}, \bibinfo {author} {\bibfnamefont {R.}~\bibnamefont {Namba}}, \ and\ \bibinfo {author} {\bibfnamefont {I.}~\bibnamefont {Obata}},\ }\href {\doibase 10.1088/1475-7516/2019/04/044} {\bibfield  {journal} {\bibinfo  {journal} {JCAP}\ }\textbf {\bibinfo {volume} {04}},\ \bibinfo {pages} {044} (\bibinfo {year} {2019}{\natexlab{a}})},\ \Eprint {http://arxiv.org/abs/1811.12371} {arXiv:1811.12371 [astro-ph.CO]} \BibitemShut {NoStop}%
\bibitem [{\citenamefont {\"Ozsoy}(2021)}]{Ozsoy:2021onx}%
  \BibitemOpen
  \bibfield  {author} {\bibinfo {author} {\bibfnamefont {O.}~\bibnamefont {\"Ozsoy}},\ }\href {\doibase 10.1103/PhysRevD.104.123523} {\bibfield  {journal} {\bibinfo  {journal} {Phys. Rev. D}\ }\textbf {\bibinfo {volume} {104}},\ \bibinfo {pages} {123523} (\bibinfo {year} {2021})},\ \Eprint {http://arxiv.org/abs/2106.14895} {arXiv:2106.14895 [astro-ph.CO]} \BibitemShut {NoStop}%
\bibitem [{\citenamefont {De~Luca}\ \emph {et~al.}(2019)\citenamefont {De~Luca}, \citenamefont {Franciolini}, \citenamefont {Kehagias}, \citenamefont {Riotto},\ and\ \citenamefont {Shiraishi}}]{DeLuca:2019jzc}%
  \BibitemOpen
  \bibfield  {author} {\bibinfo {author} {\bibfnamefont {V.}~\bibnamefont {De~Luca}}, \bibinfo {author} {\bibfnamefont {G.}~\bibnamefont {Franciolini}}, \bibinfo {author} {\bibfnamefont {A.}~\bibnamefont {Kehagias}}, \bibinfo {author} {\bibfnamefont {A.}~\bibnamefont {Riotto}}, \ and\ \bibinfo {author} {\bibfnamefont {M.}~\bibnamefont {Shiraishi}},\ }\href {\doibase 10.1103/PhysRevD.100.063535} {\bibfield  {journal} {\bibinfo  {journal} {Phys. Rev. D}\ }\textbf {\bibinfo {volume} {100}},\ \bibinfo {pages} {063535} (\bibinfo {year} {2019})},\ \Eprint {http://arxiv.org/abs/1908.00366} {arXiv:1908.00366 [astro-ph.CO]} \BibitemShut {NoStop}%
\bibitem [{\citenamefont {Shiraishi}\ \emph {et~al.}(2011{\natexlab{a}})\citenamefont {Shiraishi}, \citenamefont {Nitta},\ and\ \citenamefont {Yokoyama}}]{Shiraishi:2011st}%
  \BibitemOpen
  \bibfield  {author} {\bibinfo {author} {\bibfnamefont {M.}~\bibnamefont {Shiraishi}}, \bibinfo {author} {\bibfnamefont {D.}~\bibnamefont {Nitta}}, \ and\ \bibinfo {author} {\bibfnamefont {S.}~\bibnamefont {Yokoyama}},\ }\href {\doibase 10.1143/PTP.126.937} {\bibfield  {journal} {\bibinfo  {journal} {Prog. Theor. Phys.}\ }\textbf {\bibinfo {volume} {126}},\ \bibinfo {pages} {937} (\bibinfo {year} {2011}{\natexlab{a}})},\ \Eprint {http://arxiv.org/abs/1108.0175} {arXiv:1108.0175 [astro-ph.CO]} \BibitemShut {NoStop}%
\bibitem [{\citenamefont {Alexander}\ and\ \citenamefont {Martin}(2005)}]{Alexander:2004wk}%
  \BibitemOpen
  \bibfield  {author} {\bibinfo {author} {\bibfnamefont {S.}~\bibnamefont {Alexander}}\ and\ \bibinfo {author} {\bibfnamefont {J.}~\bibnamefont {Martin}},\ }\href {\doibase 10.1103/PhysRevD.71.063526} {\bibfield  {journal} {\bibinfo  {journal} {Phys. Rev. D}\ }\textbf {\bibinfo {volume} {71}},\ \bibinfo {pages} {063526} (\bibinfo {year} {2005})},\ \Eprint {http://arxiv.org/abs/hep-th/0410230} {arXiv:hep-th/0410230} \BibitemShut {NoStop}%
\bibitem [{\citenamefont {Soda}\ \emph {et~al.}(2011)\citenamefont {Soda}, \citenamefont {Kodama},\ and\ \citenamefont {Nozawa}}]{Soda:2011am}%
  \BibitemOpen
  \bibfield  {author} {\bibinfo {author} {\bibfnamefont {J.}~\bibnamefont {Soda}}, \bibinfo {author} {\bibfnamefont {H.}~\bibnamefont {Kodama}}, \ and\ \bibinfo {author} {\bibfnamefont {M.}~\bibnamefont {Nozawa}},\ }\href {\doibase 10.1007/JHEP08(2011)067} {\bibfield  {journal} {\bibinfo  {journal} {JHEP}\ }\textbf {\bibinfo {volume} {08}},\ \bibinfo {pages} {067} (\bibinfo {year} {2011})},\ \Eprint {http://arxiv.org/abs/1106.3228} {arXiv:1106.3228 [hep-th]} \BibitemShut {NoStop}%
\bibitem [{\citenamefont {Bartolo}\ and\ \citenamefont {Orlando}(2017)}]{Bartolo:2017szm}%
  \BibitemOpen
  \bibfield  {author} {\bibinfo {author} {\bibfnamefont {N.}~\bibnamefont {Bartolo}}\ and\ \bibinfo {author} {\bibfnamefont {G.}~\bibnamefont {Orlando}},\ }\href {\doibase 10.1088/1475-7516/2017/07/034} {\bibfield  {journal} {\bibinfo  {journal} {JCAP}\ }\textbf {\bibinfo {volume} {07}},\ \bibinfo {pages} {034} (\bibinfo {year} {2017})},\ \Eprint {http://arxiv.org/abs/1706.04627} {arXiv:1706.04627 [astro-ph.CO]} \BibitemShut {NoStop}%
\bibitem [{\citenamefont {Bartolo}\ \emph {et~al.}(2019)\citenamefont {Bartolo}, \citenamefont {Orlando},\ and\ \citenamefont {Shiraishi}}]{Bartolo:2018elp}%
  \BibitemOpen
  \bibfield  {author} {\bibinfo {author} {\bibfnamefont {N.}~\bibnamefont {Bartolo}}, \bibinfo {author} {\bibfnamefont {G.}~\bibnamefont {Orlando}}, \ and\ \bibinfo {author} {\bibfnamefont {M.}~\bibnamefont {Shiraishi}},\ }\href {\doibase 10.1088/1475-7516/2019/01/050} {\bibfield  {journal} {\bibinfo  {journal} {JCAP}\ }\textbf {\bibinfo {volume} {01}},\ \bibinfo {pages} {050} (\bibinfo {year} {2019})},\ \Eprint {http://arxiv.org/abs/1809.11170} {arXiv:1809.11170 [astro-ph.CO]} \BibitemShut {NoStop}%
\bibitem [{\citenamefont {Mylova}(2019)}]{Mylova:2019jrj}%
  \BibitemOpen
  \bibfield  {author} {\bibinfo {author} {\bibfnamefont {M.}~\bibnamefont {Mylova}},\ }\href@noop {} {\  (\bibinfo {year} {2019})},\ \Eprint {http://arxiv.org/abs/1912.00800} {arXiv:1912.00800 [gr-qc]} \BibitemShut {NoStop}%
\bibitem [{\citenamefont {Bartolo}\ \emph {et~al.}(2021)\citenamefont {Bartolo}, \citenamefont {Caloni}, \citenamefont {Orlando},\ and\ \citenamefont {Ricciardone}}]{Bartolo:2020gsh}%
  \BibitemOpen
  \bibfield  {author} {\bibinfo {author} {\bibfnamefont {N.}~\bibnamefont {Bartolo}}, \bibinfo {author} {\bibfnamefont {L.}~\bibnamefont {Caloni}}, \bibinfo {author} {\bibfnamefont {G.}~\bibnamefont {Orlando}}, \ and\ \bibinfo {author} {\bibfnamefont {A.}~\bibnamefont {Ricciardone}},\ }\href {\doibase 10.1088/1475-7516/2021/03/073} {\bibfield  {journal} {\bibinfo  {journal} {JCAP}\ }\textbf {\bibinfo {volume} {03}},\ \bibinfo {pages} {073} (\bibinfo {year} {2021})},\ \Eprint {http://arxiv.org/abs/2008.01715} {arXiv:2008.01715 [astro-ph.CO]} \BibitemShut {NoStop}%
\bibitem [{\citenamefont {Christodoulidis}\ \emph {et~al.}(2024)\citenamefont {Christodoulidis}, \citenamefont {Gong}, \citenamefont {Lin}, \citenamefont {Mylova},\ and\ \citenamefont {Sasaki}}]{Christodoulidis:2024ric}%
  \BibitemOpen
  \bibfield  {author} {\bibinfo {author} {\bibfnamefont {P.}~\bibnamefont {Christodoulidis}}, \bibinfo {author} {\bibfnamefont {J.-O.}\ \bibnamefont {Gong}}, \bibinfo {author} {\bibfnamefont {W.-C.}\ \bibnamefont {Lin}}, \bibinfo {author} {\bibfnamefont {M.}~\bibnamefont {Mylova}}, \ and\ \bibinfo {author} {\bibfnamefont {M.}~\bibnamefont {Sasaki}},\ }\href@noop {} {\  (\bibinfo {year} {2024})},\ \Eprint {http://arxiv.org/abs/2409.09935} {arXiv:2409.09935 [hep-th]} \BibitemShut {NoStop}%
\bibitem [{\citenamefont {Huang}\ \emph {et~al.}(2013)\citenamefont {Huang}, \citenamefont {Wang}, \citenamefont {Yousefi},\ and\ \citenamefont {Zhu}}]{Huang:2013epa}%
  \BibitemOpen
  \bibfield  {author} {\bibinfo {author} {\bibfnamefont {Y.}~\bibnamefont {Huang}}, \bibinfo {author} {\bibfnamefont {A.}~\bibnamefont {Wang}}, \bibinfo {author} {\bibfnamefont {R.}~\bibnamefont {Yousefi}}, \ and\ \bibinfo {author} {\bibfnamefont {T.}~\bibnamefont {Zhu}},\ }\href {\doibase 10.1103/PhysRevD.88.023523} {\bibfield  {journal} {\bibinfo  {journal} {Phys. Rev. D}\ }\textbf {\bibinfo {volume} {88}},\ \bibinfo {pages} {023523} (\bibinfo {year} {2013})},\ \Eprint {http://arxiv.org/abs/1304.1556} {arXiv:1304.1556 [hep-th]} \BibitemShut {NoStop}%
\bibitem [{\citenamefont {Dom\`enech}\ \emph {et~al.}(2017)\citenamefont {Dom\`enech}, \citenamefont {Hiramatsu}, \citenamefont {Lin}, \citenamefont {Sasaki}, \citenamefont {Shiraishi},\ and\ \citenamefont {Wang}}]{Domenech:2017kno}%
  \BibitemOpen
  \bibfield  {author} {\bibinfo {author} {\bibfnamefont {G.}~\bibnamefont {Dom\`enech}}, \bibinfo {author} {\bibfnamefont {T.}~\bibnamefont {Hiramatsu}}, \bibinfo {author} {\bibfnamefont {C.}~\bibnamefont {Lin}}, \bibinfo {author} {\bibfnamefont {M.}~\bibnamefont {Sasaki}}, \bibinfo {author} {\bibfnamefont {M.}~\bibnamefont {Shiraishi}}, \ and\ \bibinfo {author} {\bibfnamefont {Y.}~\bibnamefont {Wang}},\ }\href {\doibase 10.1088/1475-7516/2017/05/034} {\bibfield  {journal} {\bibinfo  {journal} {JCAP}\ }\textbf {\bibinfo {volume} {05}},\ \bibinfo {pages} {034} (\bibinfo {year} {2017})},\ \Eprint {http://arxiv.org/abs/1701.05554} {arXiv:1701.05554 [astro-ph.CO]} \BibitemShut {NoStop}%
\bibitem [{\citenamefont {Fujita}\ \emph {et~al.}(2019{\natexlab{b}})\citenamefont {Fujita}, \citenamefont {Kuroyanagi}, \citenamefont {Mizuno},\ and\ \citenamefont {Mukohyama}}]{Fujita:2018ehq}%
  \BibitemOpen
  \bibfield  {author} {\bibinfo {author} {\bibfnamefont {T.}~\bibnamefont {Fujita}}, \bibinfo {author} {\bibfnamefont {S.}~\bibnamefont {Kuroyanagi}}, \bibinfo {author} {\bibfnamefont {S.}~\bibnamefont {Mizuno}}, \ and\ \bibinfo {author} {\bibfnamefont {S.}~\bibnamefont {Mukohyama}},\ }\href {\doibase 10.1016/j.physletb.2018.12.025} {\bibfield  {journal} {\bibinfo  {journal} {Phys. Lett. B}\ }\textbf {\bibinfo {volume} {789}},\ \bibinfo {pages} {215} (\bibinfo {year} {2019}{\natexlab{b}})},\ \Eprint {http://arxiv.org/abs/1808.02381} {arXiv:1808.02381 [gr-qc]} \BibitemShut {NoStop}%
\bibitem [{\citenamefont {Fujita}\ \emph {et~al.}(2020)\citenamefont {Fujita}, \citenamefont {Mizuno},\ and\ \citenamefont {Mukohyama}}]{Fujita:2019tov}%
  \BibitemOpen
  \bibfield  {author} {\bibinfo {author} {\bibfnamefont {T.}~\bibnamefont {Fujita}}, \bibinfo {author} {\bibfnamefont {S.}~\bibnamefont {Mizuno}}, \ and\ \bibinfo {author} {\bibfnamefont {S.}~\bibnamefont {Mukohyama}},\ }\href {\doibase 10.1088/1475-7516/2020/01/023} {\bibfield  {journal} {\bibinfo  {journal} {JCAP}\ }\textbf {\bibinfo {volume} {01}},\ \bibinfo {pages} {023} (\bibinfo {year} {2020})},\ \Eprint {http://arxiv.org/abs/1909.07563} {arXiv:1909.07563 [astro-ph.CO]} \BibitemShut {NoStop}%
\bibitem [{\citenamefont {Gao}\ \emph {et~al.}(2011)\citenamefont {Gao}, \citenamefont {Kobayashi}, \citenamefont {Yamaguchi},\ and\ \citenamefont {Yokoyama}}]{Gao:2011vs}%
  \BibitemOpen
  \bibfield  {author} {\bibinfo {author} {\bibfnamefont {X.}~\bibnamefont {Gao}}, \bibinfo {author} {\bibfnamefont {T.}~\bibnamefont {Kobayashi}}, \bibinfo {author} {\bibfnamefont {M.}~\bibnamefont {Yamaguchi}}, \ and\ \bibinfo {author} {\bibfnamefont {J.}~\bibnamefont {Yokoyama}},\ }\href {\doibase 10.1103/PhysRevLett.107.211301} {\bibfield  {journal} {\bibinfo  {journal} {Phys. Rev. Lett.}\ }\textbf {\bibinfo {volume} {107}},\ \bibinfo {pages} {211301} (\bibinfo {year} {2011})},\ \Eprint {http://arxiv.org/abs/1108.3513} {arXiv:1108.3513 [astro-ph.CO]} \BibitemShut {NoStop}%
\bibitem [{\citenamefont {Gao}\ \emph {et~al.}(2013)\citenamefont {Gao}, \citenamefont {Kobayashi}, \citenamefont {Shiraishi}, \citenamefont {Yamaguchi}, \citenamefont {Yokoyama},\ and\ \citenamefont {Yokoyama}}]{Gao:2012ib}%
  \BibitemOpen
  \bibfield  {author} {\bibinfo {author} {\bibfnamefont {X.}~\bibnamefont {Gao}}, \bibinfo {author} {\bibfnamefont {T.}~\bibnamefont {Kobayashi}}, \bibinfo {author} {\bibfnamefont {M.}~\bibnamefont {Shiraishi}}, \bibinfo {author} {\bibfnamefont {M.}~\bibnamefont {Yamaguchi}}, \bibinfo {author} {\bibfnamefont {J.}~\bibnamefont {Yokoyama}}, \ and\ \bibinfo {author} {\bibfnamefont {S.}~\bibnamefont {Yokoyama}},\ }\href {\doibase 10.1093/ptep/ptt031} {\bibfield  {journal} {\bibinfo  {journal} {PTEP}\ }\textbf {\bibinfo {volume} {2013}},\ \bibinfo {pages} {053E03} (\bibinfo {year} {2013})},\ \Eprint {http://arxiv.org/abs/1207.0588} {arXiv:1207.0588 [astro-ph.CO]} \BibitemShut {NoStop}%
\bibitem [{\citenamefont {Ozsoy}\ \emph {et~al.}(2019)\citenamefont {Ozsoy}, \citenamefont {Mylova}, \citenamefont {Parameswaran}, \citenamefont {Powell}, \citenamefont {Tasinato},\ and\ \citenamefont {Zavala}}]{Ozsoy:2019slf}%
  \BibitemOpen
  \bibfield  {author} {\bibinfo {author} {\bibfnamefont {O.}~\bibnamefont {Ozsoy}}, \bibinfo {author} {\bibfnamefont {M.}~\bibnamefont {Mylova}}, \bibinfo {author} {\bibfnamefont {S.}~\bibnamefont {Parameswaran}}, \bibinfo {author} {\bibfnamefont {C.}~\bibnamefont {Powell}}, \bibinfo {author} {\bibfnamefont {G.}~\bibnamefont {Tasinato}}, \ and\ \bibinfo {author} {\bibfnamefont {I.}~\bibnamefont {Zavala}},\ }\href {\doibase 10.1088/1475-7516/2019/09/036} {\bibfield  {journal} {\bibinfo  {journal} {JCAP}\ }\textbf {\bibinfo {volume} {09}},\ \bibinfo {pages} {036} (\bibinfo {year} {2019})},\ \Eprint {http://arxiv.org/abs/1902.04976} {arXiv:1902.04976 [hep-th]} \BibitemShut {NoStop}%
\bibitem [{\citenamefont {Namjoo}\ \emph {et~al.}(2013)\citenamefont {Namjoo}, \citenamefont {Firouzjahi},\ and\ \citenamefont {Sasaki}}]{Namjoo:2012aa}%
  \BibitemOpen
  \bibfield  {author} {\bibinfo {author} {\bibfnamefont {M.~H.}\ \bibnamefont {Namjoo}}, \bibinfo {author} {\bibfnamefont {H.}~\bibnamefont {Firouzjahi}}, \ and\ \bibinfo {author} {\bibfnamefont {M.}~\bibnamefont {Sasaki}},\ }\href {\doibase 10.1209/0295-5075/101/39001} {\bibfield  {journal} {\bibinfo  {journal} {EPL}\ }\textbf {\bibinfo {volume} {101}},\ \bibinfo {pages} {39001} (\bibinfo {year} {2013})},\ \Eprint {http://arxiv.org/abs/1210.3692} {arXiv:1210.3692 [astro-ph.CO]} \BibitemShut {NoStop}%
\bibitem [{\citenamefont {Martin}\ \emph {et~al.}(2013)\citenamefont {Martin}, \citenamefont {Motohashi},\ and\ \citenamefont {Suyama}}]{Martin:2012pe}%
  \BibitemOpen
  \bibfield  {author} {\bibinfo {author} {\bibfnamefont {J.}~\bibnamefont {Martin}}, \bibinfo {author} {\bibfnamefont {H.}~\bibnamefont {Motohashi}}, \ and\ \bibinfo {author} {\bibfnamefont {T.}~\bibnamefont {Suyama}},\ }\href {\doibase 10.1103/PhysRevD.87.023514} {\bibfield  {journal} {\bibinfo  {journal} {Phys. Rev. D}\ }\textbf {\bibinfo {volume} {87}},\ \bibinfo {pages} {023514} (\bibinfo {year} {2013})},\ \Eprint {http://arxiv.org/abs/1211.0083} {arXiv:1211.0083 [astro-ph.CO]} \BibitemShut {NoStop}%
\bibitem [{\citenamefont {Shiraishi}\ \emph {et~al.}(2011{\natexlab{b}})\citenamefont {Shiraishi}, \citenamefont {Nitta}, \citenamefont {Yokoyama}, \citenamefont {Ichiki},\ and\ \citenamefont {Takahashi}}]{Shiraishi:2011dh}%
  \BibitemOpen
  \bibfield  {author} {\bibinfo {author} {\bibfnamefont {M.}~\bibnamefont {Shiraishi}}, \bibinfo {author} {\bibfnamefont {D.}~\bibnamefont {Nitta}}, \bibinfo {author} {\bibfnamefont {S.}~\bibnamefont {Yokoyama}}, \bibinfo {author} {\bibfnamefont {K.}~\bibnamefont {Ichiki}}, \ and\ \bibinfo {author} {\bibfnamefont {K.}~\bibnamefont {Takahashi}},\ }\href {\doibase 10.1103/PhysRevD.83.123003} {\bibfield  {journal} {\bibinfo  {journal} {Phys. Rev. D}\ }\textbf {\bibinfo {volume} {83}},\ \bibinfo {pages} {123003} (\bibinfo {year} {2011}{\natexlab{b}})},\ \Eprint {http://arxiv.org/abs/1103.4103} {arXiv:1103.4103 [astro-ph.CO]} \BibitemShut {NoStop}%
\bibitem [{\citenamefont {Shiraishi}(2012)}]{Shiraishi:2012sn}%
  \BibitemOpen
  \bibfield  {author} {\bibinfo {author} {\bibfnamefont {M.}~\bibnamefont {Shiraishi}},\ }\href {\doibase 10.1088/1475-7516/2012/06/015} {\bibfield  {journal} {\bibinfo  {journal} {JCAP}\ }\textbf {\bibinfo {volume} {06}},\ \bibinfo {pages} {015} (\bibinfo {year} {2012})},\ \Eprint {http://arxiv.org/abs/1202.2847} {arXiv:1202.2847 [astro-ph.CO]} \BibitemShut {NoStop}%
\bibitem [{\citenamefont {Gong}\ \emph {et~al.}(2023)\citenamefont {Gong}, \citenamefont {Mylova},\ and\ \citenamefont {Sasaki}}]{Gong:2023kpe}%
  \BibitemOpen
  \bibfield  {author} {\bibinfo {author} {\bibfnamefont {J.-O.}\ \bibnamefont {Gong}}, \bibinfo {author} {\bibfnamefont {M.}~\bibnamefont {Mylova}}, \ and\ \bibinfo {author} {\bibfnamefont {M.}~\bibnamefont {Sasaki}},\ }\href {\doibase 10.1007/JHEP10(2023)140} {\bibfield  {journal} {\bibinfo  {journal} {JHEP}\ }\textbf {\bibinfo {volume} {10}},\ \bibinfo {pages} {140} (\bibinfo {year} {2023})},\ \Eprint {http://arxiv.org/abs/2303.05178} {arXiv:2303.05178 [hep-th]} \BibitemShut {NoStop}%
\bibitem [{\citenamefont {Kanno}\ and\ \citenamefont {Sasaki}(2022)}]{Kanno:2022mkx}%
  \BibitemOpen
  \bibfield  {author} {\bibinfo {author} {\bibfnamefont {S.}~\bibnamefont {Kanno}}\ and\ \bibinfo {author} {\bibfnamefont {M.}~\bibnamefont {Sasaki}},\ }\href {\doibase 10.1007/JHEP08(2022)210} {\bibfield  {journal} {\bibinfo  {journal} {JHEP}\ }\textbf {\bibinfo {volume} {08}},\ \bibinfo {pages} {210} (\bibinfo {year} {2022})},\ \Eprint {http://arxiv.org/abs/2206.03667} {arXiv:2206.03667 [hep-th]} \BibitemShut {NoStop}%
\bibitem [{\citenamefont {Berezhiani}\ and\ \citenamefont {Khoury}(2014)}]{Berezhiani:2014kga}%
  \BibitemOpen
  \bibfield  {author} {\bibinfo {author} {\bibfnamefont {L.}~\bibnamefont {Berezhiani}}\ and\ \bibinfo {author} {\bibfnamefont {J.}~\bibnamefont {Khoury}},\ }\href {\doibase 10.1088/1475-7516/2014/09/018} {\bibfield  {journal} {\bibinfo  {journal} {JCAP}\ }\textbf {\bibinfo {volume} {09}},\ \bibinfo {pages} {018} (\bibinfo {year} {2014})},\ \Eprint {http://arxiv.org/abs/1406.2689} {arXiv:1406.2689 [hep-th]} \BibitemShut {NoStop}%
\bibitem [{\citenamefont {Akama}\ \emph {et~al.}(2020)\citenamefont {Akama}, \citenamefont {Hirano},\ and\ \citenamefont {Kobayashi}}]{Akama:2020jko}%
  \BibitemOpen
  \bibfield  {author} {\bibinfo {author} {\bibfnamefont {S.}~\bibnamefont {Akama}}, \bibinfo {author} {\bibfnamefont {S.}~\bibnamefont {Hirano}}, \ and\ \bibinfo {author} {\bibfnamefont {T.}~\bibnamefont {Kobayashi}},\ }\href {\doibase 10.1103/PhysRevD.102.023513} {\bibfield  {journal} {\bibinfo  {journal} {Phys. Rev. D}\ }\textbf {\bibinfo {volume} {102}},\ \bibinfo {pages} {023513} (\bibinfo {year} {2020})},\ \Eprint {http://arxiv.org/abs/2003.10686} {arXiv:2003.10686 [gr-qc]} \BibitemShut {NoStop}%
\bibitem [{\citenamefont {Naskar}\ and\ \citenamefont {Pal}(2022)}]{Naskar:2020vkd}%
  \BibitemOpen
  \bibfield  {author} {\bibinfo {author} {\bibfnamefont {A.}~\bibnamefont {Naskar}}\ and\ \bibinfo {author} {\bibfnamefont {S.}~\bibnamefont {Pal}},\ }\href {\doibase 10.1140/epjc/s10052-022-10869-x} {\bibfield  {journal} {\bibinfo  {journal} {Eur. Phys. J. C}\ }\textbf {\bibinfo {volume} {82}},\ \bibinfo {pages} {900} (\bibinfo {year} {2022})},\ \Eprint {http://arxiv.org/abs/2003.14066} {arXiv:2003.14066 [astro-ph.CO]} \BibitemShut {NoStop}%
\bibitem [{\citenamefont {Peng}\ \emph {et~al.}(2024)\citenamefont {Peng}, \citenamefont {Fang},\ and\ \citenamefont {Guo}}]{Peng:2024eok}%
  \BibitemOpen
  \bibfield  {author} {\bibinfo {author} {\bibfnamefont {Z.-Z.}\ \bibnamefont {Peng}}, \bibinfo {author} {\bibfnamefont {C.-J.}\ \bibnamefont {Fang}}, \ and\ \bibinfo {author} {\bibfnamefont {Z.-K.}\ \bibnamefont {Guo}},\ }\href {\doibase 10.1103/PhysRevD.110.043538} {\bibfield  {journal} {\bibinfo  {journal} {Phys. Rev. D}\ }\textbf {\bibinfo {volume} {110}},\ \bibinfo {pages} {043538} (\bibinfo {year} {2024})},\ \Eprint {http://arxiv.org/abs/2403.04617} {arXiv:2403.04617 [astro-ph.CO]} \BibitemShut {NoStop}%
\bibitem [{\citenamefont {Duivenvoorden}\ \emph {et~al.}(2020)\citenamefont {Duivenvoorden}, \citenamefont {Meerburg},\ and\ \citenamefont {Freese}}]{Duivenvoorden:2019ses}%
  \BibitemOpen
  \bibfield  {author} {\bibinfo {author} {\bibfnamefont {A.~J.}\ \bibnamefont {Duivenvoorden}}, \bibinfo {author} {\bibfnamefont {P.~D.}\ \bibnamefont {Meerburg}}, \ and\ \bibinfo {author} {\bibfnamefont {K.}~\bibnamefont {Freese}},\ }\href {\doibase 10.1103/PhysRevD.102.023521} {\bibfield  {journal} {\bibinfo  {journal} {Phys. Rev. D}\ }\textbf {\bibinfo {volume} {102}},\ \bibinfo {pages} {023521} (\bibinfo {year} {2020})},\ \Eprint {http://arxiv.org/abs/1911.11349} {arXiv:1911.11349 [astro-ph.CO]} \BibitemShut {NoStop}%
\bibitem [{\citenamefont {Meerburg}\ \emph {et~al.}(2016)\citenamefont {Meerburg}, \citenamefont {Meyers}, \citenamefont {van Engelen},\ and\ \citenamefont {Ali-Ha\"\i{}moud}}]{Meerburg:2016ecv}%
  \BibitemOpen
  \bibfield  {author} {\bibinfo {author} {\bibfnamefont {P.~D.}\ \bibnamefont {Meerburg}}, \bibinfo {author} {\bibfnamefont {J.}~\bibnamefont {Meyers}}, \bibinfo {author} {\bibfnamefont {A.}~\bibnamefont {van Engelen}}, \ and\ \bibinfo {author} {\bibfnamefont {Y.}~\bibnamefont {Ali-Ha\"\i{}moud}},\ }\href {\doibase 10.1103/PhysRevD.93.123511} {\bibfield  {journal} {\bibinfo  {journal} {Phys. Rev. D}\ }\textbf {\bibinfo {volume} {93}},\ \bibinfo {pages} {123511} (\bibinfo {year} {2016})},\ \Eprint {http://arxiv.org/abs/1603.02243} {arXiv:1603.02243 [astro-ph.CO]} \BibitemShut {NoStop}%
\bibitem [{\citenamefont {Adshead}\ and\ \citenamefont {Lim}(2010)}]{Adshead:2009bz}%
  \BibitemOpen
  \bibfield  {author} {\bibinfo {author} {\bibfnamefont {P.}~\bibnamefont {Adshead}}\ and\ \bibinfo {author} {\bibfnamefont {E.~A.}\ \bibnamefont {Lim}},\ }\href {\doibase 10.1103/PhysRevD.82.024023} {\bibfield  {journal} {\bibinfo  {journal} {Phys. Rev. D}\ }\textbf {\bibinfo {volume} {82}},\ \bibinfo {pages} {024023} (\bibinfo {year} {2010})},\ \Eprint {http://arxiv.org/abs/0912.1615} {arXiv:0912.1615 [astro-ph.CO]} \BibitemShut {NoStop}%
\bibitem [{\citenamefont {Naskar}\ and\ \citenamefont {Pal}(2020)}]{Naskar:2019shl}%
  \BibitemOpen
  \bibfield  {author} {\bibinfo {author} {\bibfnamefont {A.}~\bibnamefont {Naskar}}\ and\ \bibinfo {author} {\bibfnamefont {S.}~\bibnamefont {Pal}},\ }\href {\doibase 10.1140/epjc/s10052-020-08735-9} {\bibfield  {journal} {\bibinfo  {journal} {Eur. Phys. J. C}\ }\textbf {\bibinfo {volume} {80}},\ \bibinfo {pages} {1158} (\bibinfo {year} {2020})},\ \Eprint {http://arxiv.org/abs/1906.08558} {arXiv:1906.08558 [astro-ph.CO]} \BibitemShut {NoStop}%
\bibitem [{\citenamefont {Cabass}\ \emph {et~al.}(2022{\natexlab{a}})\citenamefont {Cabass}, \citenamefont {Pajer}, \citenamefont {Stefanyszyn},\ and\ \citenamefont {Supe\l{}}}]{Cabass:2021fnw}%
  \BibitemOpen
  \bibfield  {author} {\bibinfo {author} {\bibfnamefont {G.}~\bibnamefont {Cabass}}, \bibinfo {author} {\bibfnamefont {E.}~\bibnamefont {Pajer}}, \bibinfo {author} {\bibfnamefont {D.}~\bibnamefont {Stefanyszyn}}, \ and\ \bibinfo {author} {\bibfnamefont {J.}~\bibnamefont {Supe\l{}}},\ }\href {\doibase 10.1007/JHEP05(2022)077} {\bibfield  {journal} {\bibinfo  {journal} {JHEP}\ }\textbf {\bibinfo {volume} {05}},\ \bibinfo {pages} {077} (\bibinfo {year} {2022}{\natexlab{a}})},\ \Eprint {http://arxiv.org/abs/2109.10189} {arXiv:2109.10189 [hep-th]} \BibitemShut {NoStop}%
\bibitem [{\citenamefont {Cabass}\ \emph {et~al.}(2022{\natexlab{b}})\citenamefont {Cabass}, \citenamefont {Stefanyszyn}, \citenamefont {Supe\l{}},\ and\ \citenamefont {Thavanesan}}]{Cabass:2022jda}%
  \BibitemOpen
  \bibfield  {author} {\bibinfo {author} {\bibfnamefont {G.}~\bibnamefont {Cabass}}, \bibinfo {author} {\bibfnamefont {D.}~\bibnamefont {Stefanyszyn}}, \bibinfo {author} {\bibfnamefont {J.}~\bibnamefont {Supe\l{}}}, \ and\ \bibinfo {author} {\bibfnamefont {A.}~\bibnamefont {Thavanesan}},\ }\href {\doibase 10.1007/JHEP10(2022)154} {\bibfield  {journal} {\bibinfo  {journal} {JHEP}\ }\textbf {\bibinfo {volume} {10}},\ \bibinfo {pages} {154} (\bibinfo {year} {2022}{\natexlab{b}})},\ \Eprint {http://arxiv.org/abs/2209.00677} {arXiv:2209.00677 [hep-th]} \BibitemShut {NoStop}%
\bibitem [{\citenamefont {Bordin}\ and\ \citenamefont {Cabass}(2020)}]{Bordin:2020eui}%
  \BibitemOpen
  \bibfield  {author} {\bibinfo {author} {\bibfnamefont {L.}~\bibnamefont {Bordin}}\ and\ \bibinfo {author} {\bibfnamefont {G.}~\bibnamefont {Cabass}},\ }\href {\doibase 10.1088/1475-7516/2020/07/014} {\bibfield  {journal} {\bibinfo  {journal} {JCAP}\ }\textbf {\bibinfo {volume} {07}},\ \bibinfo {pages} {014} (\bibinfo {year} {2020})},\ \Eprint {http://arxiv.org/abs/2004.00619} {arXiv:2004.00619 [astro-ph.CO]} \BibitemShut {NoStop}%
\bibitem [{\citenamefont {Cabass}(2021)}]{Cabass:2021iii}%
  \BibitemOpen
  \bibfield  {author} {\bibinfo {author} {\bibfnamefont {G.}~\bibnamefont {Cabass}},\ }\href {\doibase 10.1088/1475-7516/2021/12/001} {\bibfield  {journal} {\bibinfo  {journal} {JCAP}\ }\textbf {\bibinfo {volume} {12}},\ \bibinfo {pages} {001} (\bibinfo {year} {2021})},\ \Eprint {http://arxiv.org/abs/2103.09816} {arXiv:2103.09816 [hep-th]} \BibitemShut {NoStop}%
\bibitem [{\citenamefont {Pajer}(2021)}]{Pajer:2020wxk}%
  \BibitemOpen
  \bibfield  {author} {\bibinfo {author} {\bibfnamefont {E.}~\bibnamefont {Pajer}},\ }\href {\doibase 10.1088/1475-7516/2021/01/023} {\bibfield  {journal} {\bibinfo  {journal} {JCAP}\ }\textbf {\bibinfo {volume} {01}},\ \bibinfo {pages} {023} (\bibinfo {year} {2021})},\ \Eprint {http://arxiv.org/abs/2010.12818} {arXiv:2010.12818 [hep-th]} \BibitemShut {NoStop}%
\bibitem [{\citenamefont {Baumann}\ \emph {et~al.}(2021)\citenamefont {Baumann}, \citenamefont {Duaso~Pueyo}, \citenamefont {Joyce}, \citenamefont {Lee},\ and\ \citenamefont {Pimentel}}]{Baumann:2020dch}%
  \BibitemOpen
  \bibfield  {author} {\bibinfo {author} {\bibfnamefont {D.}~\bibnamefont {Baumann}}, \bibinfo {author} {\bibfnamefont {C.}~\bibnamefont {Duaso~Pueyo}}, \bibinfo {author} {\bibfnamefont {A.}~\bibnamefont {Joyce}}, \bibinfo {author} {\bibfnamefont {H.}~\bibnamefont {Lee}}, \ and\ \bibinfo {author} {\bibfnamefont {G.~L.}\ \bibnamefont {Pimentel}},\ }\href {\doibase 10.21468/SciPostPhys.11.3.071} {\bibfield  {journal} {\bibinfo  {journal} {SciPost Phys.}\ }\textbf {\bibinfo {volume} {11}},\ \bibinfo {pages} {071} (\bibinfo {year} {2021})},\ \Eprint {http://arxiv.org/abs/2005.04234} {arXiv:2005.04234 [hep-th]} \BibitemShut {NoStop}%
\bibitem [{\citenamefont {Naskar}\ and\ \citenamefont {Pal}(2018)}]{Naskar:2018rmu}%
  \BibitemOpen
  \bibfield  {author} {\bibinfo {author} {\bibfnamefont {A.}~\bibnamefont {Naskar}}\ and\ \bibinfo {author} {\bibfnamefont {S.}~\bibnamefont {Pal}},\ }\href {\doibase 10.1103/PhysRevD.98.083520} {\bibfield  {journal} {\bibinfo  {journal} {Phys. Rev. D}\ }\textbf {\bibinfo {volume} {98}},\ \bibinfo {pages} {083520} (\bibinfo {year} {2018})},\ \Eprint {http://arxiv.org/abs/1806.08178} {arXiv:1806.08178 [astro-ph.CO]} \BibitemShut {NoStop}%
\bibitem [{\citenamefont {Kamionkowski}\ and\ \citenamefont {Kovetz}(2016)}]{Kamionkowski:2015yta}%
  \BibitemOpen
  \bibfield  {author} {\bibinfo {author} {\bibfnamefont {M.}~\bibnamefont {Kamionkowski}}\ and\ \bibinfo {author} {\bibfnamefont {E.~D.}\ \bibnamefont {Kovetz}},\ }\href {\doibase 10.1146/annurev-astro-081915-023433} {\bibfield  {journal} {\bibinfo  {journal} {Ann. Rev. Astron. Astrophys.}\ }\textbf {\bibinfo {volume} {54}},\ \bibinfo {pages} {227} (\bibinfo {year} {2016})},\ \Eprint {http://arxiv.org/abs/1510.06042} {arXiv:1510.06042 [astro-ph.CO]} \BibitemShut {NoStop}%
\bibitem [{\citenamefont {Jeong}\ and\ \citenamefont {Schmidt}(2012)}]{Jeong:2012nu}%
  \BibitemOpen
  \bibfield  {author} {\bibinfo {author} {\bibfnamefont {D.}~\bibnamefont {Jeong}}\ and\ \bibinfo {author} {\bibfnamefont {F.}~\bibnamefont {Schmidt}},\ }\href {\doibase 10.1103/PhysRevD.86.083512} {\bibfield  {journal} {\bibinfo  {journal} {Phys. Rev. D}\ }\textbf {\bibinfo {volume} {86}},\ \bibinfo {pages} {083512} (\bibinfo {year} {2012})},\ \Eprint {http://arxiv.org/abs/1205.1512} {arXiv:1205.1512 [astro-ph.CO]} \BibitemShut {NoStop}%
\bibitem [{\citenamefont {Schmidt}\ and\ \citenamefont {Jeong}(2012)}]{Schmidt:2012nw}%
  \BibitemOpen
  \bibfield  {author} {\bibinfo {author} {\bibfnamefont {F.}~\bibnamefont {Schmidt}}\ and\ \bibinfo {author} {\bibfnamefont {D.}~\bibnamefont {Jeong}},\ }\href {\doibase 10.1103/PhysRevD.86.083513} {\bibfield  {journal} {\bibinfo  {journal} {Phys. Rev. D}\ }\textbf {\bibinfo {volume} {86}},\ \bibinfo {pages} {083513} (\bibinfo {year} {2012})},\ \Eprint {http://arxiv.org/abs/1205.1514} {arXiv:1205.1514 [astro-ph.CO]} \BibitemShut {NoStop}%
\bibitem [{\citenamefont {Dimastrogiovanni}\ \emph {et~al.}(2014)\citenamefont {Dimastrogiovanni}, \citenamefont {Fasiello}, \citenamefont {Jeong},\ and\ \citenamefont {Kamionkowski}}]{Dimastrogiovanni:2014ina}%
  \BibitemOpen
  \bibfield  {author} {\bibinfo {author} {\bibfnamefont {E.}~\bibnamefont {Dimastrogiovanni}}, \bibinfo {author} {\bibfnamefont {M.}~\bibnamefont {Fasiello}}, \bibinfo {author} {\bibfnamefont {D.}~\bibnamefont {Jeong}}, \ and\ \bibinfo {author} {\bibfnamefont {M.}~\bibnamefont {Kamionkowski}},\ }\href {\doibase 10.1088/1475-7516/2014/12/050} {\bibfield  {journal} {\bibinfo  {journal} {JCAP}\ }\textbf {\bibinfo {volume} {12}},\ \bibinfo {pages} {050} (\bibinfo {year} {2014})},\ \Eprint {http://arxiv.org/abs/1407.8204} {arXiv:1407.8204 [astro-ph.CO]} \BibitemShut {NoStop}%
\bibitem [{\citenamefont {Philcox}\ \emph {et~al.}(2024)\citenamefont {Philcox}, \citenamefont {K\"onig}, \citenamefont {Alexander},\ and\ \citenamefont {Spergel}}]{Philcox:2023uor}%
  \BibitemOpen
  \bibfield  {author} {\bibinfo {author} {\bibfnamefont {O.~H.~E.}\ \bibnamefont {Philcox}}, \bibinfo {author} {\bibfnamefont {M.~J.}\ \bibnamefont {K\"onig}}, \bibinfo {author} {\bibfnamefont {S.}~\bibnamefont {Alexander}}, \ and\ \bibinfo {author} {\bibfnamefont {D.~N.}\ \bibnamefont {Spergel}},\ }\href {\doibase 10.1103/PhysRevD.109.063541} {\bibfield  {journal} {\bibinfo  {journal} {Phys. Rev. D}\ }\textbf {\bibinfo {volume} {109}},\ \bibinfo {pages} {063541} (\bibinfo {year} {2024})},\ \Eprint {http://arxiv.org/abs/2309.08653} {arXiv:2309.08653 [astro-ph.CO]} \BibitemShut {NoStop}%
\bibitem [{\citenamefont {Ade}\ \emph {et~al.}(2019)\citenamefont {Ade} \emph {et~al.}}]{SimonsObservatory:2018koc}%
  \BibitemOpen
  \bibfield  {author} {\bibinfo {author} {\bibfnamefont {P.}~\bibnamefont {Ade}} \emph {et~al.} (\bibinfo {collaboration} {Simons Observatory}),\ }\href {\doibase 10.1088/1475-7516/2019/02/056} {\bibfield  {journal} {\bibinfo  {journal} {JCAP}\ }\textbf {\bibinfo {volume} {02}},\ \bibinfo {pages} {056} (\bibinfo {year} {2019})},\ \Eprint {http://arxiv.org/abs/1808.07445} {arXiv:1808.07445 [astro-ph.CO]} \BibitemShut {NoStop}%
\bibitem [{\citenamefont {Allys}\ \emph {et~al.}(2023)\citenamefont {Allys} \emph {et~al.}}]{LiteBIRD:2022cnt}%
  \BibitemOpen
  \bibfield  {author} {\bibinfo {author} {\bibfnamefont {E.}~\bibnamefont {Allys}} \emph {et~al.} (\bibinfo {collaboration} {LiteBIRD}),\ }\href {\doibase 10.1093/ptep/ptac150} {\bibfield  {journal} {\bibinfo  {journal} {PTEP}\ }\textbf {\bibinfo {volume} {2023}},\ \bibinfo {pages} {042F01} (\bibinfo {year} {2023})},\ \Eprint {http://arxiv.org/abs/2202.02773} {arXiv:2202.02773 [astro-ph.IM]} \BibitemShut {NoStop}%
\bibitem [{\citenamefont {Shiraishi}\ \emph {et~al.}(2015)\citenamefont {Shiraishi}, \citenamefont {Liguori},\ and\ \citenamefont {Fergusson}}]{Shiraishi:2014ila}%
  \BibitemOpen
  \bibfield  {author} {\bibinfo {author} {\bibfnamefont {M.}~\bibnamefont {Shiraishi}}, \bibinfo {author} {\bibfnamefont {M.}~\bibnamefont {Liguori}}, \ and\ \bibinfo {author} {\bibfnamefont {J.~R.}\ \bibnamefont {Fergusson}},\ }\href {\doibase 10.1088/1475-7516/2015/01/007} {\bibfield  {journal} {\bibinfo  {journal} {JCAP}\ }\textbf {\bibinfo {volume} {01}},\ \bibinfo {pages} {007} (\bibinfo {year} {2015})},\ \Eprint {http://arxiv.org/abs/1409.0265} {arXiv:1409.0265 [astro-ph.CO]} \BibitemShut {NoStop}%
\bibitem [{\citenamefont {Ade}\ \emph {et~al.}(2016{\natexlab{a}})\citenamefont {Ade} \emph {et~al.}}]{Planck:2015zrl}%
  \BibitemOpen
  \bibfield  {author} {\bibinfo {author} {\bibfnamefont {P.~A.~R.}\ \bibnamefont {Ade}} \emph {et~al.} (\bibinfo {collaboration} {Planck}),\ }\href {\doibase 10.1051/0004-6361/201525821} {\bibfield  {journal} {\bibinfo  {journal} {Astron. Astrophys.}\ }\textbf {\bibinfo {volume} {594}},\ \bibinfo {pages} {A19} (\bibinfo {year} {2016}{\natexlab{a}})},\ \Eprint {http://arxiv.org/abs/1502.01594} {arXiv:1502.01594 [astro-ph.CO]} \BibitemShut {NoStop}%
\bibitem [{\citenamefont {Ade}\ \emph {et~al.}(2016{\natexlab{b}})\citenamefont {Ade} \emph {et~al.}}]{Planck:2015zfm}%
  \BibitemOpen
  \bibfield  {author} {\bibinfo {author} {\bibfnamefont {P.~A.~R.}\ \bibnamefont {Ade}} \emph {et~al.} (\bibinfo {collaboration} {Planck}),\ }\href {\doibase 10.1051/0004-6361/201525836} {\bibfield  {journal} {\bibinfo  {journal} {Astron. Astrophys.}\ }\textbf {\bibinfo {volume} {594}},\ \bibinfo {pages} {A17} (\bibinfo {year} {2016}{\natexlab{b}})},\ \Eprint {http://arxiv.org/abs/1502.01592} {arXiv:1502.01592 [astro-ph.CO]} \BibitemShut {NoStop}%
\bibitem [{\citenamefont {Akrami}\ \emph {et~al.}(2020{\natexlab{b}})\citenamefont {Akrami} \emph {et~al.}}]{Planck:2019kim}%
  \BibitemOpen
  \bibfield  {author} {\bibinfo {author} {\bibfnamefont {Y.}~\bibnamefont {Akrami}} \emph {et~al.} (\bibinfo {collaboration} {Planck}),\ }\href {\doibase 10.1051/0004-6361/201935891} {\bibfield  {journal} {\bibinfo  {journal} {Astron. Astrophys.}\ }\textbf {\bibinfo {volume} {641}},\ \bibinfo {pages} {A9} (\bibinfo {year} {2020}{\natexlab{b}})},\ \Eprint {http://arxiv.org/abs/1905.05697} {arXiv:1905.05697 [astro-ph.CO]} \BibitemShut {NoStop}%
\bibitem [{\citenamefont {Shiraishi}(2019)}]{Shiraishi:2019yux}%
  \BibitemOpen
  \bibfield  {author} {\bibinfo {author} {\bibfnamefont {M.}~\bibnamefont {Shiraishi}},\ }\href {\doibase 10.3389/fspas.2019.00049} {\bibfield  {journal} {\bibinfo  {journal} {Front. Astron. Space Sci.}\ }\textbf {\bibinfo {volume} {6}},\ \bibinfo {pages} {49} (\bibinfo {year} {2019})},\ \Eprint {http://arxiv.org/abs/1905.12485} {arXiv:1905.12485 [astro-ph.CO]} \BibitemShut {NoStop}%
\bibitem [{\citenamefont {Shiraishi}\ \emph {et~al.}(2012)\citenamefont {Shiraishi}, \citenamefont {Nitta}, \citenamefont {Yokoyama},\ and\ \citenamefont {Ichiki}}]{Shiraishi:2012rm}%
  \BibitemOpen
  \bibfield  {author} {\bibinfo {author} {\bibfnamefont {M.}~\bibnamefont {Shiraishi}}, \bibinfo {author} {\bibfnamefont {D.}~\bibnamefont {Nitta}}, \bibinfo {author} {\bibfnamefont {S.}~\bibnamefont {Yokoyama}}, \ and\ \bibinfo {author} {\bibfnamefont {K.}~\bibnamefont {Ichiki}},\ }\href {\doibase 10.1088/1475-7516/2012/03/041} {\bibfield  {journal} {\bibinfo  {journal} {JCAP}\ }\textbf {\bibinfo {volume} {03}},\ \bibinfo {pages} {041} (\bibinfo {year} {2012})},\ \Eprint {http://arxiv.org/abs/1201.0376} {arXiv:1201.0376 [astro-ph.CO]} \BibitemShut {NoStop}%
\bibitem [{\citenamefont {Shiraishi}(2013)}]{Shiraishi:2013vha}%
  \BibitemOpen
  \bibfield  {author} {\bibinfo {author} {\bibfnamefont {M.}~\bibnamefont {Shiraishi}},\ }\href {\doibase 10.1088/1475-7516/2013/11/006} {\bibfield  {journal} {\bibinfo  {journal} {JCAP}\ }\textbf {\bibinfo {volume} {11}},\ \bibinfo {pages} {006} (\bibinfo {year} {2013})},\ \Eprint {http://arxiv.org/abs/1308.2531} {arXiv:1308.2531 [astro-ph.CO]} \BibitemShut {NoStop}%
\bibitem [{\citenamefont {Shiraishi}\ \emph {et~al.}(2011{\natexlab{c}})\citenamefont {Shiraishi}, \citenamefont {Nitta}, \citenamefont {Yokoyama}, \citenamefont {Ichiki},\ and\ \citenamefont {Takahashi}}]{Shiraishi:2010kd}%
  \BibitemOpen
  \bibfield  {author} {\bibinfo {author} {\bibfnamefont {M.}~\bibnamefont {Shiraishi}}, \bibinfo {author} {\bibfnamefont {D.}~\bibnamefont {Nitta}}, \bibinfo {author} {\bibfnamefont {S.}~\bibnamefont {Yokoyama}}, \bibinfo {author} {\bibfnamefont {K.}~\bibnamefont {Ichiki}}, \ and\ \bibinfo {author} {\bibfnamefont {K.}~\bibnamefont {Takahashi}},\ }\href {\doibase 10.1143/PTP.125.795} {\bibfield  {journal} {\bibinfo  {journal} {Prog. Theor. Phys.}\ }\textbf {\bibinfo {volume} {125}},\ \bibinfo {pages} {795} (\bibinfo {year} {2011}{\natexlab{c}})},\ \Eprint {http://arxiv.org/abs/1012.1079} {arXiv:1012.1079 [astro-ph.CO]} \BibitemShut {NoStop}%
\bibitem [{\citenamefont {Paoletti}\ \emph {et~al.}(2024)\citenamefont {Paoletti} \emph {et~al.}}]{LiteBIRD:2024twk}%
  \BibitemOpen
  \bibfield  {author} {\bibinfo {author} {\bibfnamefont {D.}~\bibnamefont {Paoletti}} \emph {et~al.} (\bibinfo {collaboration} {LiteBIRD}),\ }\href {\doibase 10.1088/1475-7516/2024/07/086} {\bibfield  {journal} {\bibinfo  {journal} {JCAP}\ }\textbf {\bibinfo {volume} {07}},\ \bibinfo {pages} {086} (\bibinfo {year} {2024})},\ \Eprint {http://arxiv.org/abs/2403.16763} {arXiv:2403.16763 [astro-ph.CO]} \BibitemShut {NoStop}%
\bibitem [{\citenamefont {Tahara}\ and\ \citenamefont {Yokoyama}(2018)}]{Tahara:2017wud}%
  \BibitemOpen
  \bibfield  {author} {\bibinfo {author} {\bibfnamefont {H.~W.~H.}\ \bibnamefont {Tahara}}\ and\ \bibinfo {author} {\bibfnamefont {J.}~\bibnamefont {Yokoyama}},\ }\href {\doibase 10.1093/ptep/ptx185} {\bibfield  {journal} {\bibinfo  {journal} {PTEP}\ }\textbf {\bibinfo {volume} {2018}},\ \bibinfo {pages} {013E03} (\bibinfo {year} {2018})},\ \Eprint {http://arxiv.org/abs/1704.08904} {arXiv:1704.08904 [astro-ph.CO]} \BibitemShut {NoStop}%
\bibitem [{\citenamefont {Coulton}\ \emph {et~al.}(2020)\citenamefont {Coulton}, \citenamefont {Meerburg}, \citenamefont {Baker}, \citenamefont {Hotinli}, \citenamefont {Duivenvoorden},\ and\ \citenamefont {van Engelen}}]{Coulton:2019odk}%
  \BibitemOpen
  \bibfield  {author} {\bibinfo {author} {\bibfnamefont {W.~R.}\ \bibnamefont {Coulton}}, \bibinfo {author} {\bibfnamefont {P.~D.}\ \bibnamefont {Meerburg}}, \bibinfo {author} {\bibfnamefont {D.~G.}\ \bibnamefont {Baker}}, \bibinfo {author} {\bibfnamefont {S.}~\bibnamefont {Hotinli}}, \bibinfo {author} {\bibfnamefont {A.~J.}\ \bibnamefont {Duivenvoorden}}, \ and\ \bibinfo {author} {\bibfnamefont {A.}~\bibnamefont {van Engelen}},\ }\href {\doibase 10.1103/PhysRevD.101.123504} {\bibfield  {journal} {\bibinfo  {journal} {Phys. Rev. D}\ }\textbf {\bibinfo {volume} {101}},\ \bibinfo {pages} {123504} (\bibinfo {year} {2020})},\ \Eprint {http://arxiv.org/abs/1912.07619} {arXiv:1912.07619 [astro-ph.CO]} \BibitemShut {NoStop}%
\bibitem [{\citenamefont {Akrami}\ \emph {et~al.}(2020{\natexlab{c}})\citenamefont {Akrami} \emph {et~al.}}]{Planck:2020olo}%
  \BibitemOpen
  \bibfield  {author} {\bibinfo {author} {\bibfnamefont {Y.}~\bibnamefont {Akrami}} \emph {et~al.} (\bibinfo {collaboration} {Planck}),\ }\href {\doibase 10.1051/0004-6361/202038073} {\bibfield  {journal} {\bibinfo  {journal} {Astron. Astrophys.}\ }\textbf {\bibinfo {volume} {643}},\ \bibinfo {pages} {A42} (\bibinfo {year} {2020}{\natexlab{c}})},\ \Eprint {http://arxiv.org/abs/2007.04997} {arXiv:2007.04997 [astro-ph.CO]} \BibitemShut {NoStop}%
\bibitem [{\citenamefont {Philcox}(2023{\natexlab{a}})}]{Philcox:2023uwe}%
  \BibitemOpen
  \bibfield  {author} {\bibinfo {author} {\bibfnamefont {O.~H.~E.}\ \bibnamefont {Philcox}},\ }\href {\doibase 10.1103/PhysRevD.107.123516} {\bibfield  {journal} {\bibinfo  {journal} {Phys. Rev. D}\ }\textbf {\bibinfo {volume} {107}},\ \bibinfo {pages} {123516} (\bibinfo {year} {2023}{\natexlab{a}})},\ \Eprint {http://arxiv.org/abs/2303.08828} {arXiv:2303.08828 [astro-ph.CO]} \BibitemShut {NoStop}%
\bibitem [{\citenamefont {Philcox}(2023{\natexlab{b}})}]{Philcox:2023psd}%
  \BibitemOpen
  \bibfield  {author} {\bibinfo {author} {\bibfnamefont {O.~H.~E.}\ \bibnamefont {Philcox}},\ }\href {\doibase 10.1103/PhysRevD.108.063506} {\bibfield  {journal} {\bibinfo  {journal} {Phys. Rev. D}\ }\textbf {\bibinfo {volume} {108}},\ \bibinfo {pages} {063506} (\bibinfo {year} {2023}{\natexlab{b}})},\ \Eprint {http://arxiv.org/abs/2306.03915} {arXiv:2306.03915 [astro-ph.CO]} \BibitemShut {NoStop}%
\bibitem [{\citenamefont {{Philcox}}(2023)}]{PolyBin}%
  \BibitemOpen
  \bibfield  {author} {\bibinfo {author} {\bibfnamefont {O.~H.~E.}\ \bibnamefont {{Philcox}}},\ }\href@noop {} {\enquote {\bibinfo {title} {{PolyBin: Binned polyspectrum estimation on the full sky}},}\ }\bibinfo {howpublished} {Astrophysics Source Code Library, record ascl:2307.020} (\bibinfo {year} {2023}),\ \Eprint {http://arxiv.org/abs/2307.020} {ascl:2307.020} \BibitemShut {NoStop}%
\bibitem [{\citenamefont {Philcox}(2025{\natexlab{a}})}]{Philcox4pt2}%
  \BibitemOpen
  \bibfield  {author} {\bibinfo {author} {\bibfnamefont {O.~H.~E.}\ \bibnamefont {Philcox}},\ }\href@noop {} {\  (\bibinfo {year} {2025}{\natexlab{a}})},\ \Eprint {http://arxiv.org/abs/2502.05258} {arXiv:2502.05258 [astro-ph.CO]} \BibitemShut {NoStop}%
\bibitem [{\citenamefont {Shiraishi}\ and\ \citenamefont {Sekiguchi}(2014)}]{Shiraishi:2013wua}%
  \BibitemOpen
  \bibfield  {author} {\bibinfo {author} {\bibfnamefont {M.}~\bibnamefont {Shiraishi}}\ and\ \bibinfo {author} {\bibfnamefont {T.}~\bibnamefont {Sekiguchi}},\ }\href {\doibase 10.1103/PhysRevD.90.103002} {\bibfield  {journal} {\bibinfo  {journal} {Phys. Rev. D}\ }\textbf {\bibinfo {volume} {90}},\ \bibinfo {pages} {103002} (\bibinfo {year} {2014})},\ \Eprint {http://arxiv.org/abs/1304.7277} {arXiv:1304.7277 [astro-ph.CO]} \BibitemShut {NoStop}%
\bibitem [{\citenamefont {Shiraishi}\ \emph {et~al.}(2018)\citenamefont {Shiraishi}, \citenamefont {Liguori},\ and\ \citenamefont {Fergusson}}]{Shiraishi:2017yrq}%
  \BibitemOpen
  \bibfield  {author} {\bibinfo {author} {\bibfnamefont {M.}~\bibnamefont {Shiraishi}}, \bibinfo {author} {\bibfnamefont {M.}~\bibnamefont {Liguori}}, \ and\ \bibinfo {author} {\bibfnamefont {J.~R.}\ \bibnamefont {Fergusson}},\ }\href {\doibase 10.1088/1475-7516/2018/01/016} {\bibfield  {journal} {\bibinfo  {journal} {JCAP}\ }\textbf {\bibinfo {volume} {01}},\ \bibinfo {pages} {016} (\bibinfo {year} {2018})},\ \Eprint {http://arxiv.org/abs/1710.06778} {arXiv:1710.06778 [astro-ph.CO]} \BibitemShut {NoStop}%
\bibitem [{\citenamefont {Philcox}\ and\ \citenamefont {Shiraishi}(2024)}]{Philcox:2023xxk}%
  \BibitemOpen
  \bibfield  {author} {\bibinfo {author} {\bibfnamefont {O.~H.~E.}\ \bibnamefont {Philcox}}\ and\ \bibinfo {author} {\bibfnamefont {M.}~\bibnamefont {Shiraishi}},\ }\href {\doibase 10.1103/PhysRevD.109.063522} {\bibfield  {journal} {\bibinfo  {journal} {Phys. Rev. D}\ }\textbf {\bibinfo {volume} {109}},\ \bibinfo {pages} {063522} (\bibinfo {year} {2024})},\ \Eprint {http://arxiv.org/abs/2312.12498} {arXiv:2312.12498 [astro-ph.CO]} \BibitemShut {NoStop}%
\bibitem [{\citenamefont {Shiraishi}(2016{\natexlab{a}})}]{Shiraishi:2016mok}%
  \BibitemOpen
  \bibfield  {author} {\bibinfo {author} {\bibfnamefont {M.}~\bibnamefont {Shiraishi}},\ }\href {\doibase 10.1103/PhysRevD.94.083503} {\bibfield  {journal} {\bibinfo  {journal} {Phys. Rev. D}\ }\textbf {\bibinfo {volume} {94}},\ \bibinfo {pages} {083503} (\bibinfo {year} {2016}{\natexlab{a}})},\ \Eprint {http://arxiv.org/abs/1608.00368} {arXiv:1608.00368 [astro-ph.CO]} \BibitemShut {NoStop}%
\bibitem [{\citenamefont {Shiraishi}\ \emph {et~al.}(2019)\citenamefont {Shiraishi}, \citenamefont {Liguori}, \citenamefont {Fergusson},\ and\ \citenamefont {Shellard}}]{Shiraishi:2019exr}%
  \BibitemOpen
  \bibfield  {author} {\bibinfo {author} {\bibfnamefont {M.}~\bibnamefont {Shiraishi}}, \bibinfo {author} {\bibfnamefont {M.}~\bibnamefont {Liguori}}, \bibinfo {author} {\bibfnamefont {J.~R.}\ \bibnamefont {Fergusson}}, \ and\ \bibinfo {author} {\bibfnamefont {E.~P.~S.}\ \bibnamefont {Shellard}},\ }\href {\doibase 10.1088/1475-7516/2019/06/046} {\bibfield  {journal} {\bibinfo  {journal} {JCAP}\ }\textbf {\bibinfo {volume} {06}},\ \bibinfo {pages} {046} (\bibinfo {year} {2019})},\ \Eprint {http://arxiv.org/abs/1904.02599} {arXiv:1904.02599 [astro-ph.CO]} \BibitemShut {NoStop}%
\bibitem [{\citenamefont {Shiraishi}\ \emph {et~al.}(2014)\citenamefont {Shiraishi}, \citenamefont {Liguori},\ and\ \citenamefont {Fergusson}}]{Shiraishi:2014roa}%
  \BibitemOpen
  \bibfield  {author} {\bibinfo {author} {\bibfnamefont {M.}~\bibnamefont {Shiraishi}}, \bibinfo {author} {\bibfnamefont {M.}~\bibnamefont {Liguori}}, \ and\ \bibinfo {author} {\bibfnamefont {J.~R.}\ \bibnamefont {Fergusson}},\ }\href {\doibase 10.1088/1475-7516/2014/05/008} {\bibfield  {journal} {\bibinfo  {journal} {JCAP}\ }\textbf {\bibinfo {volume} {05}},\ \bibinfo {pages} {008} (\bibinfo {year} {2014})},\ \Eprint {http://arxiv.org/abs/1403.4222} {arXiv:1403.4222 [astro-ph.CO]} \BibitemShut {NoStop}%
\bibitem [{\citenamefont {{Fergusson}}\ and\ \citenamefont {{Shellard}}(2009)}]{2009PhRvD..80d3510F}%
  \BibitemOpen
  \bibfield  {author} {\bibinfo {author} {\bibfnamefont {J.~R.}\ \bibnamefont {{Fergusson}}}\ and\ \bibinfo {author} {\bibfnamefont {E.~P.~S.}\ \bibnamefont {{Shellard}}},\ }\href {\doibase 10.1103/PhysRevD.80.043510} {\bibfield  {journal} {\bibinfo  {journal} {\prd}\ }\textbf {\bibinfo {volume} {80}},\ \bibinfo {eid} {043510} (\bibinfo {year} {2009})},\ \Eprint {http://arxiv.org/abs/0812.3413} {arXiv:0812.3413 [astro-ph]} \BibitemShut {NoStop}%
\bibitem [{\citenamefont {Shiraishi}\ \emph {et~al.}(2010)\citenamefont {Shiraishi}, \citenamefont {Yokoyama}, \citenamefont {Ichiki},\ and\ \citenamefont {Takahashi}}]{Shiraishi:2010sm}%
  \BibitemOpen
  \bibfield  {author} {\bibinfo {author} {\bibfnamefont {M.}~\bibnamefont {Shiraishi}}, \bibinfo {author} {\bibfnamefont {S.}~\bibnamefont {Yokoyama}}, \bibinfo {author} {\bibfnamefont {K.}~\bibnamefont {Ichiki}}, \ and\ \bibinfo {author} {\bibfnamefont {K.}~\bibnamefont {Takahashi}},\ }\href {\doibase 10.1103/PhysRevD.82.103505} {\bibfield  {journal} {\bibinfo  {journal} {Phys. Rev. D}\ }\textbf {\bibinfo {volume} {82}},\ \bibinfo {pages} {103505} (\bibinfo {year} {2010})},\ \Eprint {http://arxiv.org/abs/1003.2096} {arXiv:1003.2096 [astro-ph.CO]} \BibitemShut {NoStop}%
\bibitem [{\citenamefont {Shiraishi}(2016{\natexlab{b}})}]{Shiraishi:2016ads}%
  \BibitemOpen
  \bibfield  {author} {\bibinfo {author} {\bibfnamefont {M.}~\bibnamefont {Shiraishi}},\ }\href {\doibase 10.1142/S0217732316400034} {\bibfield  {journal} {\bibinfo  {journal} {Mod. Phys. Lett. A}\ }\textbf {\bibinfo {volume} {31}},\ \bibinfo {pages} {1640003} (\bibinfo {year} {2016}{\natexlab{b}})}\BibitemShut {NoStop}%
\bibitem [{\citenamefont {Kamionkowski}\ and\ \citenamefont {Souradeep}(2011)}]{Kamionkowski:2010rb}%
  \BibitemOpen
  \bibfield  {author} {\bibinfo {author} {\bibfnamefont {M.}~\bibnamefont {Kamionkowski}}\ and\ \bibinfo {author} {\bibfnamefont {T.}~\bibnamefont {Souradeep}},\ }\href {\doibase 10.1103/PhysRevD.83.027301} {\bibfield  {journal} {\bibinfo  {journal} {Phys. Rev. D}\ }\textbf {\bibinfo {volume} {83}},\ \bibinfo {pages} {027301} (\bibinfo {year} {2011})},\ \Eprint {http://arxiv.org/abs/1010.4304} {arXiv:1010.4304 [astro-ph.CO]} \BibitemShut {NoStop}%
\bibitem [{\citenamefont {Kulsrud}\ and\ \citenamefont {Zweibel}(2008)}]{Kulsrud:2007an}%
  \BibitemOpen
  \bibfield  {author} {\bibinfo {author} {\bibfnamefont {R.~M.}\ \bibnamefont {Kulsrud}}\ and\ \bibinfo {author} {\bibfnamefont {E.~G.}\ \bibnamefont {Zweibel}},\ }\href {\doibase 10.1088/0034-4885/71/4/046901} {\bibfield  {journal} {\bibinfo  {journal} {Rept. Prog. Phys.}\ }\textbf {\bibinfo {volume} {71}},\ \bibinfo {pages} {0046091} (\bibinfo {year} {2008})},\ \Eprint {http://arxiv.org/abs/0707.2783} {arXiv:0707.2783 [astro-ph]} \BibitemShut {NoStop}%
\bibitem [{\citenamefont {Paoletti}\ \emph {et~al.}(2009)\citenamefont {Paoletti}, \citenamefont {Finelli},\ and\ \citenamefont {Paci}}]{Paoletti:2008ck}%
  \BibitemOpen
  \bibfield  {author} {\bibinfo {author} {\bibfnamefont {D.}~\bibnamefont {Paoletti}}, \bibinfo {author} {\bibfnamefont {F.}~\bibnamefont {Finelli}}, \ and\ \bibinfo {author} {\bibfnamefont {F.}~\bibnamefont {Paci}},\ }\href {\doibase 10.1111/j.1365-2966.2009.14727.x} {\bibfield  {journal} {\bibinfo  {journal} {Mon. Not. Roy. Astron. Soc.}\ }\textbf {\bibinfo {volume} {396}},\ \bibinfo {pages} {523} (\bibinfo {year} {2009})},\ \Eprint {http://arxiv.org/abs/0811.0230} {arXiv:0811.0230 [astro-ph]} \BibitemShut {NoStop}%
\bibitem [{\citenamefont {Shaw}\ and\ \citenamefont {Lewis}(2010)}]{Shaw:2009nf}%
  \BibitemOpen
  \bibfield  {author} {\bibinfo {author} {\bibfnamefont {J.~R.}\ \bibnamefont {Shaw}}\ and\ \bibinfo {author} {\bibfnamefont {A.}~\bibnamefont {Lewis}},\ }\href {\doibase 10.1103/PhysRevD.81.043517} {\bibfield  {journal} {\bibinfo  {journal} {Phys. Rev. D}\ }\textbf {\bibinfo {volume} {81}},\ \bibinfo {pages} {043517} (\bibinfo {year} {2010})},\ \Eprint {http://arxiv.org/abs/0911.2714} {arXiv:0911.2714 [astro-ph.CO]} \BibitemShut {NoStop}%
\bibitem [{\citenamefont {Pogosian}\ and\ \citenamefont {Zucca}(2018)}]{Pogosian:2018vfr}%
  \BibitemOpen
  \bibfield  {author} {\bibinfo {author} {\bibfnamefont {L.}~\bibnamefont {Pogosian}}\ and\ \bibinfo {author} {\bibfnamefont {A.}~\bibnamefont {Zucca}},\ }\href {\doibase 10.1088/1361-6382/aac398} {\bibfield  {journal} {\bibinfo  {journal} {Class. Quant. Grav.}\ }\textbf {\bibinfo {volume} {35}},\ \bibinfo {pages} {124004} (\bibinfo {year} {2018})},\ \Eprint {http://arxiv.org/abs/1801.08936} {arXiv:1801.08936 [astro-ph.CO]} \BibitemShut {NoStop}%
\bibitem [{\citenamefont {Horndeski}(1974)}]{Horndeski:1974wa}%
  \BibitemOpen
  \bibfield  {author} {\bibinfo {author} {\bibfnamefont {G.~W.}\ \bibnamefont {Horndeski}},\ }\href {\doibase 10.1007/BF01807638} {\bibfield  {journal} {\bibinfo  {journal} {Int. J. Theor. Phys.}\ }\textbf {\bibinfo {volume} {10}},\ \bibinfo {pages} {363} (\bibinfo {year} {1974})}\BibitemShut {NoStop}%
\bibitem [{\citenamefont {Kobayashi}\ \emph {et~al.}(2011)\citenamefont {Kobayashi}, \citenamefont {Yamaguchi},\ and\ \citenamefont {Yokoyama}}]{Kobayashi:2011nu}%
  \BibitemOpen
  \bibfield  {author} {\bibinfo {author} {\bibfnamefont {T.}~\bibnamefont {Kobayashi}}, \bibinfo {author} {\bibfnamefont {M.}~\bibnamefont {Yamaguchi}}, \ and\ \bibinfo {author} {\bibfnamefont {J.}~\bibnamefont {Yokoyama}},\ }\href {\doibase 10.1143/PTP.126.511} {\bibfield  {journal} {\bibinfo  {journal} {Prog. Theor. Phys.}\ }\textbf {\bibinfo {volume} {126}},\ \bibinfo {pages} {511} (\bibinfo {year} {2011})},\ \Eprint {http://arxiv.org/abs/1105.5723} {arXiv:1105.5723 [hep-th]} \BibitemShut {NoStop}%
\bibitem [{\citenamefont {Deffayet}\ \emph {et~al.}(2011)\citenamefont {Deffayet}, \citenamefont {Gao}, \citenamefont {Steer},\ and\ \citenamefont {Zahariade}}]{Deffayet:2011gz}%
  \BibitemOpen
  \bibfield  {author} {\bibinfo {author} {\bibfnamefont {C.}~\bibnamefont {Deffayet}}, \bibinfo {author} {\bibfnamefont {X.}~\bibnamefont {Gao}}, \bibinfo {author} {\bibfnamefont {D.~A.}\ \bibnamefont {Steer}}, \ and\ \bibinfo {author} {\bibfnamefont {G.}~\bibnamefont {Zahariade}},\ }\href {\doibase 10.1103/PhysRevD.84.064039} {\bibfield  {journal} {\bibinfo  {journal} {Phys. Rev. D}\ }\textbf {\bibinfo {volume} {84}},\ \bibinfo {pages} {064039} (\bibinfo {year} {2011})},\ \Eprint {http://arxiv.org/abs/1103.3260} {arXiv:1103.3260 [hep-th]} \BibitemShut {NoStop}%
\bibitem [{\citenamefont {Agrawal}\ \emph {et~al.}(2018{\natexlab{b}})\citenamefont {Agrawal}, \citenamefont {Fujita},\ and\ \citenamefont {Komatsu}}]{Agrawal:2018mrg}%
  \BibitemOpen
  \bibfield  {author} {\bibinfo {author} {\bibfnamefont {A.}~\bibnamefont {Agrawal}}, \bibinfo {author} {\bibfnamefont {T.}~\bibnamefont {Fujita}}, \ and\ \bibinfo {author} {\bibfnamefont {E.}~\bibnamefont {Komatsu}},\ }\href {\doibase 10.1088/1475-7516/2018/06/027} {\bibfield  {journal} {\bibinfo  {journal} {JCAP}\ }\textbf {\bibinfo {volume} {06}},\ \bibinfo {pages} {027} (\bibinfo {year} {2018}{\natexlab{b}})},\ \Eprint {http://arxiv.org/abs/1802.09284} {arXiv:1802.09284 [astro-ph.CO]} \BibitemShut {NoStop}%
\bibitem [{\citenamefont {Creque-Sarbinowski}\ \emph {et~al.}(2023)\citenamefont {Creque-Sarbinowski}, \citenamefont {Alexander}, \citenamefont {Kamionkowski},\ and\ \citenamefont {Philcox}}]{CyrilCS}%
  \BibitemOpen
  \bibfield  {author} {\bibinfo {author} {\bibfnamefont {C.}~\bibnamefont {Creque-Sarbinowski}}, \bibinfo {author} {\bibfnamefont {S.}~\bibnamefont {Alexander}}, \bibinfo {author} {\bibfnamefont {M.}~\bibnamefont {Kamionkowski}}, \ and\ \bibinfo {author} {\bibfnamefont {O.}~\bibnamefont {Philcox}},\ }\href {\doibase 10.1088/1475-7516/2023/11/029} {\bibfield  {journal} {\bibinfo  {journal} {JCAP}\ }\textbf {\bibinfo {volume} {11}},\ \bibinfo {pages} {029} (\bibinfo {year} {2023})},\ \Eprint {http://arxiv.org/abs/2303.04815} {arXiv:2303.04815 [astro-ph.CO]} \BibitemShut {NoStop}%
\bibitem [{\citenamefont {Dimastrogiovanni}\ \emph {et~al.}(2018{\natexlab{b}})\citenamefont {Dimastrogiovanni}, \citenamefont {Fasiello}, \citenamefont {Hardwick}, \citenamefont {Assadullahi}, \citenamefont {Koyama},\ and\ \citenamefont {Wands}}]{Dimastrogiovanni:2018xnn}%
  \BibitemOpen
  \bibfield  {author} {\bibinfo {author} {\bibfnamefont {E.}~\bibnamefont {Dimastrogiovanni}}, \bibinfo {author} {\bibfnamefont {M.}~\bibnamefont {Fasiello}}, \bibinfo {author} {\bibfnamefont {R.~J.}\ \bibnamefont {Hardwick}}, \bibinfo {author} {\bibfnamefont {H.}~\bibnamefont {Assadullahi}}, \bibinfo {author} {\bibfnamefont {K.}~\bibnamefont {Koyama}}, \ and\ \bibinfo {author} {\bibfnamefont {D.}~\bibnamefont {Wands}},\ }\href {\doibase 10.1088/1475-7516/2018/11/029} {\bibfield  {journal} {\bibinfo  {journal} {JCAP}\ }\textbf {\bibinfo {volume} {11}},\ \bibinfo {pages} {029} (\bibinfo {year} {2018}{\natexlab{b}})},\ \Eprint {http://arxiv.org/abs/1806.05474} {arXiv:1806.05474 [astro-ph.CO]} \BibitemShut {NoStop}%
\bibitem [{\citenamefont {Mata}\ \emph {et~al.}(2013)\citenamefont {Mata}, \citenamefont {Raju},\ and\ \citenamefont {Trivedi}}]{Mata:2012bx}%
  \BibitemOpen
  \bibfield  {author} {\bibinfo {author} {\bibfnamefont {I.}~\bibnamefont {Mata}}, \bibinfo {author} {\bibfnamefont {S.}~\bibnamefont {Raju}}, \ and\ \bibinfo {author} {\bibfnamefont {S.}~\bibnamefont {Trivedi}},\ }\href {\doibase 10.1007/JHEP07(2013)015} {\bibfield  {journal} {\bibinfo  {journal} {JHEP}\ }\textbf {\bibinfo {volume} {07}},\ \bibinfo {pages} {015} (\bibinfo {year} {2013})},\ \Eprint {http://arxiv.org/abs/1211.5482} {arXiv:1211.5482 [hep-th]} \BibitemShut {NoStop}%
\bibitem [{\citenamefont {Komatsu}\ \emph {et~al.}(2005)\citenamefont {Komatsu}, \citenamefont {Spergel},\ and\ \citenamefont {Wandelt}}]{Komatsu:2003iq}%
  \BibitemOpen
  \bibfield  {author} {\bibinfo {author} {\bibfnamefont {E.}~\bibnamefont {Komatsu}}, \bibinfo {author} {\bibfnamefont {D.~N.}\ \bibnamefont {Spergel}}, \ and\ \bibinfo {author} {\bibfnamefont {B.~D.}\ \bibnamefont {Wandelt}},\ }\href {\doibase 10.1086/491724} {\bibfield  {journal} {\bibinfo  {journal} {Astrophys. J.}\ }\textbf {\bibinfo {volume} {634}},\ \bibinfo {pages} {14} (\bibinfo {year} {2005})},\ \Eprint {http://arxiv.org/abs/astro-ph/0305189} {arXiv:astro-ph/0305189} \BibitemShut {NoStop}%
\bibitem [{\citenamefont {Philcox}(2025{\natexlab{b}})}]{Philcox4pt3}%
  \BibitemOpen
  \bibfield  {author} {\bibinfo {author} {\bibfnamefont {O.~H.~E.}\ \bibnamefont {Philcox}},\ }\href@noop {} {\  (\bibinfo {year} {2025}{\natexlab{b}})},\ \Eprint {http://arxiv.org/abs/2502.06931} {arXiv:2502.06931 [astro-ph.CO]} \BibitemShut {NoStop}%
\bibitem [{\citenamefont {Carron}\ \emph {et~al.}(2022)\citenamefont {Carron}, \citenamefont {Mirmelstein},\ and\ \citenamefont {Lewis}}]{Carron:2022eyg}%
  \BibitemOpen
  \bibfield  {author} {\bibinfo {author} {\bibfnamefont {J.}~\bibnamefont {Carron}}, \bibinfo {author} {\bibfnamefont {M.}~\bibnamefont {Mirmelstein}}, \ and\ \bibinfo {author} {\bibfnamefont {A.}~\bibnamefont {Lewis}},\ }\href {\doibase 10.1088/1475-7516/2022/09/039} {\bibfield  {journal} {\bibinfo  {journal} {JCAP}\ }\textbf {\bibinfo {volume} {09}},\ \bibinfo {pages} {039} (\bibinfo {year} {2022})},\ \Eprint {http://arxiv.org/abs/2206.07773} {arXiv:2206.07773 [astro-ph.CO]} \BibitemShut {NoStop}%
\bibitem [{\citenamefont {Coulton}\ and\ \citenamefont {Spergel}(2019)}]{Coulton:2019bnz}%
  \BibitemOpen
  \bibfield  {author} {\bibinfo {author} {\bibfnamefont {W.~R.}\ \bibnamefont {Coulton}}\ and\ \bibinfo {author} {\bibfnamefont {D.~N.}\ \bibnamefont {Spergel}},\ }\href {\doibase 10.1088/1475-7516/2019/10/056} {\bibfield  {journal} {\bibinfo  {journal} {JCAP}\ }\textbf {\bibinfo {volume} {10}},\ \bibinfo {pages} {056} (\bibinfo {year} {2019})},\ \Eprint {http://arxiv.org/abs/1901.04515} {arXiv:1901.04515 [astro-ph.CO]} \BibitemShut {NoStop}%
\bibitem [{\citenamefont {{Smith}}\ \emph {et~al.}(2015)\citenamefont {{Smith}}, \citenamefont {{Senatore}},\ and\ \citenamefont {{Zaldarriaga}}}]{2015arXiv150200635S}%
  \BibitemOpen
  \bibfield  {author} {\bibinfo {author} {\bibfnamefont {K.~M.}\ \bibnamefont {{Smith}}}, \bibinfo {author} {\bibfnamefont {L.}~\bibnamefont {{Senatore}}}, \ and\ \bibinfo {author} {\bibfnamefont {M.}~\bibnamefont {{Zaldarriaga}}},\ }\href@noop {} {\bibfield  {journal} {\bibinfo  {journal} {arXiv e-prints}\ ,\ \bibinfo {eid} {arXiv:1502.00635}} (\bibinfo {year} {2015})},\ \Eprint {http://arxiv.org/abs/1502.00635} {arXiv:1502.00635 [astro-ph.CO]} \BibitemShut {NoStop}%
\bibitem [{\citenamefont {Bucher}\ \emph {et~al.}(2016)\citenamefont {Bucher}, \citenamefont {Racine},\ and\ \citenamefont {van Tent}}]{Bucher:2015ura}%
  \BibitemOpen
  \bibfield  {author} {\bibinfo {author} {\bibfnamefont {M.}~\bibnamefont {Bucher}}, \bibinfo {author} {\bibfnamefont {B.}~\bibnamefont {Racine}}, \ and\ \bibinfo {author} {\bibfnamefont {B.}~\bibnamefont {van Tent}},\ }\href {\doibase 10.1088/1475-7516/2016/05/055} {\bibfield  {journal} {\bibinfo  {journal} {JCAP}\ }\textbf {\bibinfo {volume} {05}},\ \bibinfo {pages} {055} (\bibinfo {year} {2016})},\ \Eprint {http://arxiv.org/abs/1509.08107} {arXiv:1509.08107 [astro-ph.CO]} \BibitemShut {NoStop}%
\bibitem [{\citenamefont {Bucher}\ \emph {et~al.}(2010)\citenamefont {Bucher}, \citenamefont {Van~Tent},\ and\ \citenamefont {Carvalho}}]{Bucher:2009nm}%
  \BibitemOpen
  \bibfield  {author} {\bibinfo {author} {\bibfnamefont {M.}~\bibnamefont {Bucher}}, \bibinfo {author} {\bibfnamefont {B.}~\bibnamefont {Van~Tent}}, \ and\ \bibinfo {author} {\bibfnamefont {C.~S.}\ \bibnamefont {Carvalho}},\ }\href {\doibase 10.1111/j.1365-2966.2010.17089.x} {\bibfield  {journal} {\bibinfo  {journal} {Mon. Not. Roy. Astron. Soc.}\ }\textbf {\bibinfo {volume} {407}},\ \bibinfo {pages} {2193} (\bibinfo {year} {2010})},\ \Eprint {http://arxiv.org/abs/0911.1642} {arXiv:0911.1642 [astro-ph.CO]} \BibitemShut {NoStop}%
\bibitem [{\citenamefont {{Smith}}\ and\ \citenamefont {{Zaldarriaga}}(2011)}]{2011MNRAS.417....2S}%
  \BibitemOpen
  \bibfield  {author} {\bibinfo {author} {\bibfnamefont {K.~M.}\ \bibnamefont {{Smith}}}\ and\ \bibinfo {author} {\bibfnamefont {M.}~\bibnamefont {{Zaldarriaga}}},\ }\href {\doibase 10.1111/j.1365-2966.2010.18175.x} {\bibfield  {journal} {\bibinfo  {journal} {\mnras}\ }\textbf {\bibinfo {volume} {417}},\ \bibinfo {pages} {2} (\bibinfo {year} {2011})},\ \Eprint {http://arxiv.org/abs/astro-ph/0612571} {arXiv:astro-ph/0612571 [astro-ph]} \BibitemShut {NoStop}%
\bibitem [{\citenamefont {Philcox}(2025{\natexlab{c}})}]{Philcox4pt1}%
  \BibitemOpen
  \bibfield  {author} {\bibinfo {author} {\bibfnamefont {O.~H.~E.}\ \bibnamefont {Philcox}},\ }\href@noop {} {\  (\bibinfo {year} {2025}{\natexlab{c}})},\ \Eprint {http://arxiv.org/abs/2502.04434} {arXiv:2502.04434 [astro-ph.CO]} \BibitemShut {NoStop}%
\bibitem [{\citenamefont {G\'orski}\ \emph {et~al.}(2005)\citenamefont {G\'orski}, \citenamefont {Hivon}, \citenamefont {Banday}, \citenamefont {Wandelt}, \citenamefont {Hansen}, \citenamefont {Reinecke},\ and\ \citenamefont {Bartelman}}]{Gorski:2004by}%
  \BibitemOpen
  \bibfield  {author} {\bibinfo {author} {\bibfnamefont {K.~M.}\ \bibnamefont {G\'orski}}, \bibinfo {author} {\bibfnamefont {E.}~\bibnamefont {Hivon}}, \bibinfo {author} {\bibfnamefont {A.~J.}\ \bibnamefont {Banday}}, \bibinfo {author} {\bibfnamefont {B.~D.}\ \bibnamefont {Wandelt}}, \bibinfo {author} {\bibfnamefont {F.~K.}\ \bibnamefont {Hansen}}, \bibinfo {author} {\bibfnamefont {M.}~\bibnamefont {Reinecke}}, \ and\ \bibinfo {author} {\bibfnamefont {M.}~\bibnamefont {Bartelman}},\ }\href {\doibase 10.1086/427976} {\bibfield  {journal} {\bibinfo  {journal} {Astrophys. J.}\ }\textbf {\bibinfo {volume} {622}},\ \bibinfo {pages} {759} (\bibinfo {year} {2005})},\ \Eprint {http://arxiv.org/abs/astro-ph/0409513} {arXiv:astro-ph/0409513} \BibitemShut {NoStop}%
\bibitem [{\citenamefont {Goldberg}\ and\ \citenamefont {Spergel}(1999)}]{Goldberg:1999xm}%
  \BibitemOpen
  \bibfield  {author} {\bibinfo {author} {\bibfnamefont {D.~M.}\ \bibnamefont {Goldberg}}\ and\ \bibinfo {author} {\bibfnamefont {D.~N.}\ \bibnamefont {Spergel}},\ }\href {\doibase 10.1103/PhysRevD.59.103002} {\bibfield  {journal} {\bibinfo  {journal} {Phys. Rev. D}\ }\textbf {\bibinfo {volume} {59}},\ \bibinfo {pages} {103002} (\bibinfo {year} {1999})},\ \Eprint {http://arxiv.org/abs/astro-ph/9811251} {arXiv:astro-ph/9811251} \BibitemShut {NoStop}%
\bibitem [{\citenamefont {Hu}(2000)}]{Hu:2000ee}%
  \BibitemOpen
  \bibfield  {author} {\bibinfo {author} {\bibfnamefont {W.}~\bibnamefont {Hu}},\ }\href {\doibase 10.1103/PhysRevD.62.043007} {\bibfield  {journal} {\bibinfo  {journal} {Phys. Rev. D}\ }\textbf {\bibinfo {volume} {62}},\ \bibinfo {pages} {043007} (\bibinfo {year} {2000})},\ \Eprint {http://arxiv.org/abs/astro-ph/0001303} {arXiv:astro-ph/0001303} \BibitemShut {NoStop}%
\bibitem [{\citenamefont {Lewis}\ \emph {et~al.}(2011)\citenamefont {Lewis}, \citenamefont {Challinor},\ and\ \citenamefont {Hanson}}]{Lewis:2011fk}%
  \BibitemOpen
  \bibfield  {author} {\bibinfo {author} {\bibfnamefont {A.}~\bibnamefont {Lewis}}, \bibinfo {author} {\bibfnamefont {A.}~\bibnamefont {Challinor}}, \ and\ \bibinfo {author} {\bibfnamefont {D.}~\bibnamefont {Hanson}},\ }\href {\doibase 10.1088/1475-7516/2011/03/018} {\bibfield  {journal} {\bibinfo  {journal} {JCAP}\ }\textbf {\bibinfo {volume} {03}},\ \bibinfo {pages} {018} (\bibinfo {year} {2011})},\ \Eprint {http://arxiv.org/abs/1101.2234} {arXiv:1101.2234 [astro-ph.CO]} \BibitemShut {NoStop}%
\bibitem [{\citenamefont {Philcox}\ and\ \citenamefont {Hill}(2025)}]{Philcox:2025lxt}%
  \BibitemOpen
  \bibfield  {author} {\bibinfo {author} {\bibfnamefont {O.~H.~E.}\ \bibnamefont {Philcox}}\ and\ \bibinfo {author} {\bibfnamefont {J.~C.}\ \bibnamefont {Hill}},\ }\href@noop {} {\  (\bibinfo {year} {2025})},\ \Eprint {http://arxiv.org/abs/2504.03826} {arXiv:2504.03826 [astro-ph.CO]} \BibitemShut {NoStop}%
\bibitem [{\citenamefont {Lacasa}\ \emph {et~al.}(2014)\citenamefont {Lacasa}, \citenamefont {P\'enin},\ and\ \citenamefont {Aghanim}}]{Lacasa:2013yya}%
  \BibitemOpen
  \bibfield  {author} {\bibinfo {author} {\bibfnamefont {F.}~\bibnamefont {Lacasa}}, \bibinfo {author} {\bibfnamefont {A.}~\bibnamefont {P\'enin}}, \ and\ \bibinfo {author} {\bibfnamefont {N.}~\bibnamefont {Aghanim}},\ }\href {\doibase 10.1093/mnras/stt2373} {\bibfield  {journal} {\bibinfo  {journal} {Mon. Not. Roy. Astron. Soc.}\ }\textbf {\bibinfo {volume} {439}},\ \bibinfo {pages} {123} (\bibinfo {year} {2014})},\ \Eprint {http://arxiv.org/abs/1312.1251} {arXiv:1312.1251 [astro-ph.CO]} \BibitemShut {NoStop}%
\bibitem [{\citenamefont {P\'enin}\ \emph {et~al.}(2014)\citenamefont {P\'enin}, \citenamefont {Lacasa},\ and\ \citenamefont {Aghanim}}]{Penin:2013zya}%
  \BibitemOpen
  \bibfield  {author} {\bibinfo {author} {\bibfnamefont {A.}~\bibnamefont {P\'enin}}, \bibinfo {author} {\bibfnamefont {F.}~\bibnamefont {Lacasa}}, \ and\ \bibinfo {author} {\bibfnamefont {N.}~\bibnamefont {Aghanim}},\ }\href {\doibase 10.1093/mnras/stt2372} {\bibfield  {journal} {\bibinfo  {journal} {Mon. Not. Roy. Astron. Soc.}\ }\textbf {\bibinfo {volume} {439}},\ \bibinfo {pages} {143} (\bibinfo {year} {2014})},\ \Eprint {http://arxiv.org/abs/1312.1252} {arXiv:1312.1252 [astro-ph.CO]} \BibitemShut {NoStop}%
\bibitem [{\citenamefont {Hill}(2018)}]{Hill:2018ypf}%
  \BibitemOpen
  \bibfield  {author} {\bibinfo {author} {\bibfnamefont {J.~C.}\ \bibnamefont {Hill}},\ }\href {\doibase 10.1103/PhysRevD.98.083542} {\bibfield  {journal} {\bibinfo  {journal} {Phys. Rev. D}\ }\textbf {\bibinfo {volume} {98}},\ \bibinfo {pages} {083542} (\bibinfo {year} {2018})},\ \Eprint {http://arxiv.org/abs/1807.07324} {arXiv:1807.07324 [astro-ph.CO]} \BibitemShut {NoStop}%
\bibitem [{\citenamefont {Coulton}\ \emph {et~al.}(2023)\citenamefont {Coulton}, \citenamefont {Miranthis},\ and\ \citenamefont {Challinor}}]{Coulton:2022wln}%
  \BibitemOpen
  \bibfield  {author} {\bibinfo {author} {\bibfnamefont {W.}~\bibnamefont {Coulton}}, \bibinfo {author} {\bibfnamefont {A.}~\bibnamefont {Miranthis}}, \ and\ \bibinfo {author} {\bibfnamefont {A.}~\bibnamefont {Challinor}},\ }\href {\doibase 10.1093/mnras/stad1305} {\bibfield  {journal} {\bibinfo  {journal} {Mon. Not. Roy. Astron. Soc.}\ }\textbf {\bibinfo {volume} {523}},\ \bibinfo {pages} {825} (\bibinfo {year} {2023})},\ \Eprint {http://arxiv.org/abs/2208.12270} {arXiv:2208.12270 [astro-ph.CO]} \BibitemShut {NoStop}%
\bibitem [{\citenamefont {M\"unchmeyer}\ and\ \citenamefont {Smith}(2019)}]{Munchmeyer:2019kng}%
  \BibitemOpen
  \bibfield  {author} {\bibinfo {author} {\bibfnamefont {M.}~\bibnamefont {M\"unchmeyer}}\ and\ \bibinfo {author} {\bibfnamefont {K.~M.}\ \bibnamefont {Smith}},\ }\href@noop {} {\  (\bibinfo {year} {2019})},\ \Eprint {http://arxiv.org/abs/1905.05846} {arXiv:1905.05846 [astro-ph.CO]} \BibitemShut {NoStop}%
\end{thebibliography}%

\end{document}